\documentclass[a4paper,11pt]{article}
\pdfoutput=1 

\usepackage{jcappub} 

\usepackage[T1]{fontenc}

\title{Voids in massive neutrino cosmologies}

\author[a,b]{Elena Massara,}
\author[c,b]{Francisco Villaescusa-Navarro,}
\author[c,b]{Matteo Viel,}
\author[c,b,d]{P. M. Sutter }

\affiliation[a]{SISSA- International School for Advanced Studies, Via Bonomea 265, I-34136 Trieste, Italy}
\affiliation[b]{INFN-National Institute for Nuclear Physics, via Valerio 2, I-34127 Trieste, Italy}
\affiliation[c]{INAF-Osservatorio Astronomico di Trieste, via Tiepolo 11, I-34143 Trieste, Italy}
\affiliation[d]{Center for Cosmology and Astro-Particle Physics, Ohio State University, Columbus, OH 43210}

\emailAdd{emassara@sissa.it}
\emailAdd{villaescusa@oats.inaf.it}
\emailAdd{viel@oats.inaf.it}
\emailAdd{sutter@oats.inaf.it}

\abstract{Cosmic voids are a promising environment to characterize neutrino-induced effects on the large-scale distribution of matter in the universe. We perform a comprehensive numerical study of the statistical properties of voids, identified both in the matter and galaxy distributions, in massive and massless neutrino cosmologies. The matter density field is obtained by running several independent $N$-body simulations with cold dark matter and neutrino particles, while the galaxy catalogs are modeled by populating the dark matter halos in simulations via a halo occupation distribution (HOD) model to reproduce the clustering properties observed by the Sloan Digital Sky Survey (SDSS) II Data Release 7. We focus on the impact of massive neutrinos on the following void statistical properties: number density, ellipticities, two-point statistics, density and velocity profiles. Considering the matter density field, we find that voids in massive neutrino cosmologies are less evolved than those in the corresponding massless neutrinos case: there is a larger number of small voids and a smaller number of large ones, their profiles are less evacuated, and they present a lower wall at the edge. Moreover, the degeneracy between $\sigma_8$ and $\Omega_{\nu}$ is broken when looking at void properties. In terms of the galaxy density field, we find that differences among cosmologies are difficult to detect because of the small number of galaxy voids in the simulations. Differences are instead present when looking at the matter density and velocity profiles around these voids.}

\begin{document}
\maketitle
%\flush bottom

%%%%%%%%%%%%%%%%%%%%%%%%%%%%%%%%%%%%%%%%%%%%%%
\section{Introduction}

$N$-body simulations  of structure formation show that the spatial distribution of matter resembles a network: the so-called {\it  cosmic web}. In this web, knots represent regions of high density, known as dark matter halos, connected by filaments, i.e. relatively thin and low density regions. Finally, filaments are usually surrounded by large regions with very low densities: voids. 

Besides the anisotropies in the cosmic microwave background, the vast majority of the constraints on the values of cosmological parameters arise from the observations of only one element of the cosmic web: the dark matter halos (or the galaxies residing within them). Much attention has been put in understanding the spatial distribution and statistics of galaxies and dark matter halos. This is because it is believed that those objects are biased tracers of the underlying matter distribution. Thus, 
by measuring their spatial correlations on large scales it is possible to constrain the values of the cosmological parameters. Additional information on the value of the cosmological parameters can be obtained by measuring the abundance of dark matter halos as a function of their mass, i.e. via the halo mass function (see for example \cite{Borgani_2000}).

Recently, it has been pointed out that cosmic voids can also be used to constrain the values of  cosmological parameters \cite{Viel_voids_2008,Stark_Lyalpha_void,Lee_2009,Lavaux_2012,Ilic_2013,Villaescusa-Navarro_2012,Pisani_2015,Zivick_2014,Cai_2014,Cai_2013,Szapudi_2014}. As a different element of the cosmic web, the information embedded into the properties of voids will complement the one obtained from the dark matter halos. That additional information can then be used to break degeneracies and to further constrain the parameters of the cosmological model. Among the different observables that can be used to retrieve cosmological information with voids there are: the distribution of void ellipticities \cite{Lee_2009}, the Alcock-Paczynski test \cite{Lavaux_2012,Cai_2013,Szapudi_2014}, the integrated Sachs-Wolfe effect \cite{Ilic_2013} and the weak lensing effect \cite{Kaiser_1992_WL}. 

However, cold dark matter and baryons are not the only massive particles present in the universe that could play a role in the formation of structure, and in particular in the evolution of voids: there are also neutrinos.
The standard model of particle physics describe neutrinos as fermions of spin 1/2, neutral and massless particles. However, neutrino oscillations experiments have pointed out that at least two out of the three neutrino families are massive (see \cite{Fogli:2005cq,Maltoni:2004ei} for review). These observations are one of the clearest evidences pointing towards the existence of physics beyond the standard model. Thus, one of the most important questions of modern physics is to know the masses and hierarchy of the three neutrino families. 

The existence of a neutrino cosmic background is one of the predictions of the Big Bang theory. The presence of a hot component in the total matter density field affects cosmological observables in many different ways. At the linear level massive neutrinos modify the matter-radiation equality time and the growth of matter perturbations \cite{LesgourguesPastor,LesgourguesBook}. Those effects are commonly used to put upper limits on the sum of the neutrino masses from a variety of cosmological observables such as the anisotropies in the cosmic microwave background (CMB), the spatial clustering of galaxies, and the clustering properties of neutral hydrogen absorbers in the Ly$\alpha$ forest \cite{Hannestad_2003,Reid,Thomas,Swanson,Saito_2010,dePutter,Xia2012,WiggleZ,Zhao2012,Costanzi,Basse,Planck_2013,Costanzi_2014,Wyman_2013,Battye_2013,Hamann_2013,Beutler_2014,Giusarma_2014,Palanque-Delabrouille:2014jca,Dell'Oro_2015,Palanque-Delabrouille:2015pga}. The upper limits on the sum of the neutrino masses obtained from cosmological observables ($\sum m_\nu<0.12$ eV at 95\% C.L. \cite{Palanque-Delabrouille:2015pga}) are much tighter than those achieved from laboratory experiments. 

Therefore, cosmological observables are currently the best probe to constrain neutrino masses and thus, it is timely to investigate the impact of massive neutrinos on the large scale structure. The impact of massive neutrinos on different non-linear observables such as the matter power spectrum, the abundance and spatial distribution of dark matter halos and the galaxy properties has been recently addressed by a number of authors \cite{Agarwal2011,Bird_2011,Wagner2012,Viel_2010,Yacine-Bird,Baldi_2013,Paco_2013,Castorina_2013,LoVerde_2014a,LoVerde_2014b,Ma,Wong,Brandbyge_2008,Brandbyge_haloes,Paco_2011,Paco_2012,LoVerde_2013,Marulli_2011,Ichiki_Takada,Costanzi_2013, Fontanot_2014,Massara_2014}. 
  
There are reasons to expect that massive neutrinos affect more strongly the properties of cosmic voids than the properties of dark matter halos. Given their large thermal velocities, neutrino clustering within dark matter halos and galaxies will be limited to the neutrinos populating the low-momentum tail of the Fermi-Dirac distribution \cite{Paco_2012}. In other words, the relative contribution of relic neutrinos to the total mass of dark matter halos is much smaller than the cosmic ratio $\Omega_\nu/\Omega_{\rm cdm}$. A consequence of this fact is that most of the properties of the dark matter halos, such as their mass function or their spatial bias, can be described in terms of the CDM density field, rather than the total matter density field \cite{Ichiki_Takada,Paco_2013,Castorina_2013,Costanzi_2013,LoVerde_2014a,LoVerde_2014b,Biagetti_2014}.
On the other hand, the large thermal velocities of the neutrinos will prevent their evacuation from cosmic voids. This should manifest as an extra mass within voids that will affect their overall evolution. It is thus expected that voids in a massive neutrino universe would be smaller and denser. This fact was firstly noted by \cite{Villaescusa-Navarro_2012}, who studied the properties of voids in massive neutrino cosmologies using the Ly$\alpha$ forest. 

In this paper we investigate for the first time the impact of massive neutrinos on cosmic voids, as identified in the spatial distribution of galaxies and in the underlying matter density field. We study the properties of voids in the spatial distribution of matter by running large box  $N$-body simulations in cosmologies with massless and massive neutrinos. The spatial distribution of galaxies is modeled using a halo occupation distribution (HOD) model and requiring that the resulting mock galaxy catalogues reproduce, for a given galaxy population, the observed number density and two-point correlation function of a particular survey. We use the publicly-available code {\sc VIDE} \cite{VIDE} to identify the voids in both the matter and galaxy distribution. The void catalogs obtained with this void finder - and watershed void finder in general, to be discussed below - depend on the resolution of the N-body simulation considered. However, in this work we are interested in the relative differences between cosmologies with and without massive neutrinos, rather than absolute results. Therefore, we compare catalogs and void properties obtained from simulations at the same resolution.

This paper is organized as follows. In Sec. \ref{sec:N-body} we describe the N-body simulations we have run for this project. The method used to identify voids in both the matter and the galaxy fields is briefly outlined in Sec. \ref{sec:finder}. A comparison between the different properties of voids detected in the matter density field in the different cosmological models is shown in Sec. \ref{sec:voids_matter}. In Sec. \ref{sec:voids_galaxies} we present the properties of voids identified in the galaxy distribution. Finally, we draw the main conclusions in Sec. \ref{sec:conclusions}.

%%%%%%%%%%%%%%%%%%%%%%%%%%%%%%%%%%%%%%%%%%%%%%
\section{N-body simulations}
\label{sec:N-body}

We ran the $N$-body simulations using the TreePM code {\sc GADGET-III} \cite{Springel_2005}. The simulation parameters are summarized in table~\ref{tab:i}. 
The values of the cosmological parameters, for all simulations with the exception of the L0s8 and of the L0s8CDM runs, are the ones found by the Planck collaboration \cite{Planck_2015}: $\Omega_{\rm m}=0.3175$, $\Omega_{\rm b}=0.049$, $\Omega_\Lambda=0.6825$, $h=0.6711$, $n_s=0.9624$, $A_s=2.13\times10^{-9}$. We have run simulations for four different cosmological models: a model with massless neutrinos and three models with massive neutrinos (three degenerate species) with masses $\sum m_\nu=0.15~{\rm eV}$, $\sum m_\nu=0.3~{\rm eV}$ and $\sum m_\nu=0.6~{\rm eV}$. In massive neutrinos cosmologies we set $\Omega_{\rm cdm}=\Omega_{\rm m}-\Omega_{\rm b}-\Omega_\nu$ where $\Omega_\nu h^2\cong\sum m_{\nu}/(94.1~{\rm eV})$. We chose these values for the neutrino masses because we want to study and isolate the effect of massive neutrinos on voids properties. In simulations with the sum of the neutrino masses lower than current tightest bounds $\sum m_\nu<0.12$ eV \cite{Palanque-Delabrouille:2015pga} the effect of massive neutrinos may not have been seen properly or it would have been largely contaminated by cosmic variance. Note further that some groups are claiming that concordance cosmology between Planck CMB data and cluster abundance or weak lensing could be achieved by allowing a non-zero neutrino mass of 0.3 eV ($\pm 0.1$ eV) (see however the discussion in the latest Planck cosmological parameter paper \cite{Planck_2015}). In order to investigate the $\sum m_\nu-\sigma_8$ degeneracy, we have also run simulations with massless neutrinos, but with a value of $\sigma_8$ equal to the one of the $\sum m_\nu=0.6$ eV model. The $\sigma_8$ value can be computed from the total matter or from the CDM power spectrum; we denote the simulations with the former as L0s8, the ones with the latter as L0s8CDM.

In this paper we investigate the properties of voids as identified in the matter and galaxy density fields, in cosmologies with massless and massive neutrinos. Our simulations can thus be divided in two different categories: 1) simulations with low resolution and large volumes, used to detect voids in the matter distribution and 2) simulations with relatively high resolution and low volumes, used to isolate voids in the spatial distribution of galaxies.

The low resolution simulations consist of ten independent realizations, for the same cosmological model, each of them following the evolution of $256^3$ CDM particles (and $256^3$ neutrinos for massive neutrinos cosmologies) within a periodic box of size 1000 Mpc$/h$. We chose to work with such a low number of particles because it is difficult to run the void finder VIDE on more particles, due to the large memory required. The high resolution simulations only contain one single realization, which traces the evolution of $512^3$ CDM particles (and $512^3$ neutrinos) in a box of 500 Mpc$/h$ size. In all simulations the gravitational softening is set to $1/40$ of the mean linear inter-particle distance.

We generate the initial conditions of the simulations at $z=99$ by displacing the particle positions, that are initially located in a regular grid, making use of the Zel'dovich approximation. In simulations with massive neutrinos we use the massive neutrinos transfer function and a mass weighted average between the CDM and baryons transfer functions to set up the initial conditions of the neutrino and the CDM particles, respectively. 
In massless neutrinos simulations we use the total matter transfer function for setting up the initial conditions of the CDM particles. Notice that, in massive neutrino cosmologies, the displacements and the peculiar velocities of the CDM and the neutrino particles in the initial conditions have been generated taken into account the scale-dependent growth present in these cosmologies.

\begin{table}
\begin{center}
\begin{tabular}{| c | c | c | c | c | c | c | c |}
\hline
Name & Box Size & $\sum m_\nu$ & $N_{\rm cdm}$ & $N_\nu$ & $\epsilon$& $\sigma_8$ & realizations\\
&  (${\rm Mpc}/h$) & (eV) & & & (${\rm kpc}/h$) & (z=0) &\\
\hline
L0 & 1000 & 0.00 & $256^3$ & $0$ & 100 &0.834 & 10\\
\hline
L0s8 & 1000 & 0.00 & $256^3$ & $0$ & 100 &0.693 & 10\\
\hline
L0s8CDM & 1000 & 0.00 & $256^3$ & $0$ & 100 &0.717 & 10\\
\hline
L15 & 1000 & 0.15 & $256^3$ & $256^3$ & 100 & 0.801 & 10\\
\hline
L30 & 1000 & 0.30 & $256^3$ & $256^3$ & 100 & 0.764 & 10\\
\hline
L60 & 1000 & 0.60 & $256^3$ & $256^3$ & 100 & 0.693 & 10\\
\hline
H0 & 500 & 0.00 & $512^3$ & $0$ & 25 & 0.834 & 1\\
\hline
H15 & 500 & 0.15 & $512^3$ & $512^3$ & 25 & 0.801 & 1\\
\hline
H30 & 500 & 0.30 & $512^3$ & $512^3$ & 25 & 0.764 & 1\\
\hline
H60 & 500 & 0.60 & $512^3$ & $512^3$ & 25 & 0.693 & 1\\
\hline
\end{tabular}
\caption{\label{tab:i}Specifications of our N-body simulation suite. The first letter of the simulation name indicates whether it is a low-resolution (L) or high-resolution (H) simulation. The values of the following cosmological parameters are the same for all the simulations (with the exception of the simulation L0s8 in which $A_s=1.473\times10^{-9}$ and L0s8CDM where $A_s=1.576\times10^{-9}$): $\Omega_{\rm m}=0.3175$, $\Omega_{\rm b}=0.049$, $\Omega_\Lambda=0.6825$, $h=0.6711$, $A_s=2.13\times10^{-9}$, $n_s=0.9624$.}
\end{center}
\end{table}

%%%%%%%%%%%%%%%%%%%%%%%%%%%%%%%%%%%%%%%%%%%%%%
\section{Void Finder}
\label{sec:finder}

We identified voids using the publicly available Void Identification
and Examination ({\sc VIDE}) toolkit~\cite{VIDE}, which uses a
substantially modified version of ZOBOV~\cite{ZOBOV} to perform a
Voronoi tessellation of the particles and then a watershed
transform to group the Voronoi cells into a hierarchical tree of
subvoids and voids~\cite{Watershed}. A final catalog of voids is
then built using the criteria that the voids must be larger than the
mean particle separation, so for our analysis we will only consider
voids larger than 4.0 Mpc/$h$ in the CDM field (low resolution simulations) and voids larger than 10 Mpc/h in the galaxy distribution (high resolution simulations). 
While voids naturally form a nested
hierarchy, we only consider top-level (i.e., parent) voids in this
work.

When identifying voids in the matter density field, we run {\sc VIDE} on top of the cold dark matter particles only, even if we are considering cosmologies with massive neutrinos. In principle we should select voids in the total matter field, including also neutrinos. However, {\sc VIDE} cannot discriminate between two different particle populations having two different masses. Therefore we must select voids in one of the two density fields and the CDM is the one mainly responsible for the evolution of the cosmic structures. It would be interesting to understand the differences arising in defining voids in the CDM or in the total matter field and which of the two is the fundamental field to identify voids. When selecting voids in the galaxy distribution we instead run {\sc VIDE} on top of the whole mock galaxy catalogue.

The output of {\sc VIDE} is a catalog which contains much information about the identified voids. In this work we use only the following:  position of the center, particle members, effective radius ($R_{\rm eff}$) and ellipticity. The effective radius $R_{\rm eff}$ is defined as the radius of a sphere containing the same volume as the watershed region that delimits the void, and the center is the volume-weighted center of all the Voronoi cells in each void.

%%%%%%%%%%%%%%%%%%%%%%%%%%%%%%%%%%%%%%%%%%%%%%
\section{Void in the matter distribution}
\label{sec:voids_matter}

In this section we present the analysis for voids identified in the matter density field, which is modeled via our low resolution $N$-body simulations. As explained in the above section, voids are selected in the matter density field by running {\sc VIDE} on top of all CDM particles, for every realization of each cosmological model. Our goal is to investigate the impact of massive neutrinos on the formation and evolution of cosmic voids. Thus, here we study some of the main properties of voids and examine how these depend on the cosmological model. The void properties analyzed are: number densities, ellipticities, two-point correlation functions, density profiles, and velocity profiles.

\subsection{Number density}
\label{subsec:number_density_matter}

\begin{figure}[tbp]
\centering %\begin{center}/\end{center} takes some additional vertical space
\includegraphics[width=.49\textwidth,clip]{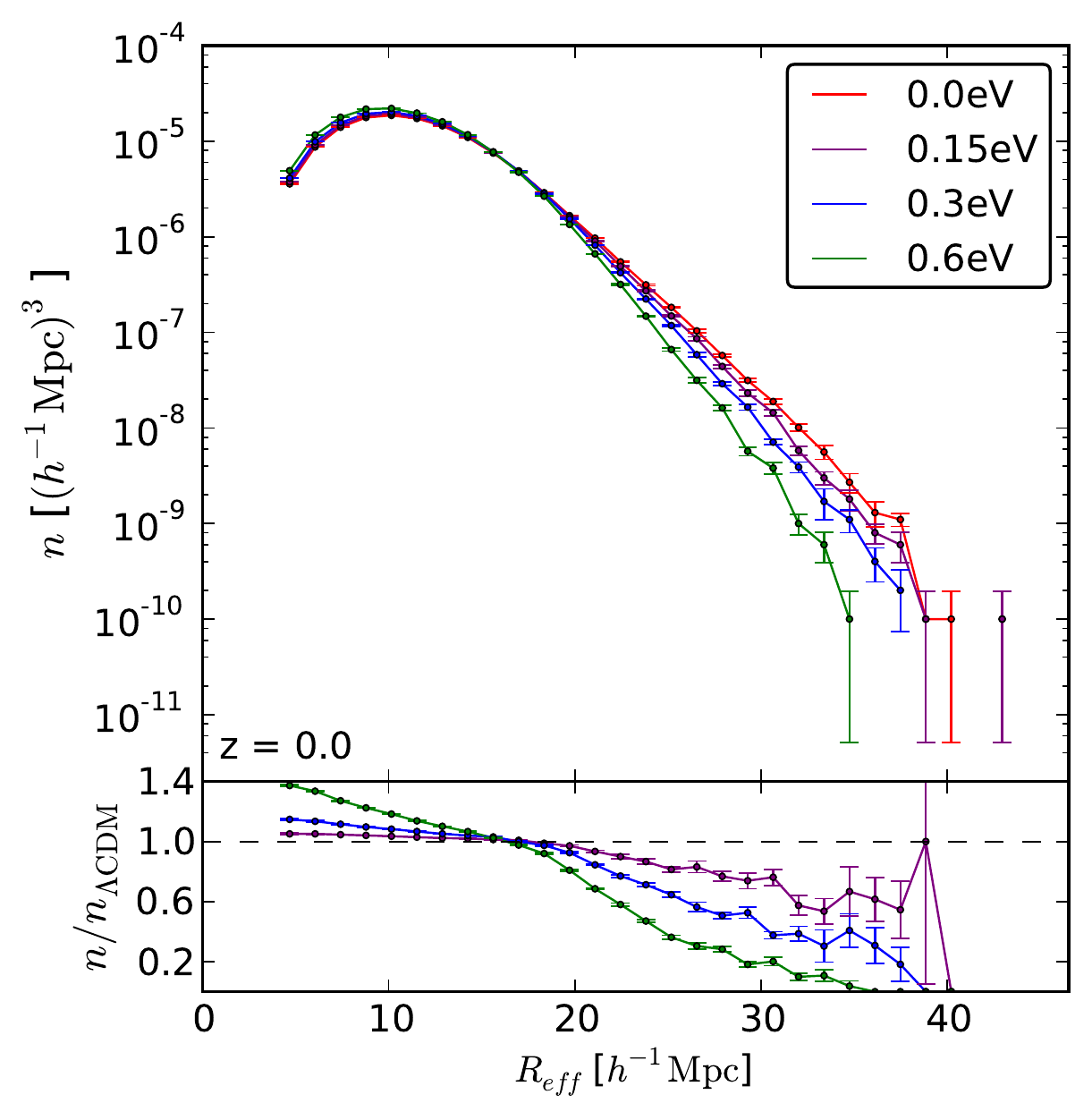}
\includegraphics[width=.49\textwidth,origin=c]{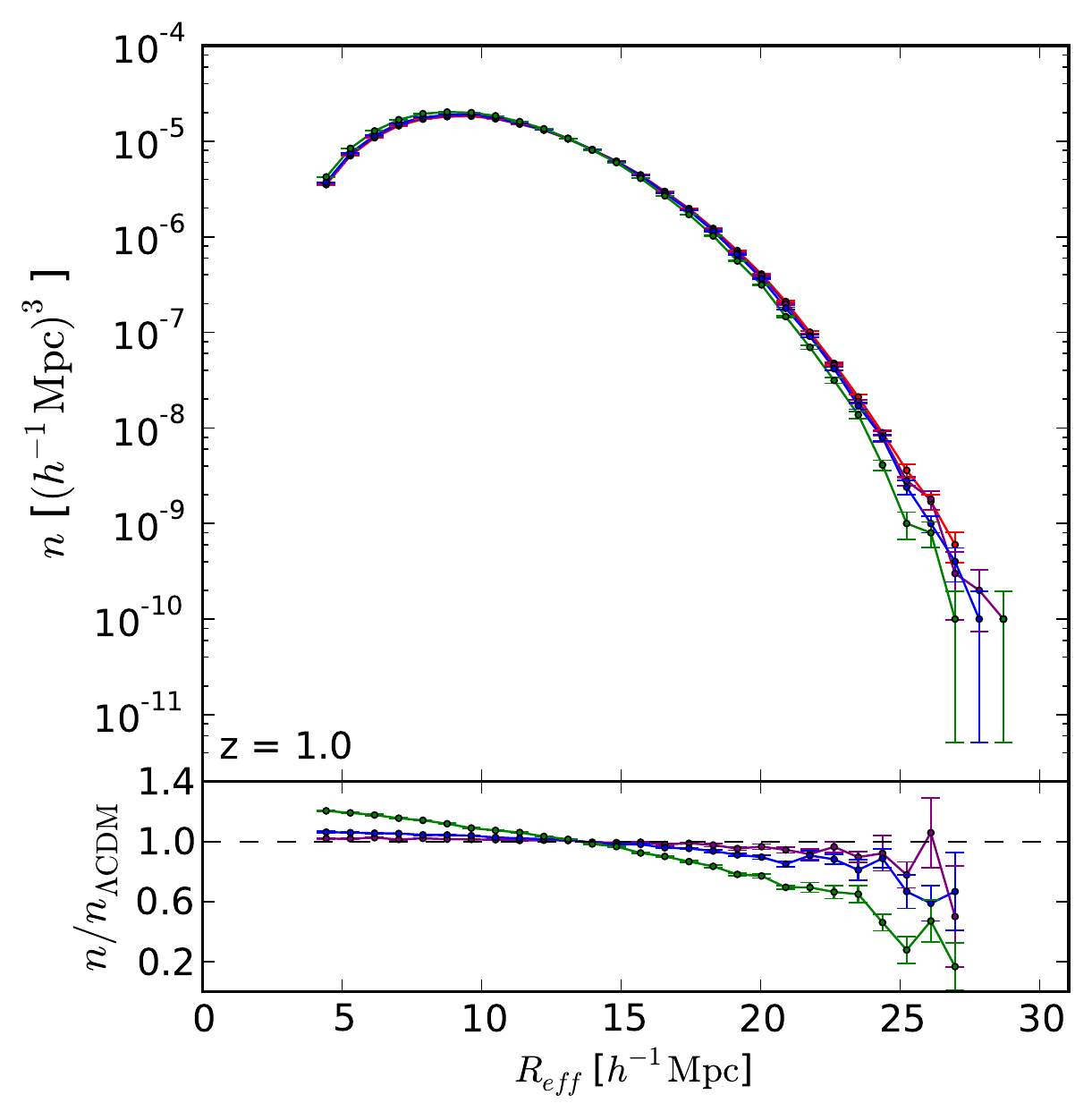}
%\hfill
% "\includegraphics" is very powerful; the graphicx package is already loaded
\caption{\label{fig:number_density_voids} Number density of voids as function of their effective radius. The top panels show the results for different cosmologies: red line is the $\Lambda$CDM cosmology, purple, blue and green lines correspond to $\sum m_\nu=0.15$ eV, $\sum m_\nu=0.3$ eV and $\sum m_\nu=0.6$ eV cosmologies, respectively. The bottom panels show the ratio between the void number density in massive and massless neutrino cosmologies. Left and right plots display results at redshift $z=0$ and $z=1$, respectively. }
\end{figure}

We compute the number density of voids at $z=0$ and $z=1$ as a function of their effective radius $R_{\rm eff}$, and show the results in Fig.~\ref{fig:number_density_voids}. Different colors indicate different cosmologies while error bars represent the scatter around the mean value obtained from the 10 independent realizations divided by $\sqrt{10}$. The bottom panel of each plot displays the ratio between the number density of voids in massive to massless neutrinos cosmologies. We find that in cosmologies with massive neutrinos the abundance of small voids is larger than in massless neutrinos cosmologies, whereas the number density of big voids is highly suppressed in 
massive neutrino cosmologies. These trends take place both at $z=0$ and $z=1$, while relative differences are smaller at $z=1$.

These results are in perfect agreement with our expectations: neutrinos have large thermal velocities, which gives rise to two main effects. On one hand it avoids their clustering within dark matter halos, and on the other hand it makes them less sensitive to void dynamics. Therefore, as a first approximation, the neutrinos density contrast can be approximated as $\delta_{\nu}=0$ and the matter density as $\delta_{\rm m} = \delta_{\rm cdm} \rho_{\rm cdm}/ \rho_{\rm m} + \delta_{\nu} \rho_{\nu}/ \rho_{\rm m} \simeq \delta_{\rm cdm} \rho_{\rm cdm}/ \rho_{\rm m}$, where the subscripts 'cdm' and 'm' stand for CDM and matter, respectively. This brings an extra mass inside the voids that will slow down their overall evolution. It is thus expected that voids in a massive neutrino universe would be smaller and denser, and therefore appear less-evolved. Moreover, the higher the neutrino mass the higher the ratio $\rho_{\nu}/ \rho_{\rm m}$ and the additional mass inside the voids. Therefore the relative differences with respect to $\Lambda$CDM increases with the neutrino mass. We also expect differences to become smaller at higher redshift. At higher redshifts voids will be denser overall, and the additional mass given by neutrinos will play a less important role and the void number density will be more similar among different cosmologies.

We conclude that the number density of voids, identified in the matter density field, is very sensitive to the neutrino masses.

\subsection{Ellipicity}

\begin{figure}[tbp]
\centering %\begin{center}/\end{center} takes some additional vertical space
\includegraphics[width=.49\textwidth,clip]{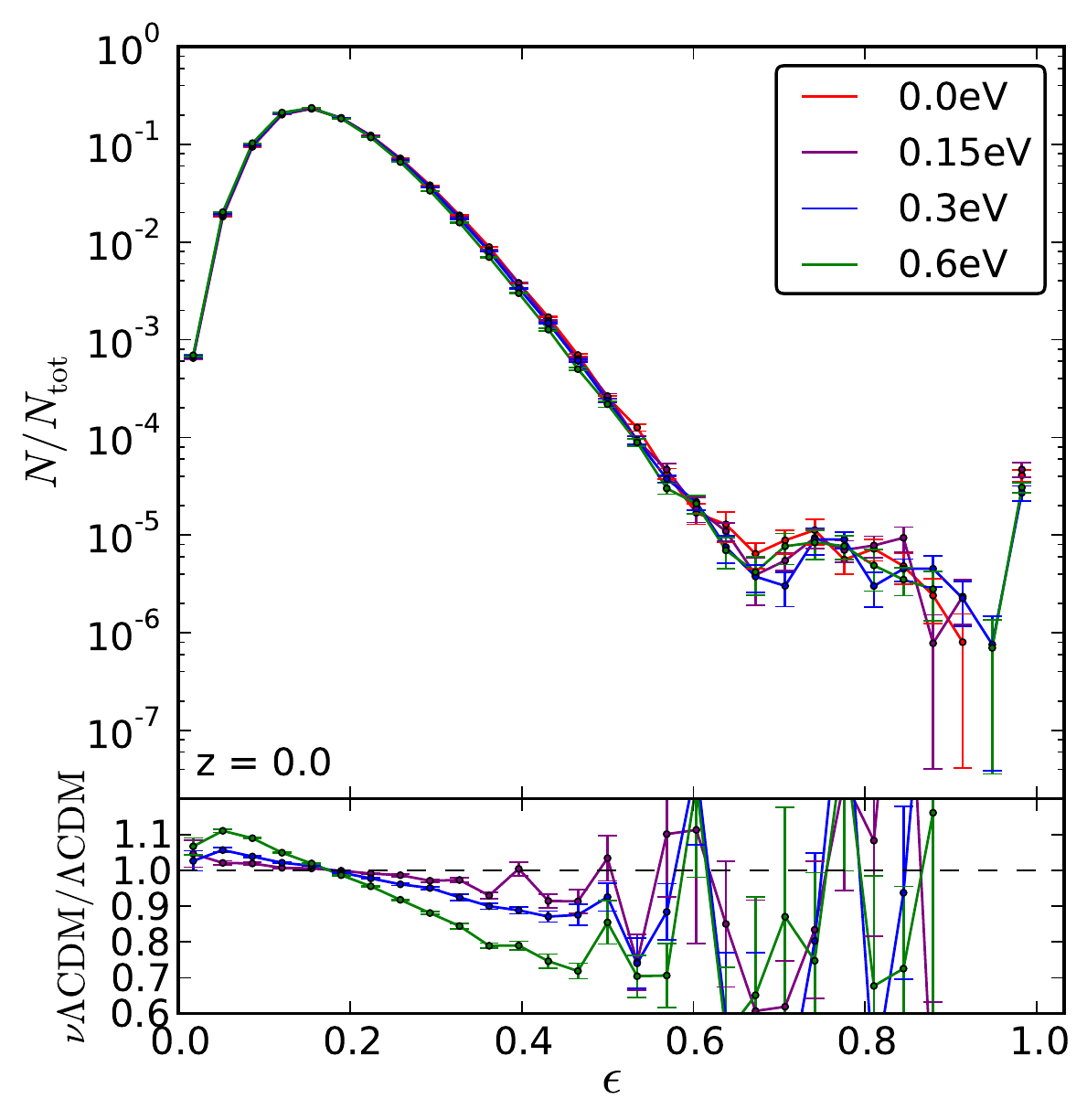}
\includegraphics[width=.49\textwidth,origin=c]{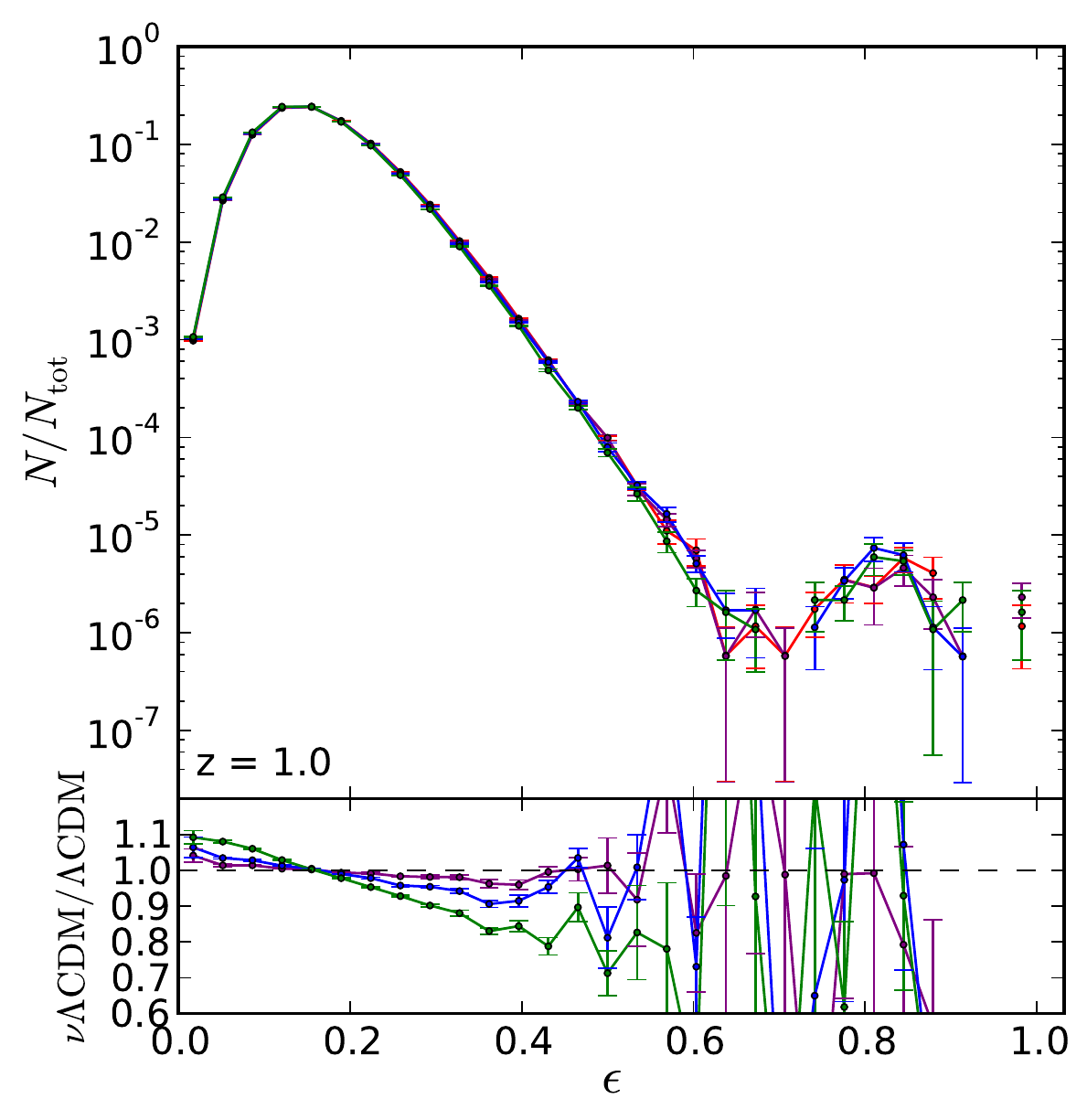}
%\hfill
% "\includegraphics" is very powerful; the graphicx package is already loaded
\caption{\label{fig:ellipticity_voids} Number of voids with a given ellipticity, normalized by the total number of voids. The top panels show results from different cosmologies: red line is the $\Lambda$CDM cosmology, purple, blue and green lines correspond to $\sum m_\nu=0.15$ eV, $\sum m_\nu=0.3$ eV and $\sum m_\nu=0.6$ eV cosmologies. The bottom panels show the ratio between the same quantity in a massive and a massless neutrino cosmologies. Left and right plots display results at redshift $z=0$ and $z=1$, respectively. }
\end{figure}

We now investigate the distribution of voids ellipticities in cosmologies with massive and massless neutrinos. For each cosmological model we compute the fraction of voids with a given ellipticity and show the results in Fig. \ref{fig:ellipticity_voids} at redshifts $z=0$ (left) and $z=1$ (right). The error bars represent the scatter around the mean value obtained from the 10 independent realizations normalized by $\sqrt{10}$. The bottom panels show the ratio between the models with massive neutrinos to the model with massless neutrinos. We find differences of the order of a few percent between $\Lambda$CDM and 0.15 eV cosmologies for void with ellipticities $\epsilon < 0.4$, where statistical error bars are relatively small. The differences increase as the sum of neutrino masses increases and it reaches  20$\%$ for the 0.6 eV cosmology for both redshifts $z=0$ and $z=1$. We find that relative differences among models slightly decrease at $z=1$.

With our interpretation that voids in massive neutrino cosmologies appear younger than those in massless neutrino cosmologies, it is straightforward to understand the results. Voids in cosmologies with massive neutrinos will be in a earlier evolutionary stage and therefore we would expect to find more voids with low ellipticities than in a $\Lambda$CDM cosmology. For the same reason, there is a deficit of voids with large ellipticites in massive neutrino cosmologies.

\subsection{Correlation function}

\begin{figure}[tbp]
\centering %\begin{center}/\end{center} takes some additional vertical space
\includegraphics[width=.49\textwidth,clip]{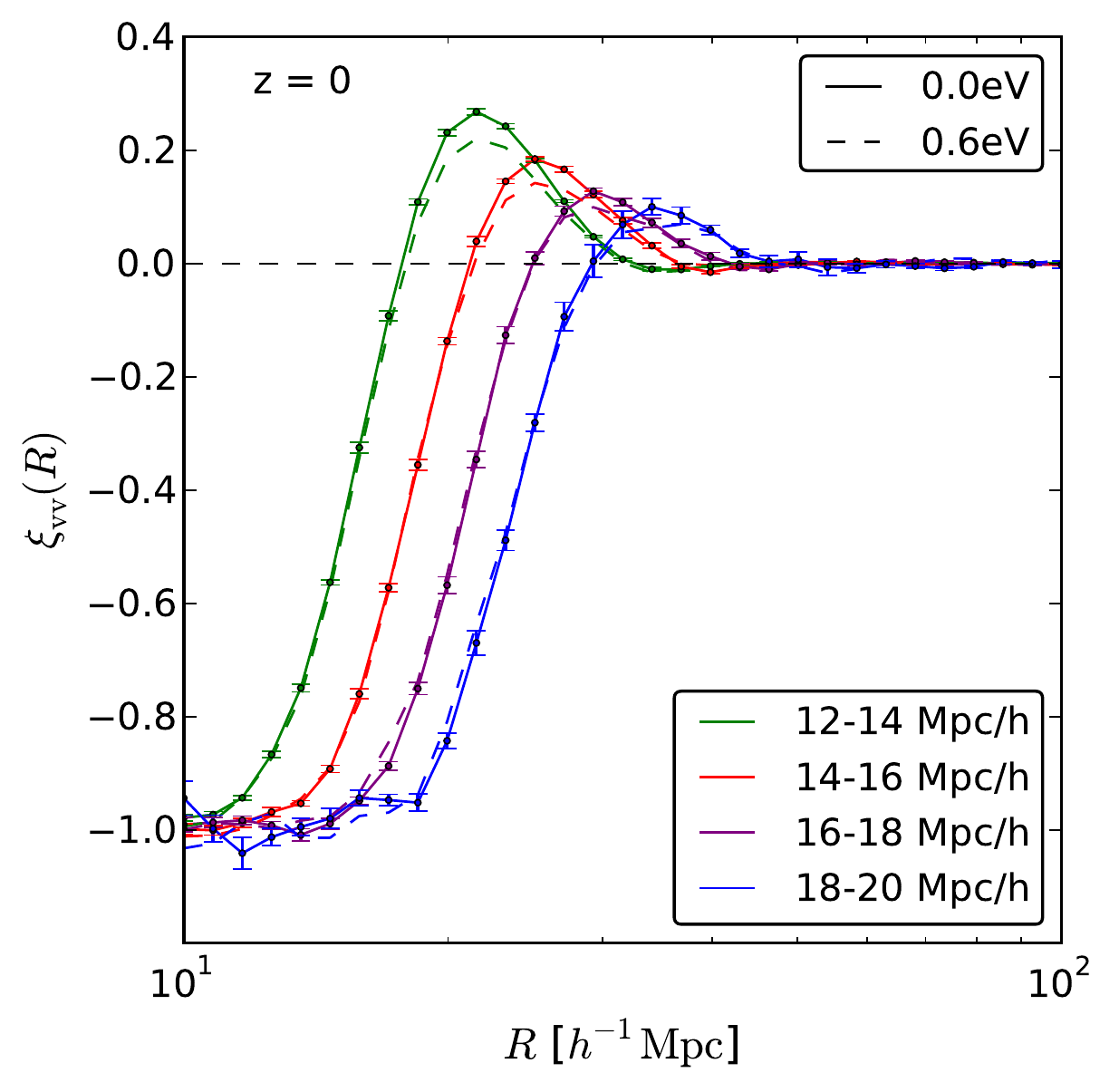}
\includegraphics[width=.49\textwidth,origin=c]{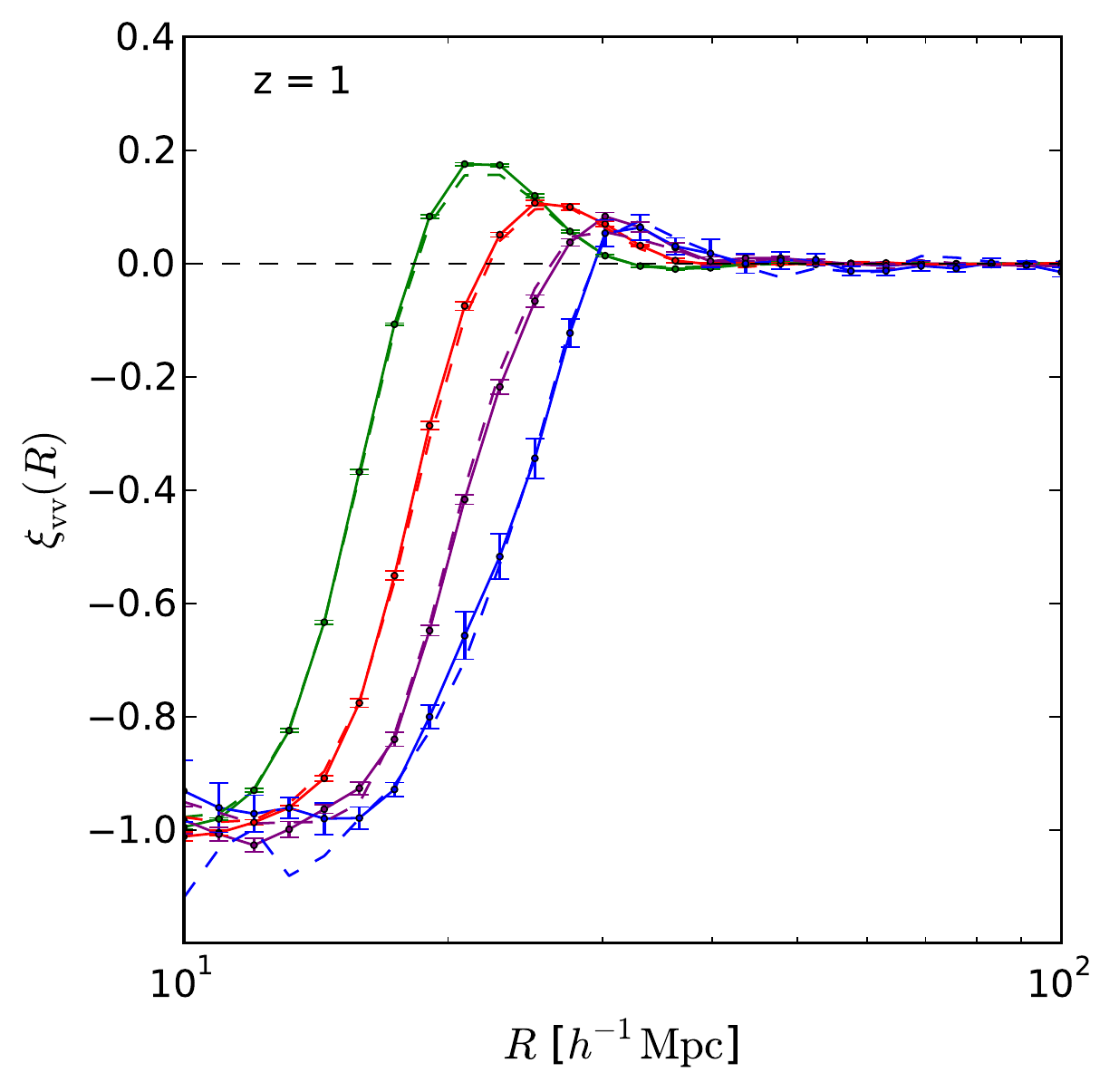}
%\hfill
% "\includegraphics" is very powerful; the graphicx package is already loaded
\caption{\label{fig:correlation_voids} Two-point correlation function of voids with an effective radius in a given range. Solid lines show the results for the $\Lambda$CDM cosmology whereas the dashed lines indicate the results for the $\sum m_\nu=0.6$ eV cosmology; different colors are related to different bins in the void radii as reported in the legend. Left and right plots display results at redshift $z=0$ and $z=1$, respectively. }
\end{figure}

The structure of the void correlation function has been well studied (for example, in \cite{Hamaus_correlation_2013}), which identified the void-exclusion scale and the effects of galaxy bias. Here we investigate the clustering properties of voids in the different cosmological models. For each cosmology, we split our void catalogues in different groups depending on their size $R_{\rm eff}$ and compute the correlation function of the void centers belonging to the same group. The correlation function is measured using the Landy-Szalay estimator \cite{Landy-Szalay}. The random catalogue contains a number of points 20 times larger than the void catalogue.

Figure~\ref{fig:correlation_voids} shows the results at redshift $z=0$ (on the left) and $z=1$ (on the right), with error bars given by the scatter around the mean value obtained from the 10 independent realizations divided by $\sqrt{10}$.
For clarity we show the correlation functions for only two cosmologies: the solid lines refer to $\Lambda$CDM and the dashed ones refer to $\sum m_\nu=0.6$ eV cosmology. Different colors indicate different ranges in $R_{\rm eff}$. The results show that, for both cosmologies, the two-point correlation function goes to zero on large scales. It increases as $R$ decreases and it reaches a maximum, whose height and position depend on the size of the voids considered. Finally, going towards smaller distances, the correlation function decreases until it reaches the lowest boundary -1. 

The behavior of the correlation function on small scales is due to exclusion effects: top-level voids are extended objects that cannot overlap. In fact, if we imagine them as spheres of radius $R_{\rm eff}$, the probability of finding two void centers at a distance smaller than the sum of their effective radii (exclusion scale) is zero, which corresponds to a value of the correlation function equal to -1. We would naively expect to have a sharp transition from the positive correlation to the exclusion regime. However, this is not the case for several reasons. First of all, the presence of different void sizes in each of the considered groups. Secondly, the void finder used in this paper does not return spherical voids, but under-dense regions with a complicated shape, and therefore it can happen that the distance between two void centers is smaller than the sum of their two effective radii. It can be seen that the exclusion scale increases with the effective radius of the voids considered.

The presence of a positive peak in the two-point correlation function arises from two different processes: the rise of the non-linear clustering of voids  going to smaller scales and the exclusion effect. We can also understand why the clustering of voids in massive neutrino cosmologies is lower with respect to the massless neutrino case if we take into account that voids in the former are younger than those in the latter. Indeed, if we compare the clustering of voids at $z=0$ and at $z=1$, we find that the clustering decreases with redshift. Thus, since voids in the massive neutrino cosmologies are effectively younger, they are expected to exhibit a lower amplitude in the 2pt-correlation function, as our results show. At higher redshifts the properties of voids in cosmologies with massive and massless neutrinos get closer, and therefore it is natural that differences in their clustering properties decrease.

\subsection{Density profile}
\label{sec:density_profile}
\begin{figure}[!tbp]
\centering %\begin{center}/\end{center} takes some additional vertical space
\includegraphics[width=.49\textwidth,clip]{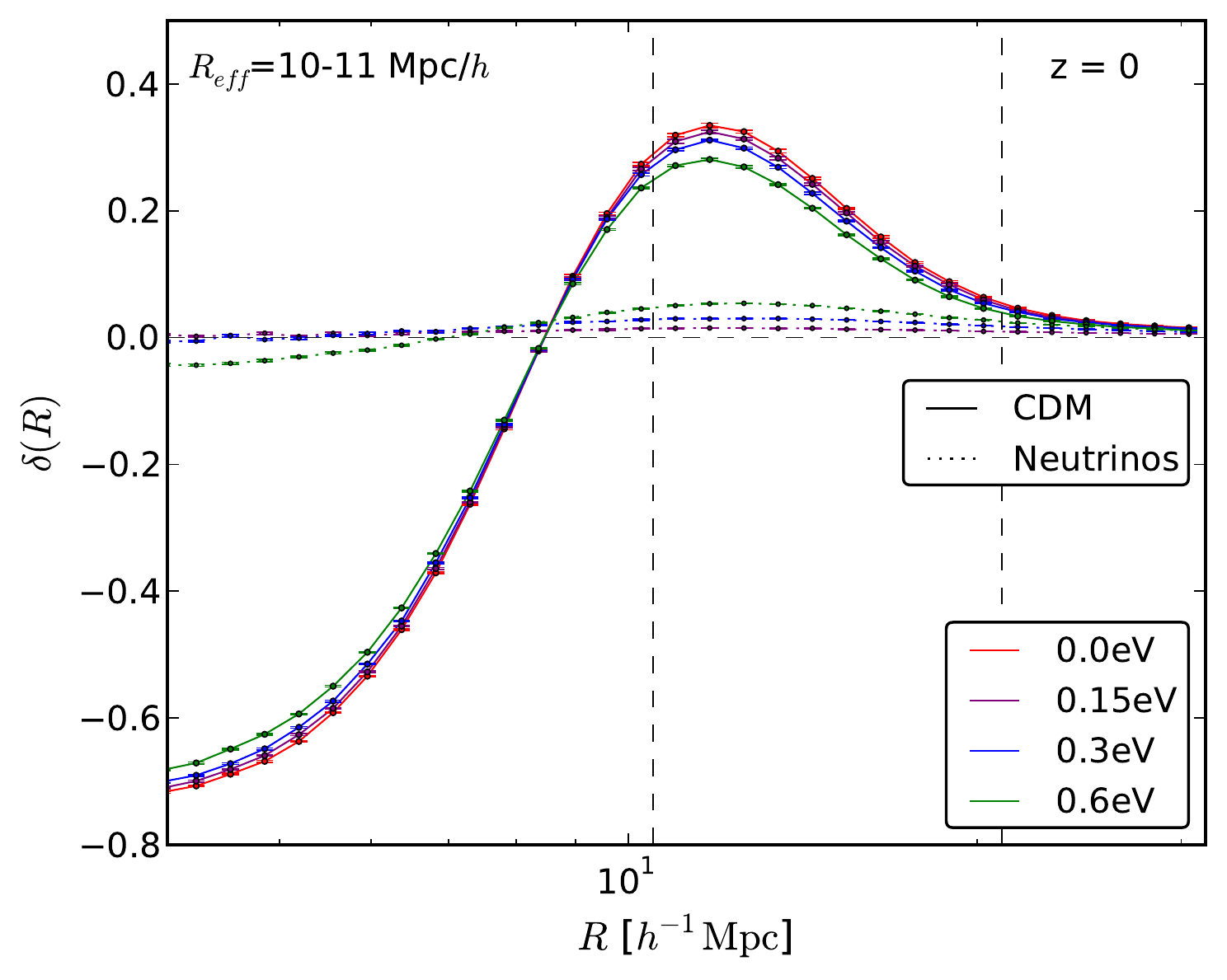}
%\hfill
\includegraphics[width=.49\textwidth,origin=c]{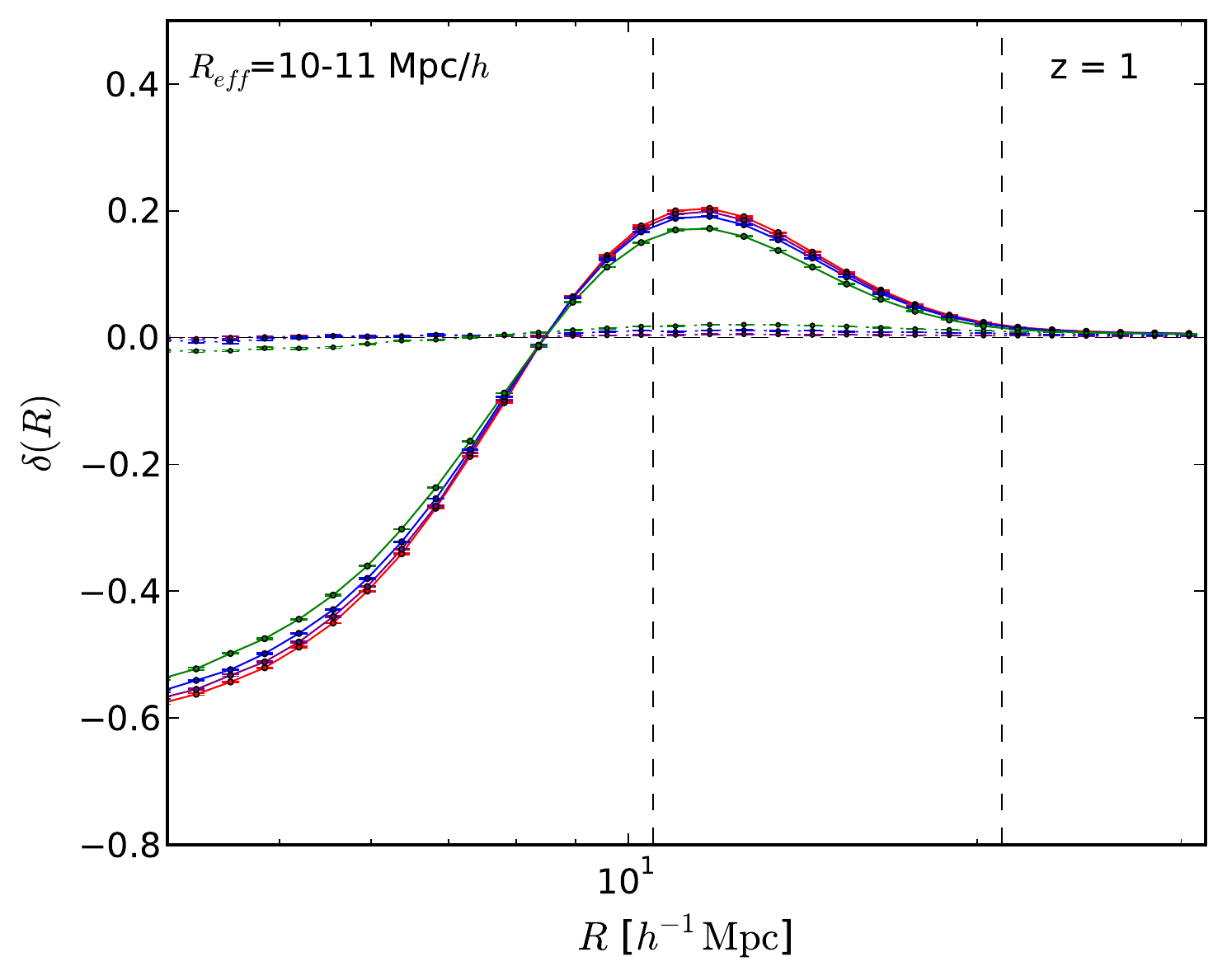}
\includegraphics[width=.49\textwidth,clip]{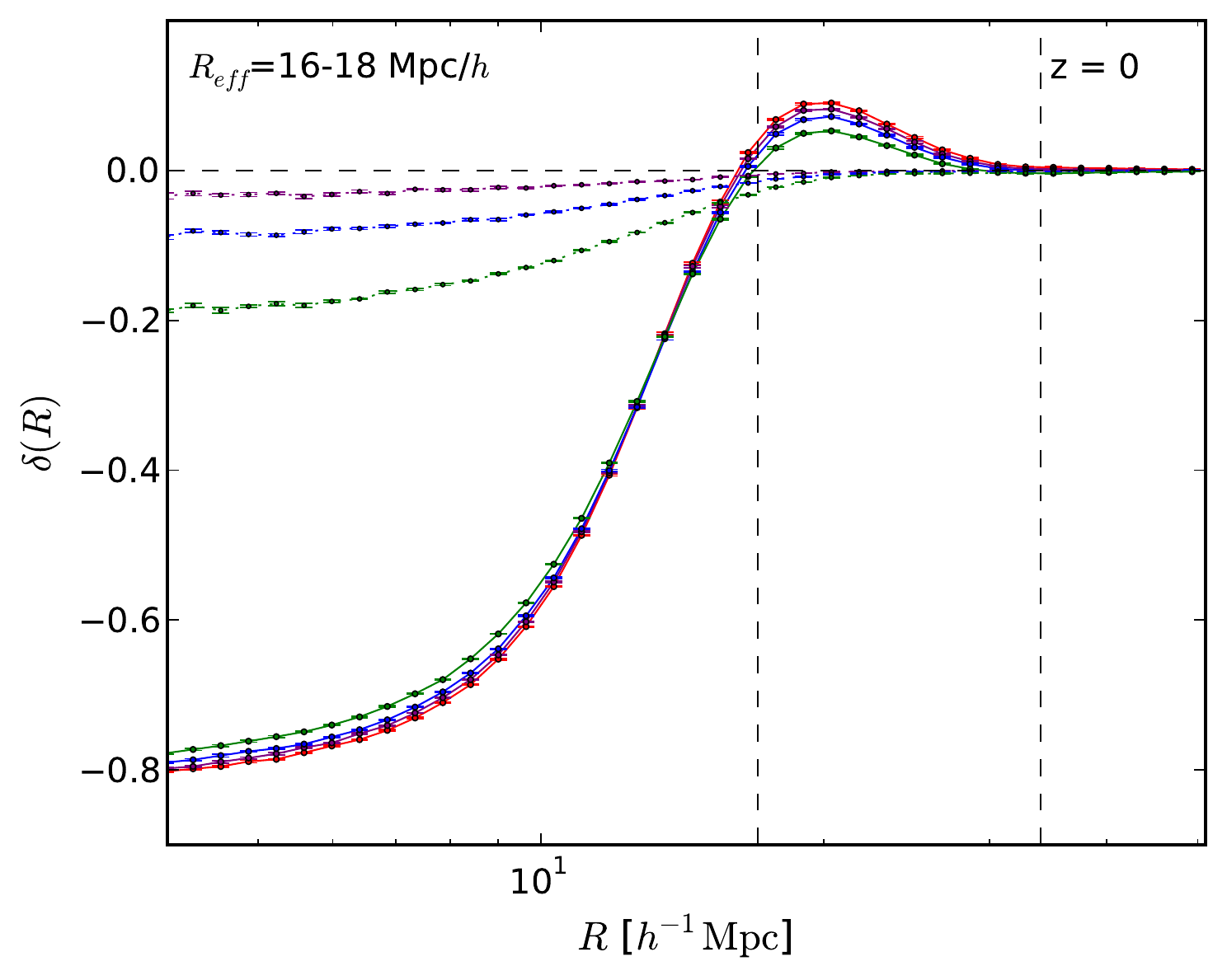}
%\hfill
\includegraphics[width=.49\textwidth,origin=c]{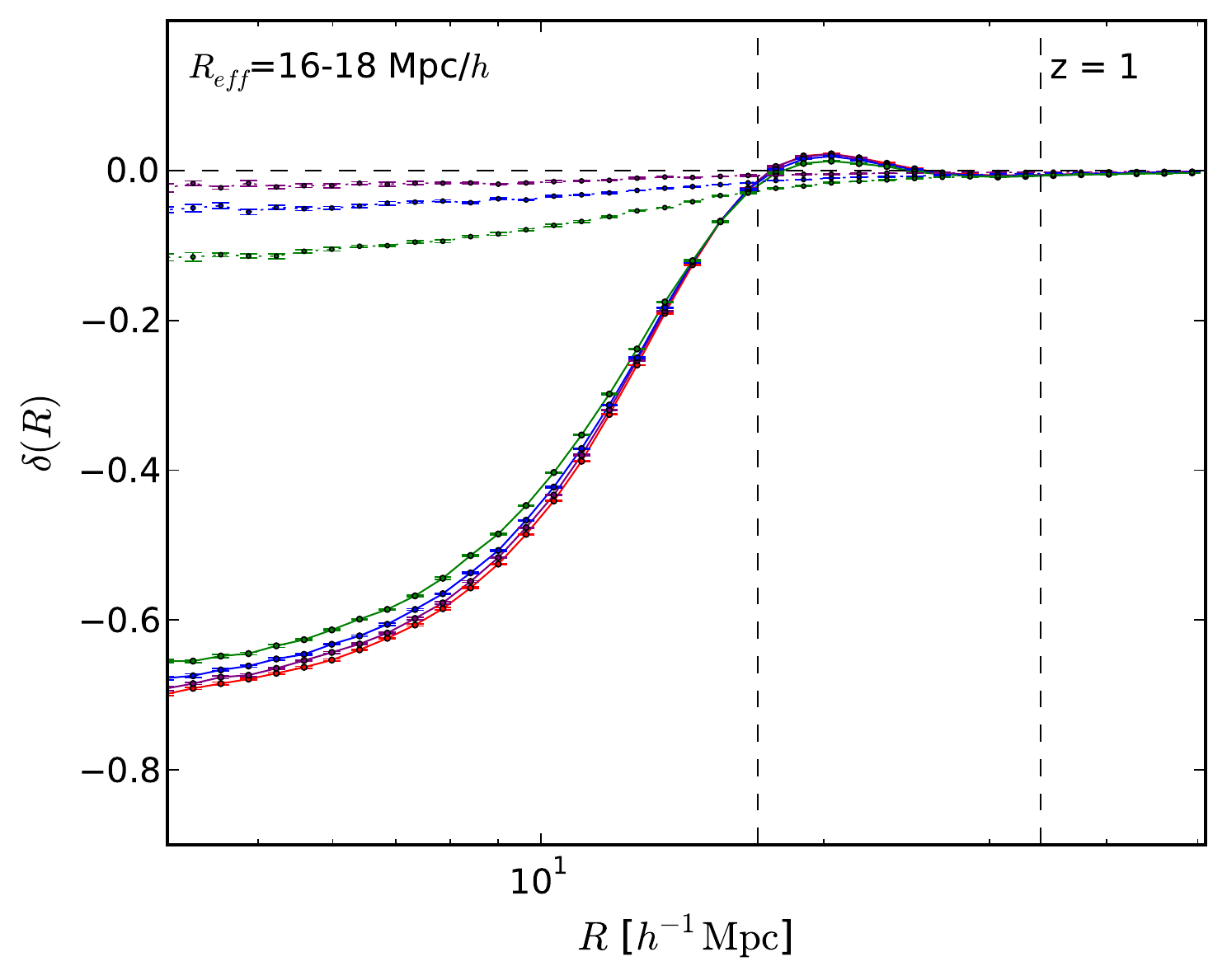}
\includegraphics[width=.49\textwidth,clip]{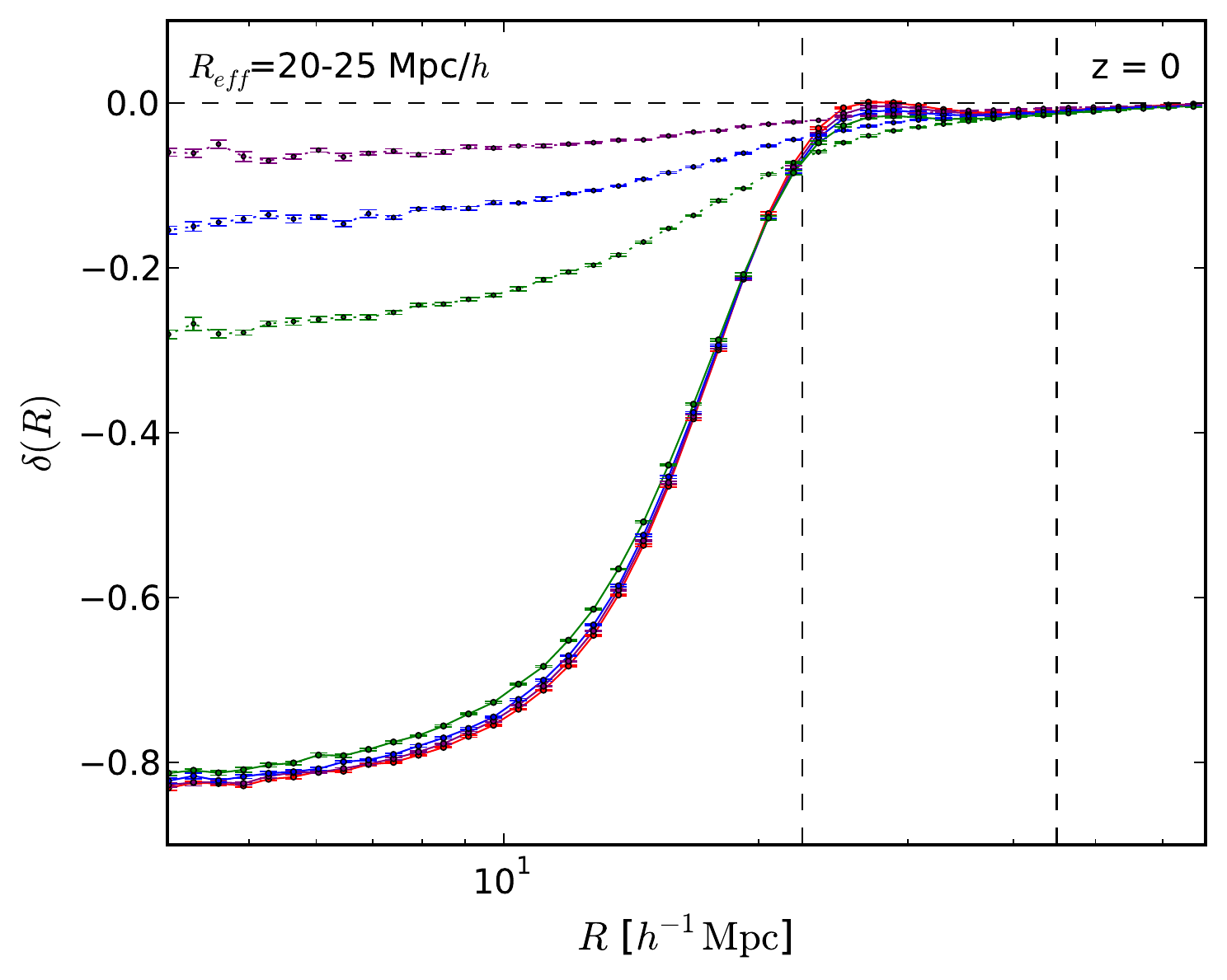}
\includegraphics[width=.49\textwidth,origin=c]{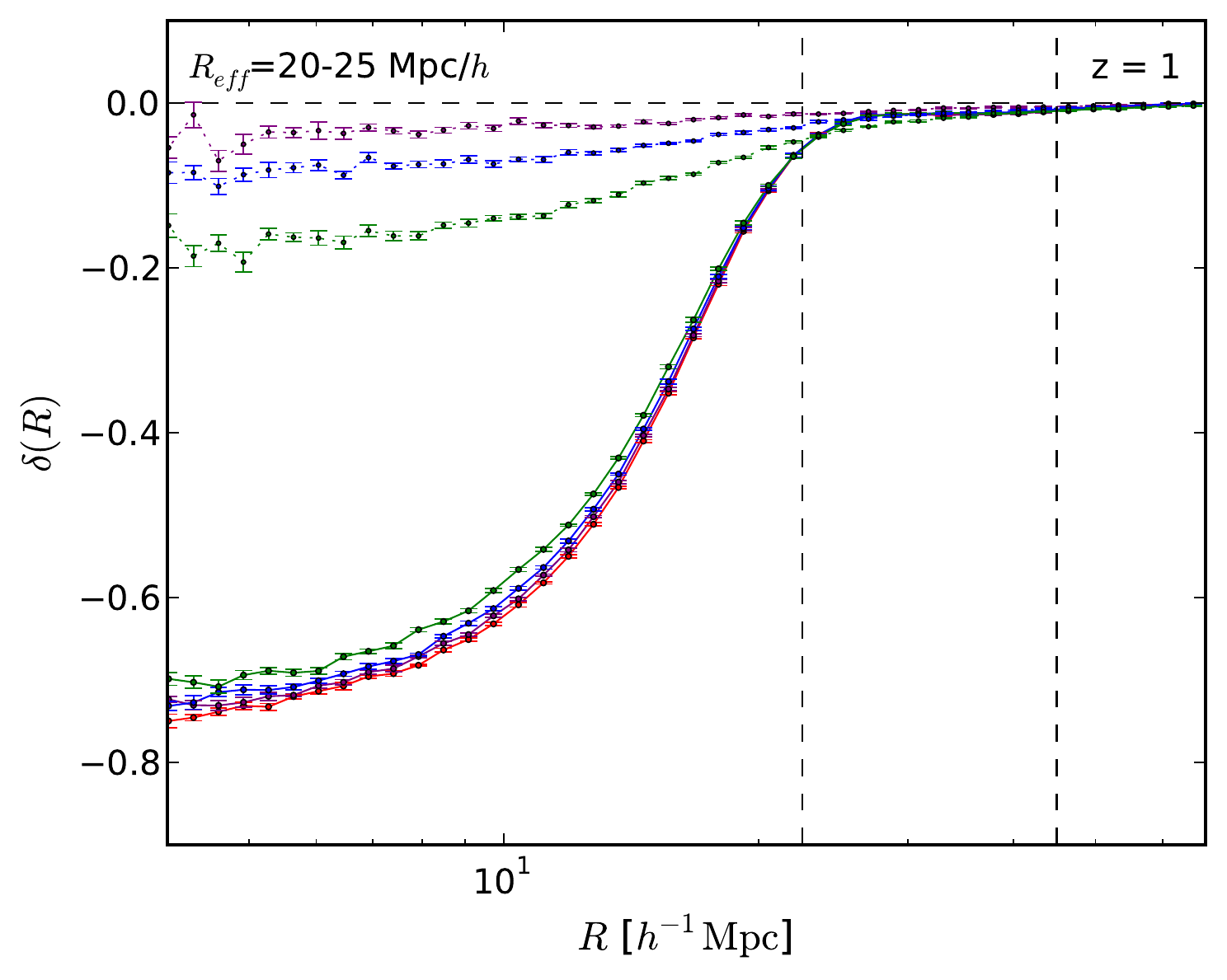}
\caption{\label{fig:cdm-nu_dens_profile} Average cold dark matter (solid lines) and neutrino (dotted lines) density profiles around voids with different sizes: $R_{\rm eff}$=10-11 Mpc/$h$ (top), $R_{\rm eff}$=16-18 Mpc/$h$ (center), $R_{\rm eff}$=20-25 Mpc/$h$ (bottom). Left and right panels show results at redshifts $z=0$ and $z=1$, respectively. Red, purple, blue and green lines show the 0.0, 0.15, 0.3 and 0.6 eV cosmologies, respectively. The dashed black lines indicate the mean value of the void radii in the selected range and two times the same quantity.}
\end{figure}

Here we investigate the impact of massive neutrinos on the shape of the void density profiles. We compute the cold dark matter, massive neutrino, and total matter density profile in voids of different sizes.

The profiles have been computed in the following way. For each simulation box, we select the voids whose effective radius lie within a certain interval. For each void, we compute the density field inside shells around the center, then we stack all the shells of the different voids to obtain the average density profile at a given distance from the void center. We repeat the procedure for all the ten realizations with the same cosmology, then we average over the realizations. The error that we present is the scatter between the different realizations normalized by $\sqrt{10}$.  We repeat this procedure for different intervals in the void radius and for all the cosmologies considered in this paper. 

Figure~\ref{fig:cdm-nu_dens_profile} shows the average cold dark matter (solid lines) and neutrino (dotted lines) density profiles around void centers with radius $R_{\rm eff}$=10-11 Mpc/$h$ (top), $R_{\rm eff}$=16-18 Mpc/$h$ (center), and $R_{\rm eff}$=20-25 Mpc/$h$ (bottom). Left and right plots show results at redshift $z=0$ and $z=1$, respectively, and different colors indicate different cosmologies. Let us focus first on the cold dark matter density profile. We can immediately notice that small voids present a compensated wall at the edge, whereas the large ones have a negative density profile, as already shown in~\cite{HSW}. Moreover, fixing the cosmology, voids present a typical behavior: large voids are emptier; i.e., they present deeper underdensities in the core than small ones. However, we have checked that this behavior can change depending on the resolution of the considered simulations. 

For voids with a given radius, we find that the height of the wall becomes higher when decreasing the neutrino mass. On the other hand, the inner part of the density profile is emptier in the massless neutrino case and it becomes denser as the neutrino mass increases. Voids evolve in time building a higher wall and becoming progressively  emptier and emptier. Again we can understand the difference between the cosmologies: the impact of the neutrino mass can be described as slowing down the evolution of the void profiles and giving a less evolved universe in these regions. 

The neutrino profile is found to follow the corresponding cold dark matter one. Around very small voids, the neutrino profile presents an overdensity at the void wall, whereas it presents troughs around very large voids. Any departure from the mean background density (both overdensities and troughs) is greater for larger neutrino masses. This corresponds to the neutrinos with lower thermal velocities, which are more susceptible to the presence of the cold dark matter gravitational field. 

\begin{figure}[!t]
\centering %\begin{center}/\end{center} takes some additional vertical space
\includegraphics[width=.455\textwidth,clip]{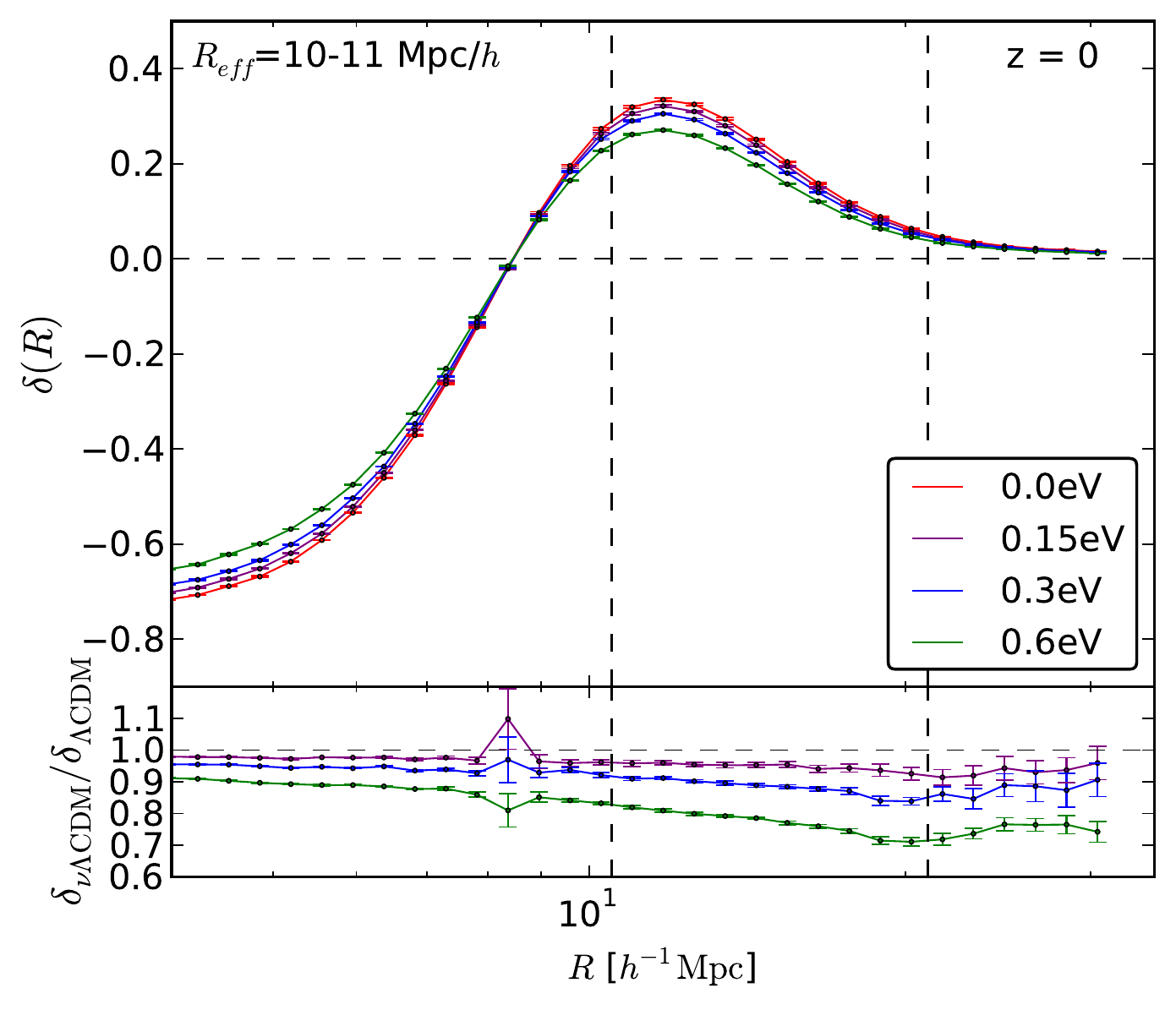}
%\hfil
\includegraphics[width=.455\textwidth,origin=c]{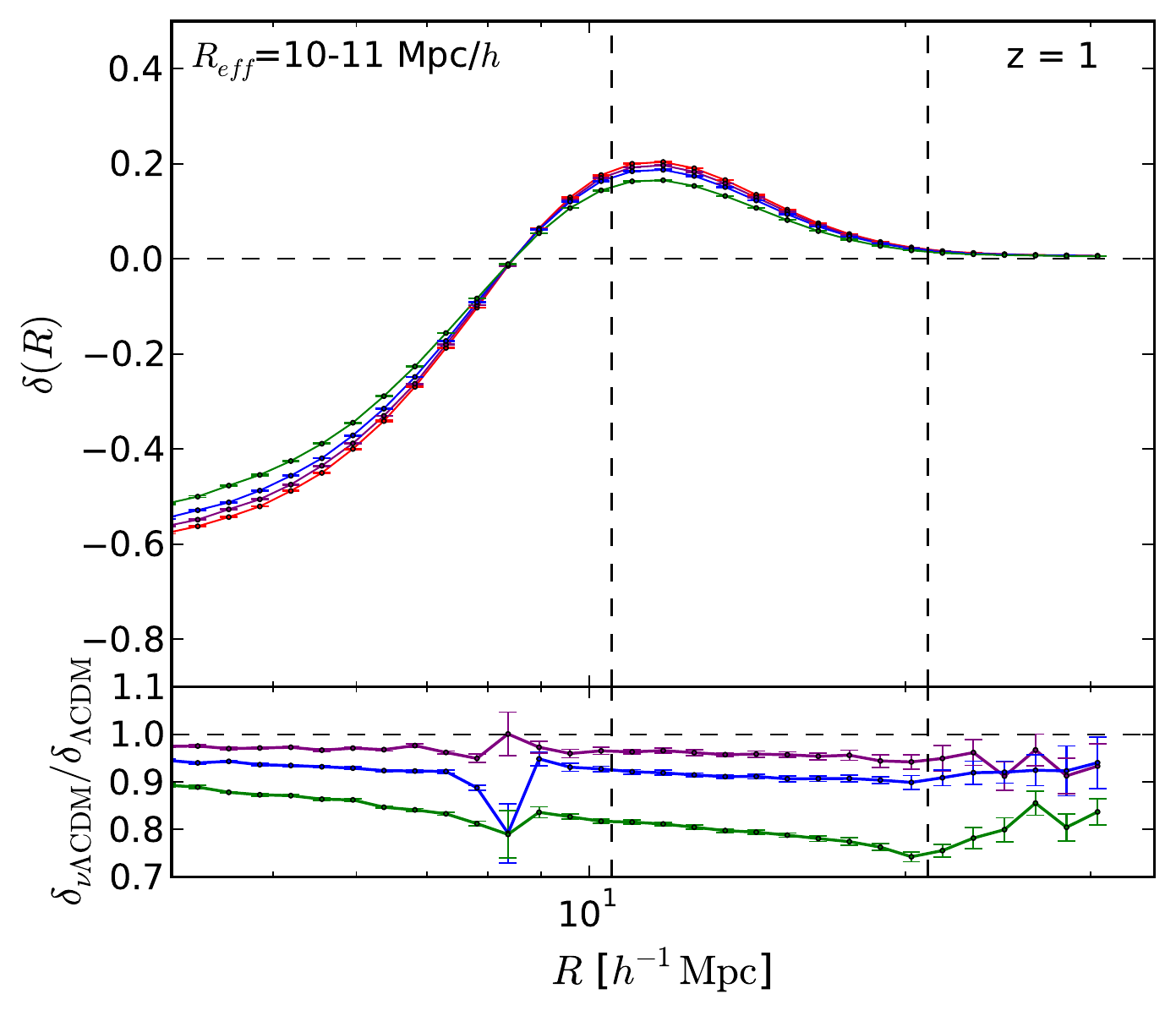}
\includegraphics[width=.455\textwidth,clip]{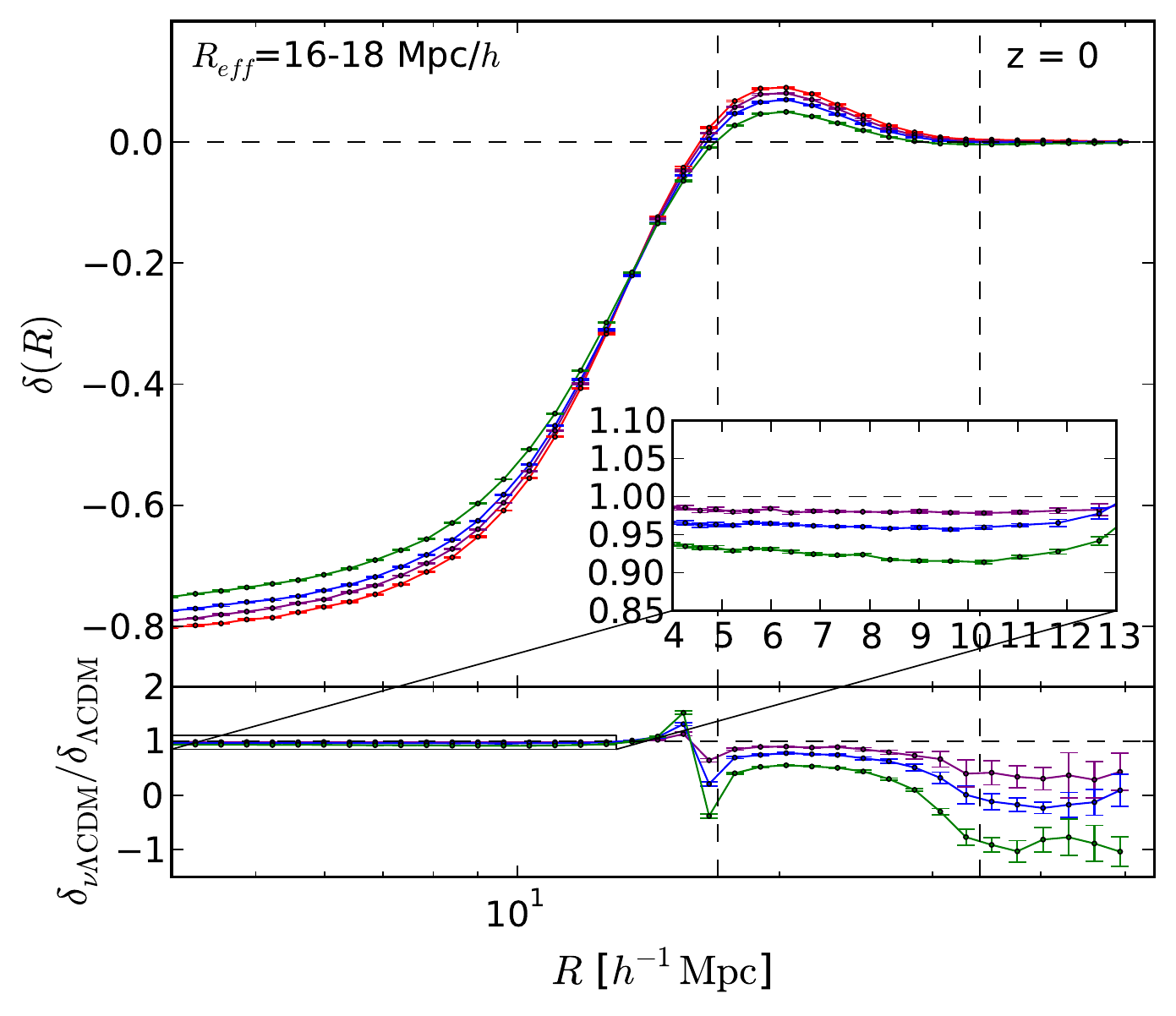}
%\hfill
\includegraphics[width=.455\textwidth,origin=c]{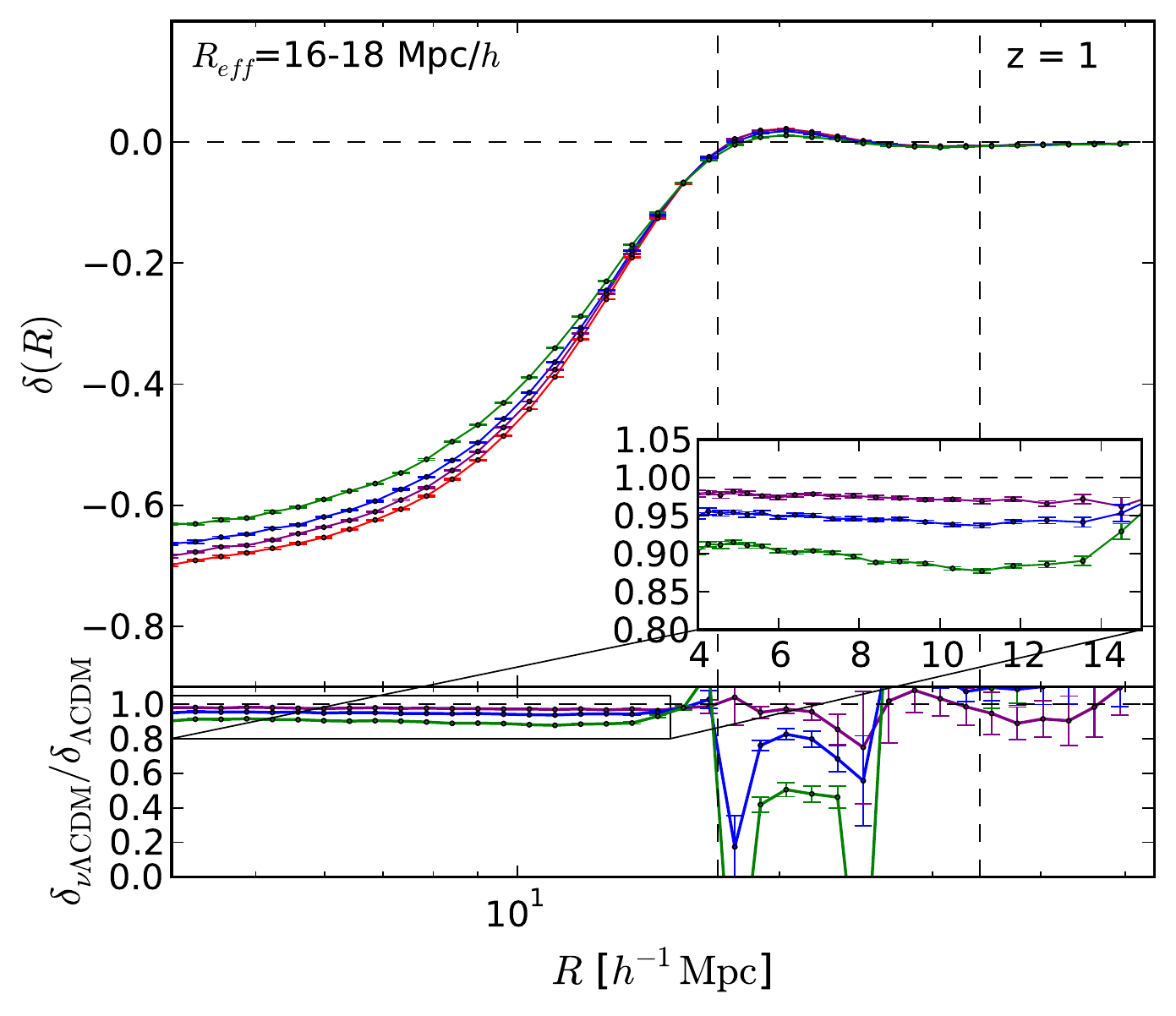}
\includegraphics[width=.455\textwidth,clip]{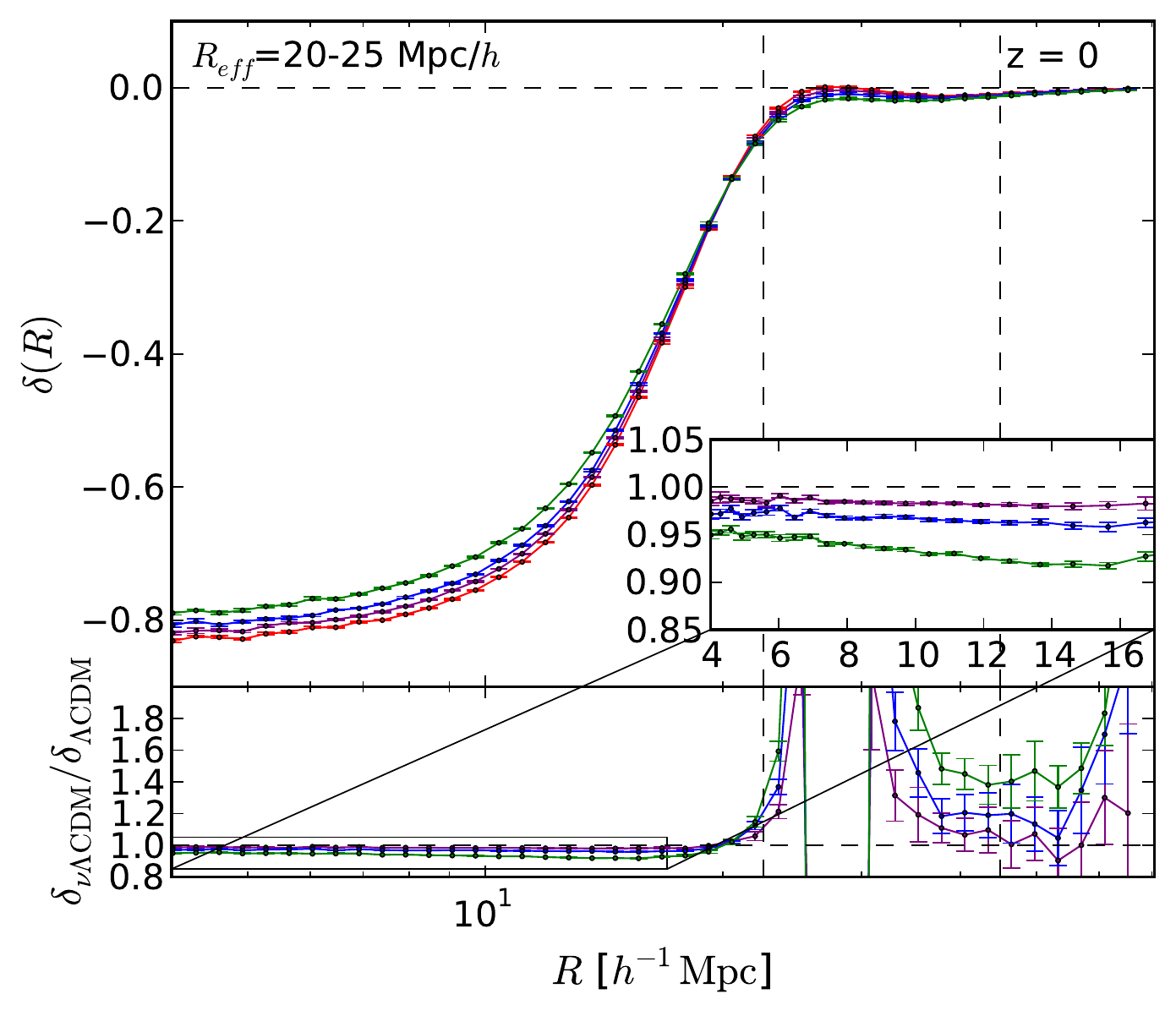}
\includegraphics[width=.455\textwidth,origin=c]{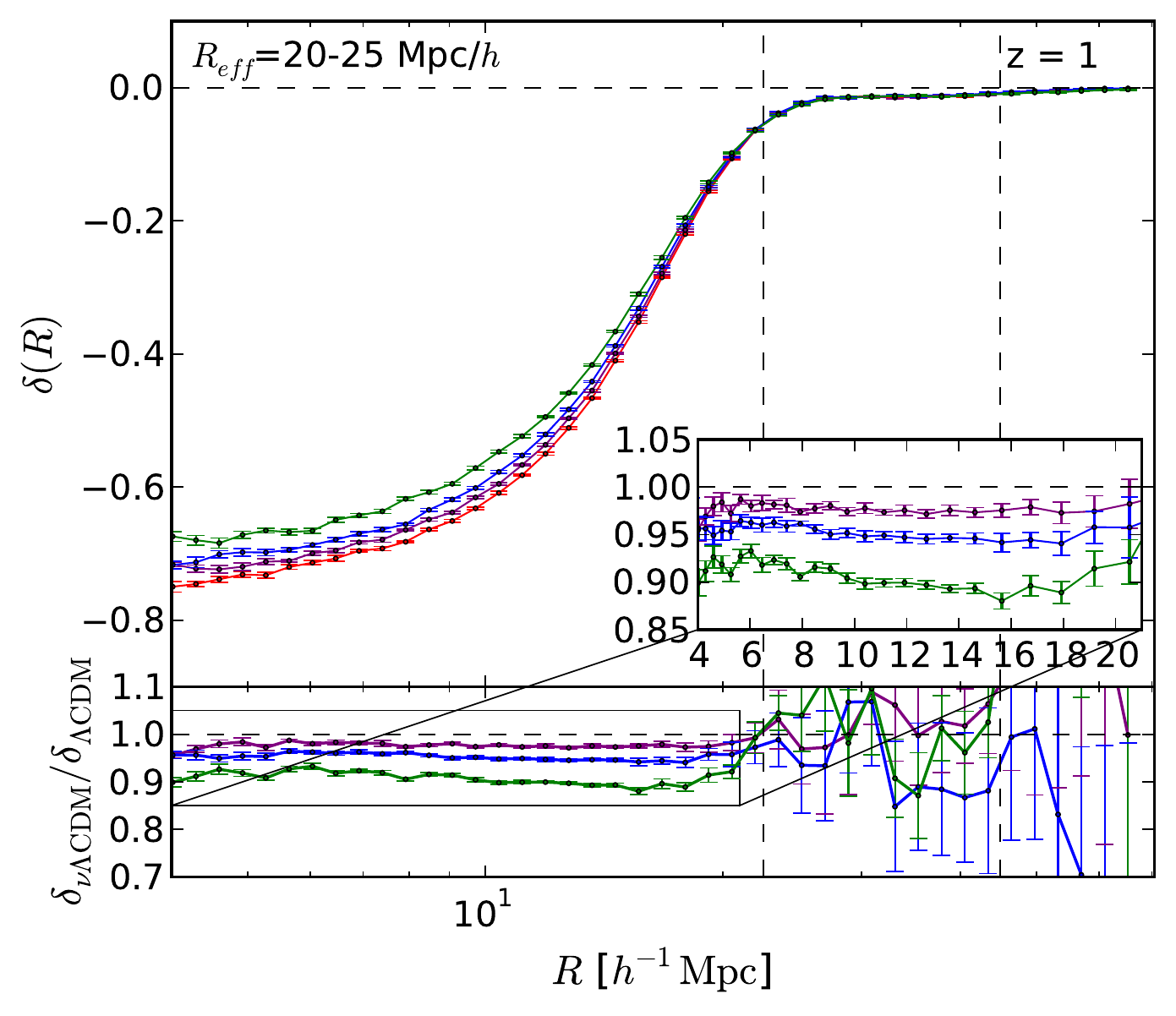}
\caption{\label{fig:matter_dens_profile} Average total matter density profiles around voids with different sizes: $R_{\rm eff}$=10-11 Mpc/$h$ (top), $R_{\rm eff}$=16-18 Mpc/$h$ (center), and $R_{\rm eff}$=20-25 Mpc/$h$ (bottom). Left and right panels show results at redshifts $z=0$ and $z=1$, respectively. Red, purple, blue and green lines show the 0.0, 0.15, 0.3 and 0.6 eV cosmologies, respectively. At the bottom of each panel we display the ratio between the results from the massive neutrino cosmologies and the $\Lambda$CDM one. The vertical dashed black lines indicate the mean value of the void radii in the selected range and two times the same quantity.}
\end{figure}

In Fig. \ref{fig:matter_dens_profile} we show the total matter density profile, which is the most relevant quantity since most of the observables depend on the total matter distribution, e.g.  weak lensing and the ISW effect. We present the results for the same void sizes as above and at $z=0$ (left) and $z=1$ (right). The bottom panel of each plots shows the ratio between the density profiles of the three massive neutrino cosmologies with respect to $\Lambda$CDM. The main features observed in the cold dark matter profiles are overall preserved, but enhanced here. In the void core, the differences between $\Lambda$CDM and the 0.15 eV cosmology are at the level of 1-3 \%, depending on the void radius, and they reach 10\% for the 0.6 eV case. Near the wall and for the very small voids, the difference is at the 5 \% level for 0.15 eV and at the 20\% level for 0.6 eV cosmologies. All the differences are more pronounced at  $z=1$.

\subsection{Velocity profile}

We now present the radial velocity profiles of the cold dark matter, neutrino, and total matter fields. The first two have been computed directly from simulations in the following way. We select voids with radii in a given range and then we compute the radial velocity profiles of particles inside a shell of radius $r$ around each void center, using
\begin{equation}
v(r) = \frac{1}{N}\sum_{i =1}^N\vec{v}(r_i) \cdot \frac{\vec{r}_i}{|\vec{r}_i|}\, ,
\end{equation}
where $N$ is the number of particles in the shell and $\vec{r}_i$ are the coordinates of the particles with respect to the void center. Next, we stack all the shells from the different voids for computing the average velocity at a given distance from the void center. Finally, we average over the ten realizations with the same cosmology. The error associated with the profiles is the scatter between the ten realizations divided by $\sqrt{10}$. The velocity of the total matter field is instead computed as the density weighted average between the velocity profiles of the cold dark matter and neutrino particles, via
\begin{equation}
v_{\rm m} (r)= \frac{v_{\rm cdm}(r)\rho_{\rm cdm}(r)+v_{\nu}(r)\rho_{\nu}(r)}{\rho_{\rm cdm}(r)+\rho_{\nu}(r)}\, ,
\label{eq:velocity}
\end{equation}
where the subscripts m, cdm and $\nu$ stand for matter, cold dark matter and neutrinos, respectively. The associated errors are computed via error propagation. Other velocity estimators have been used in literature, e.g. the Voronoi weighting estimator \cite{HSW}. However, we have checked that the relative differences between cosmologies (massive vs massless) are also reproduced by this estimator.

\begin{figure}%[!tbp]
\centering %\begin{center}/\end{center} takes some additional vertical space
\includegraphics[width=.45\textwidth,clip]{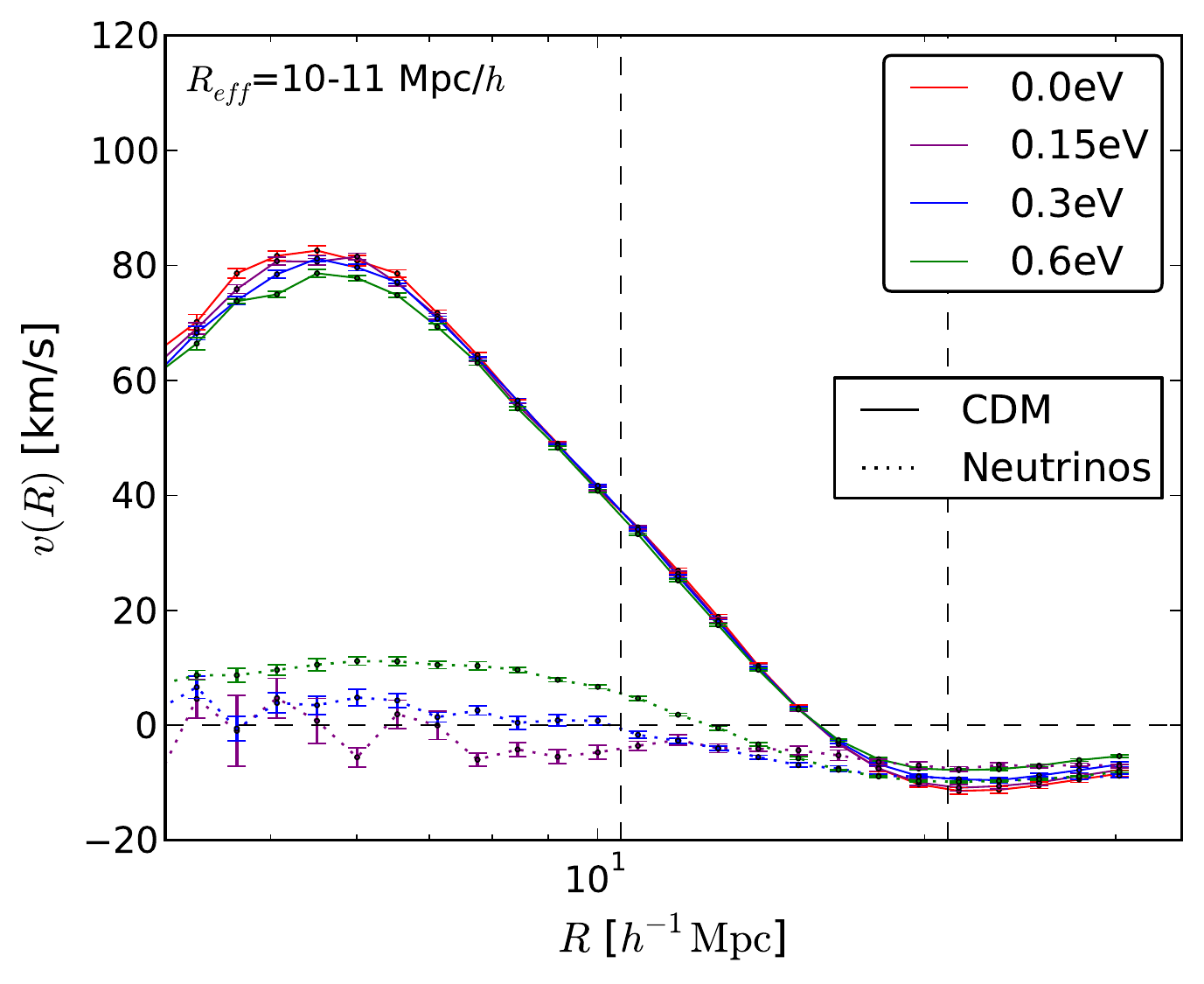}
%\hfill
\includegraphics[width=.45\textwidth,origin=c]{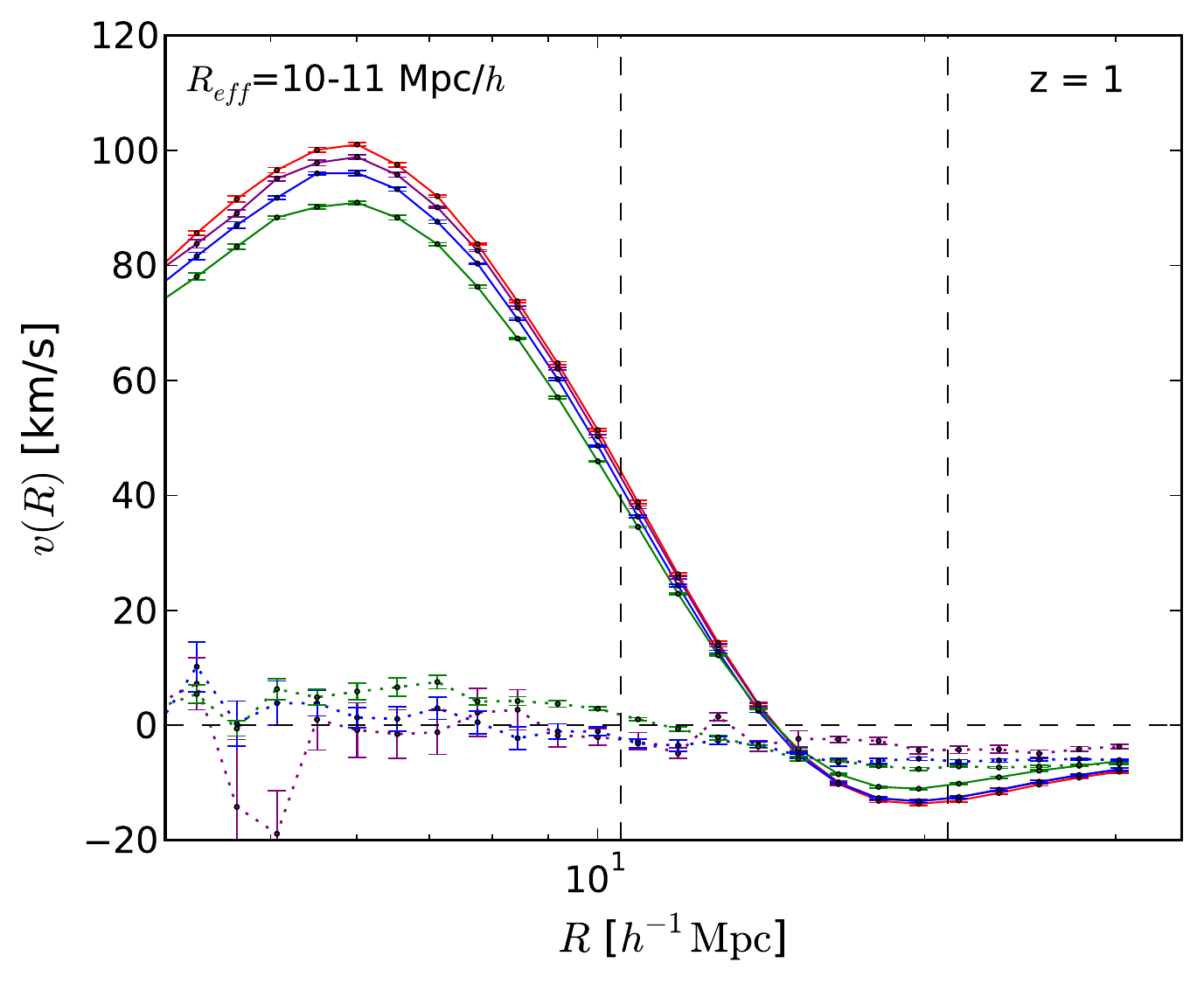}
\includegraphics[width=.45\textwidth,clip]{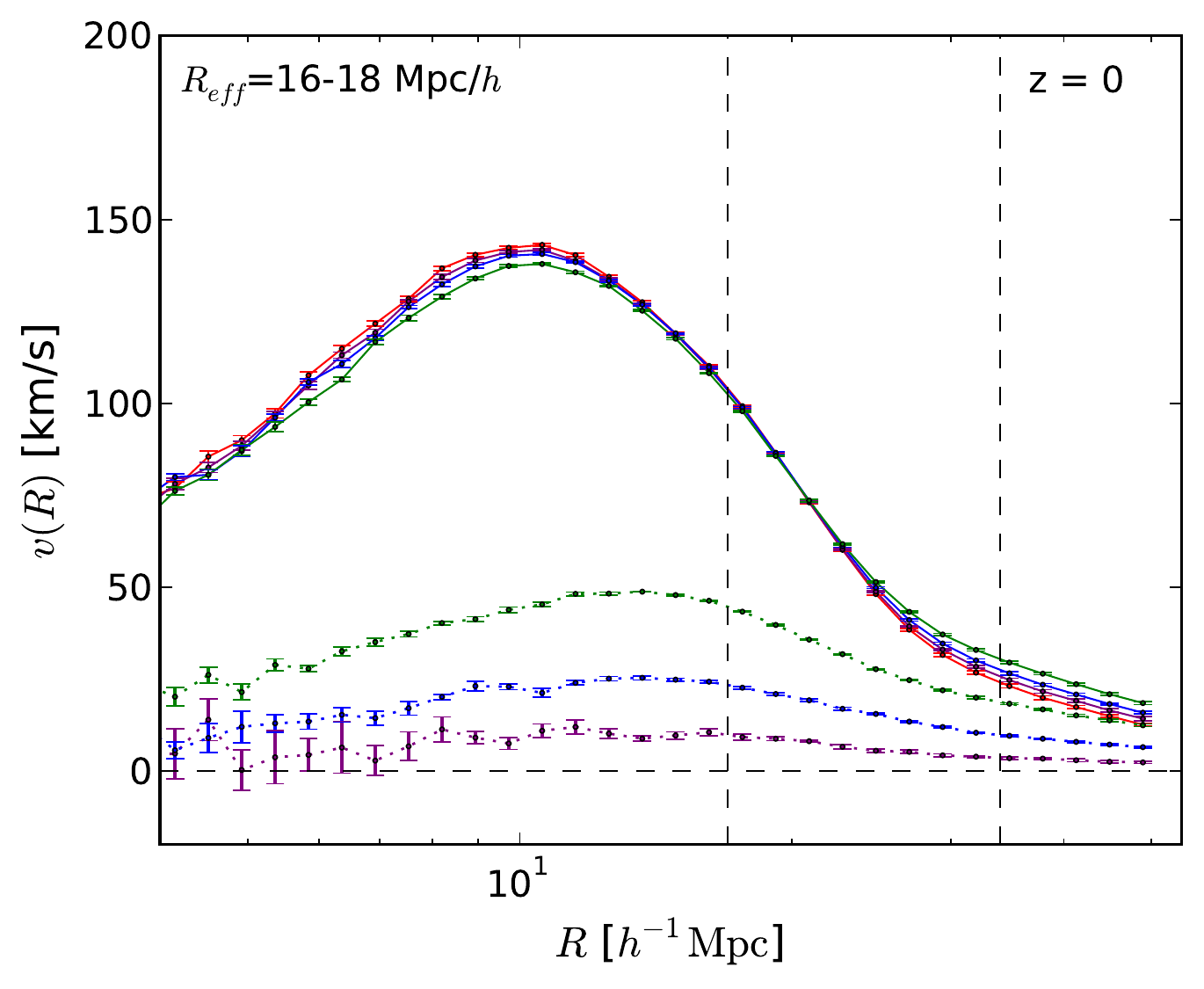}
%\hfill
\includegraphics[width=.45\textwidth,origin=c]{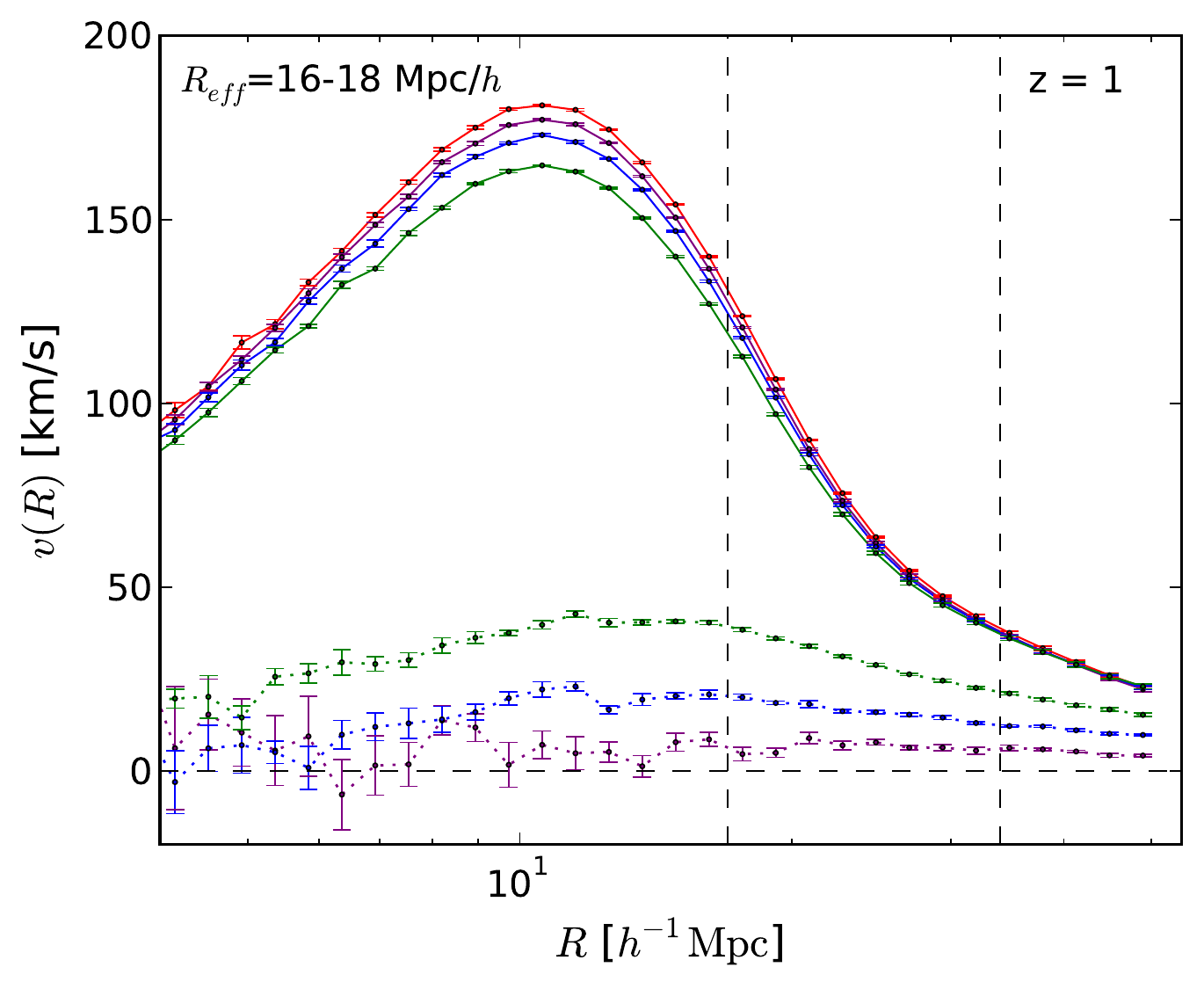}
\includegraphics[width=.45\textwidth,clip]{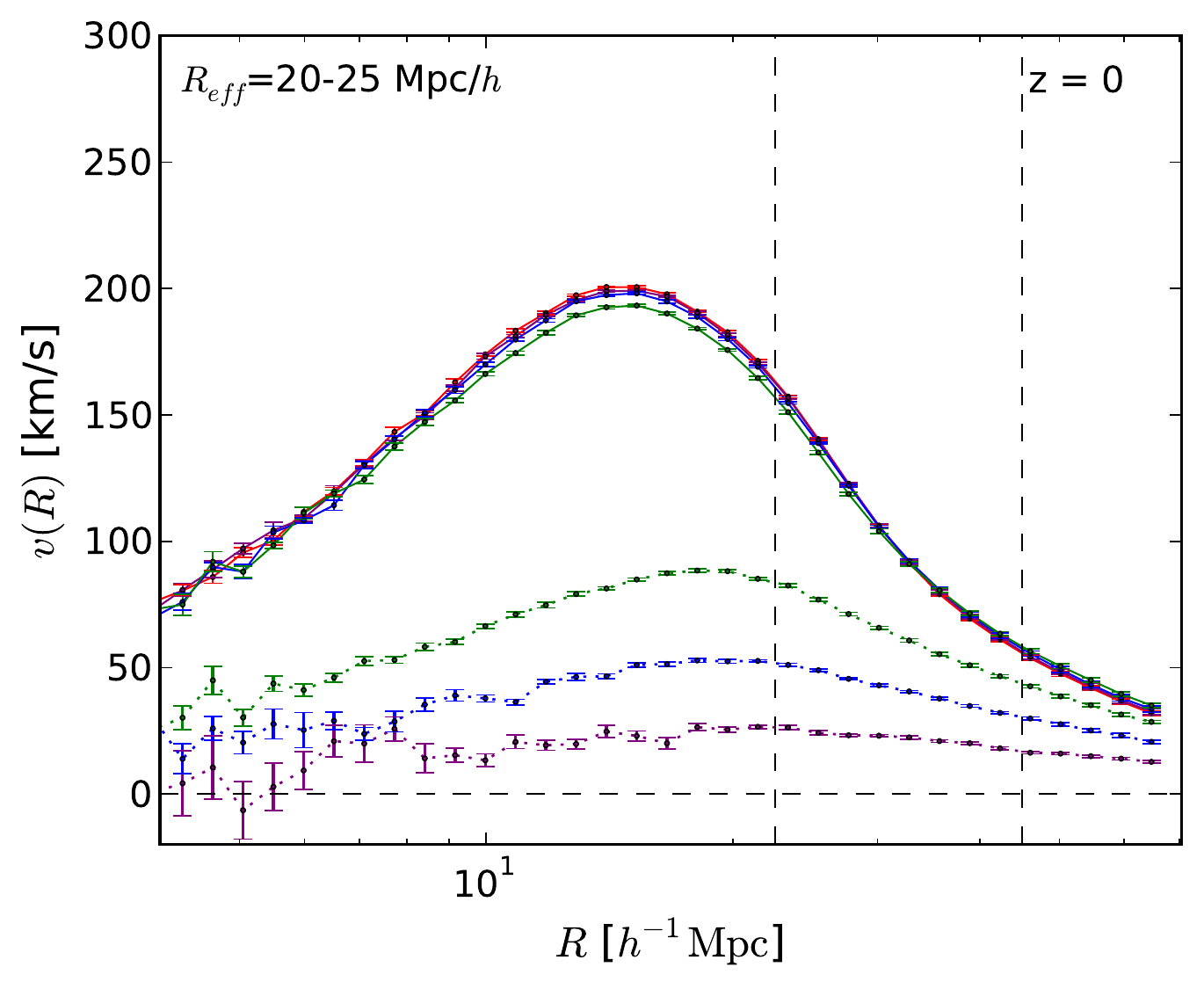}
\includegraphics[width=.45\textwidth,origin=c]{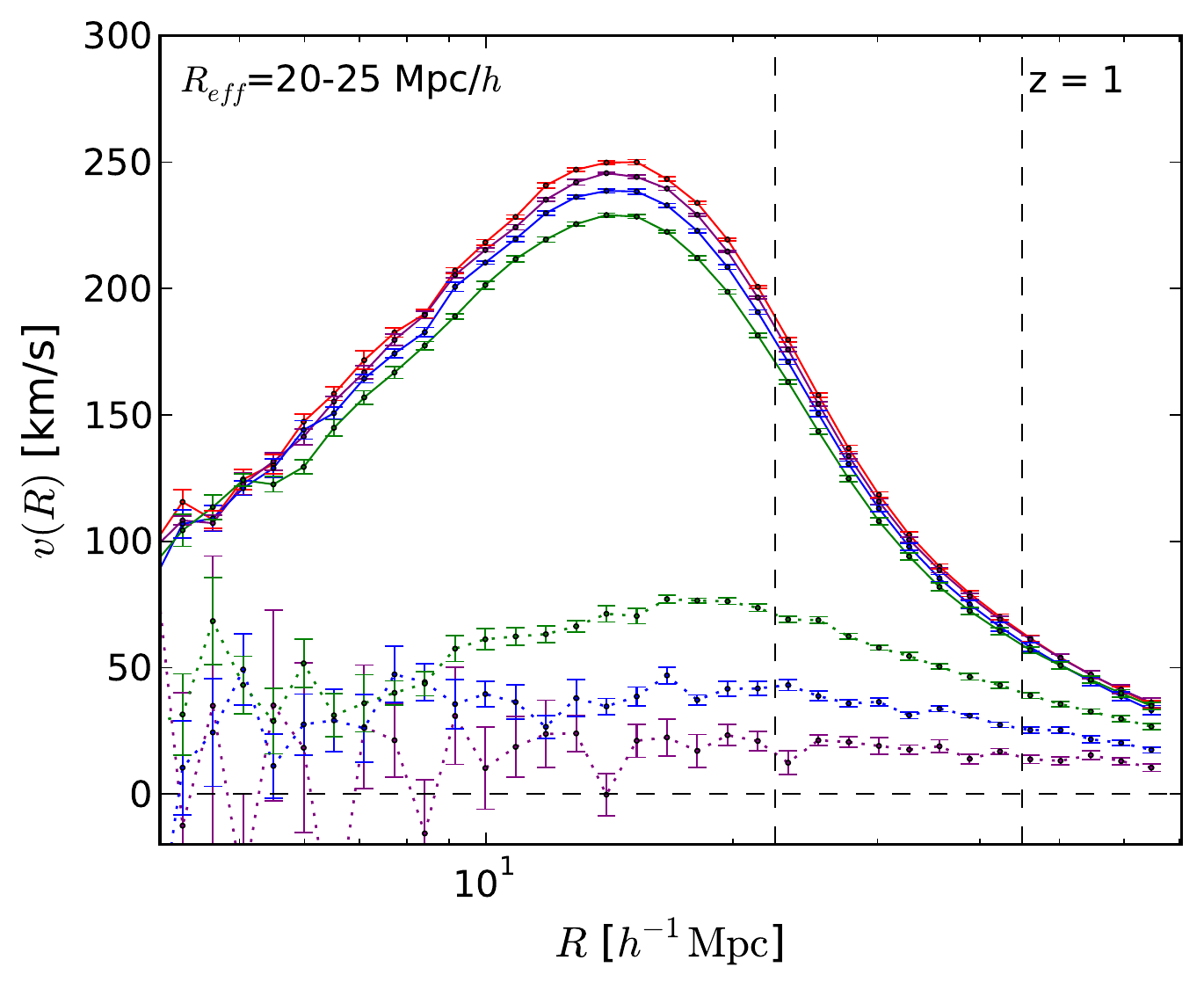}
\caption{\label{fig:velocity_cdm-nu} Cold dark matter (solid lines) and neutrinos (points) velocity profile measured from simulations around voids with different sizes: $R_{\rm eff}$=10-11 Mpc/$h$ (top), $R_{\rm eff}$=16-18 Mpc/$h$ (center), and $R_{\rm eff}$=20-25 Mpc/$h$ (bottom). Left and right panels are computed at redshift $z=0$ and $z=1$, respectively. 
Red lines indicate the $\Lambda$CDM cosmology, and purple, blue and green lines show the 0.15, 0.3 and 0.6 eV cosmologies. The dashed black lines indicate the mean value of the void radii in the selected range and two times the same quantity.}
\end{figure}

\begin{figure}%[!tbp]
\centering %\begin{center}/\end{center} takes some additional vertical space
\includegraphics[width=.45\textwidth,clip]{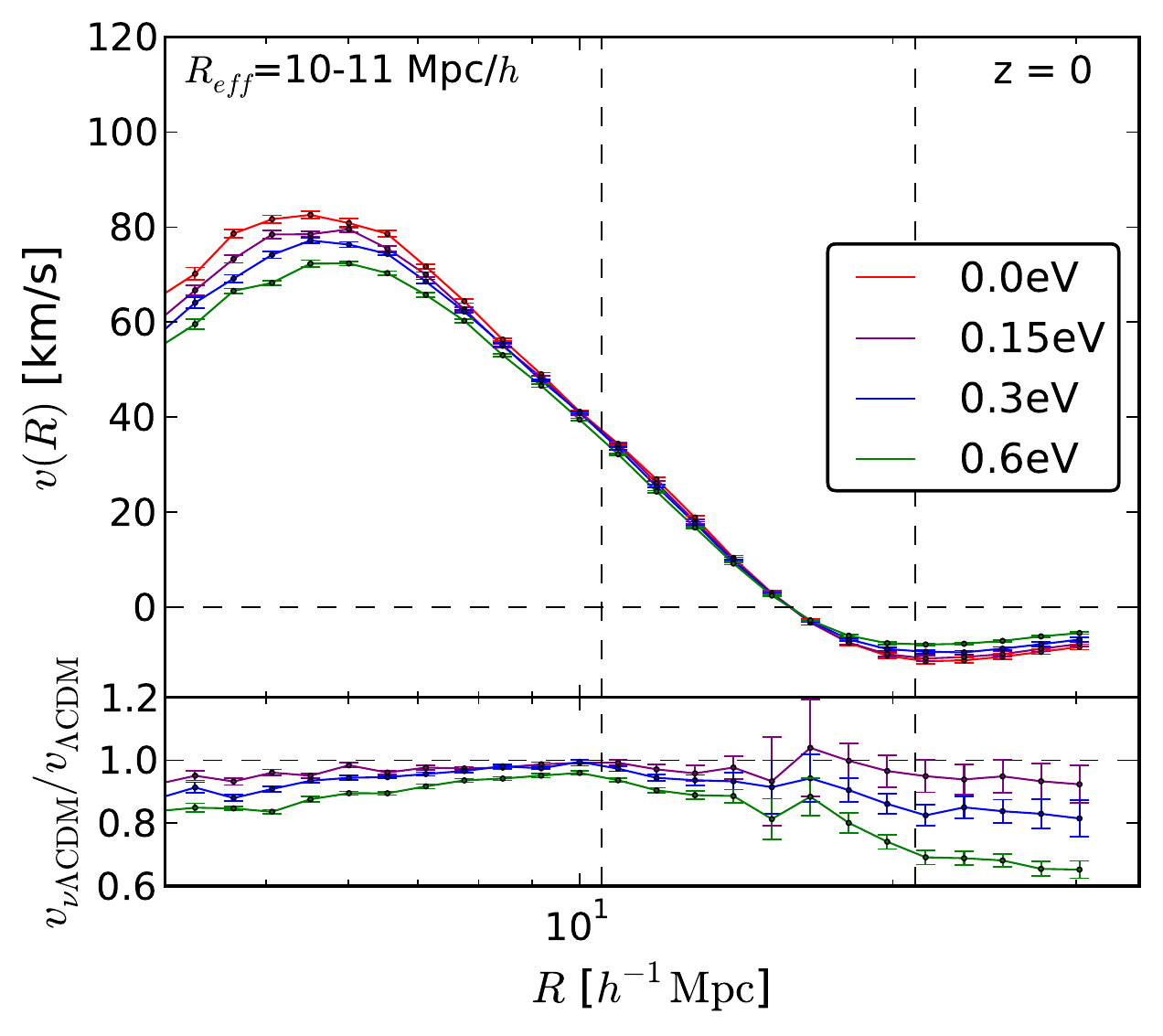}
%\hfill
\includegraphics[width=.45\textwidth,origin=c]{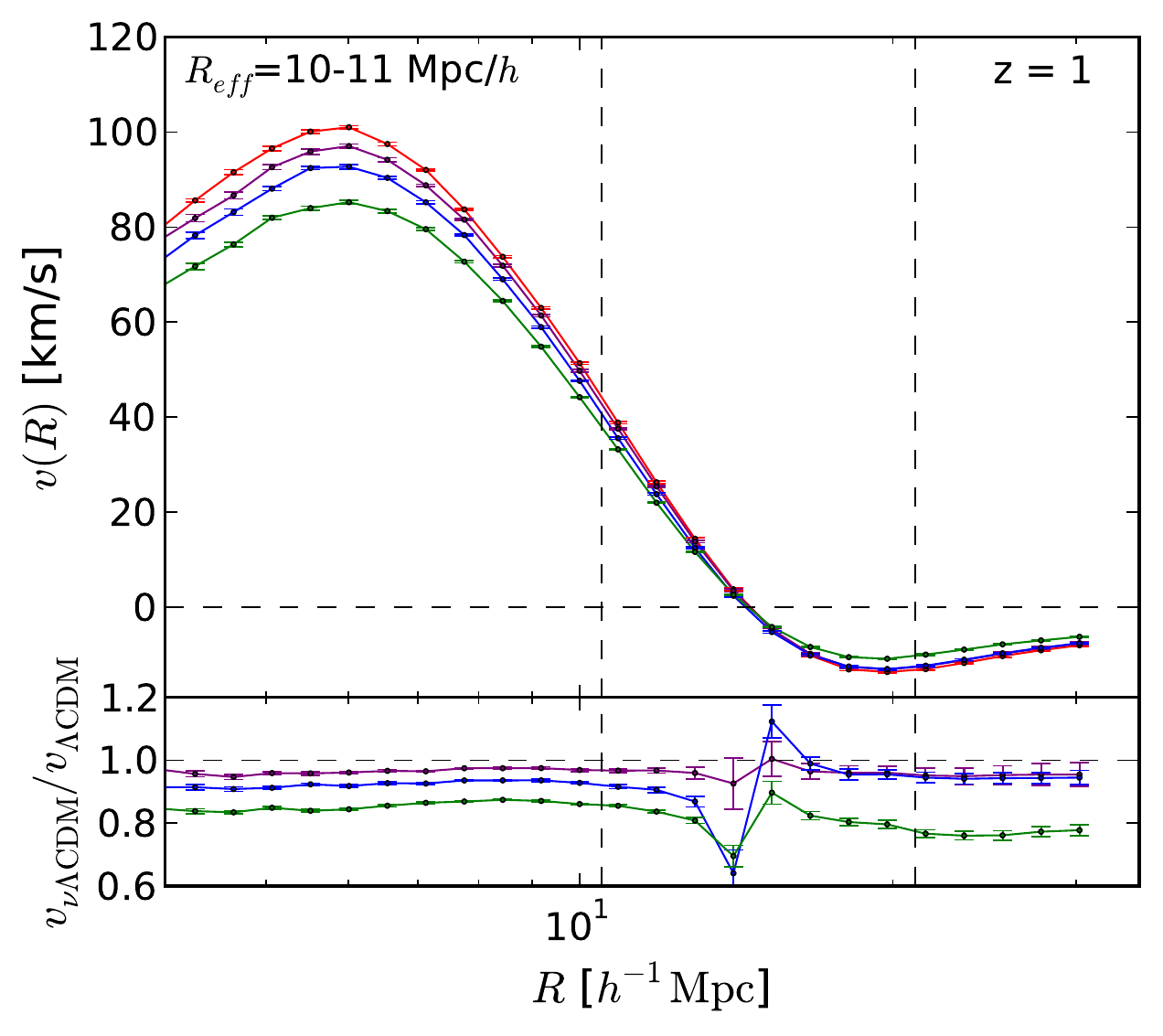}
\includegraphics[width=.45\textwidth,clip]{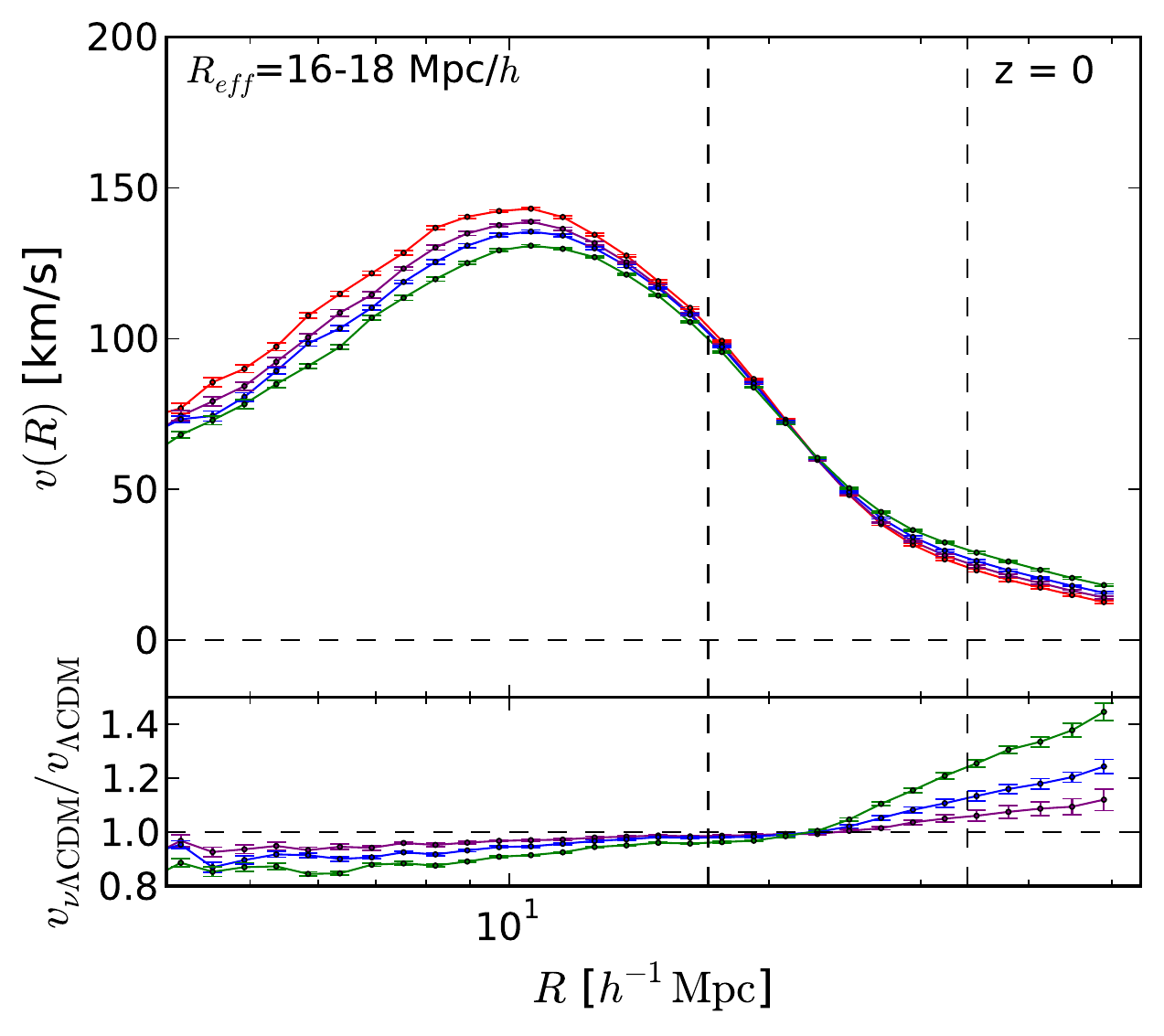}
%\hfill
\includegraphics[width=.45\textwidth,origin=c]{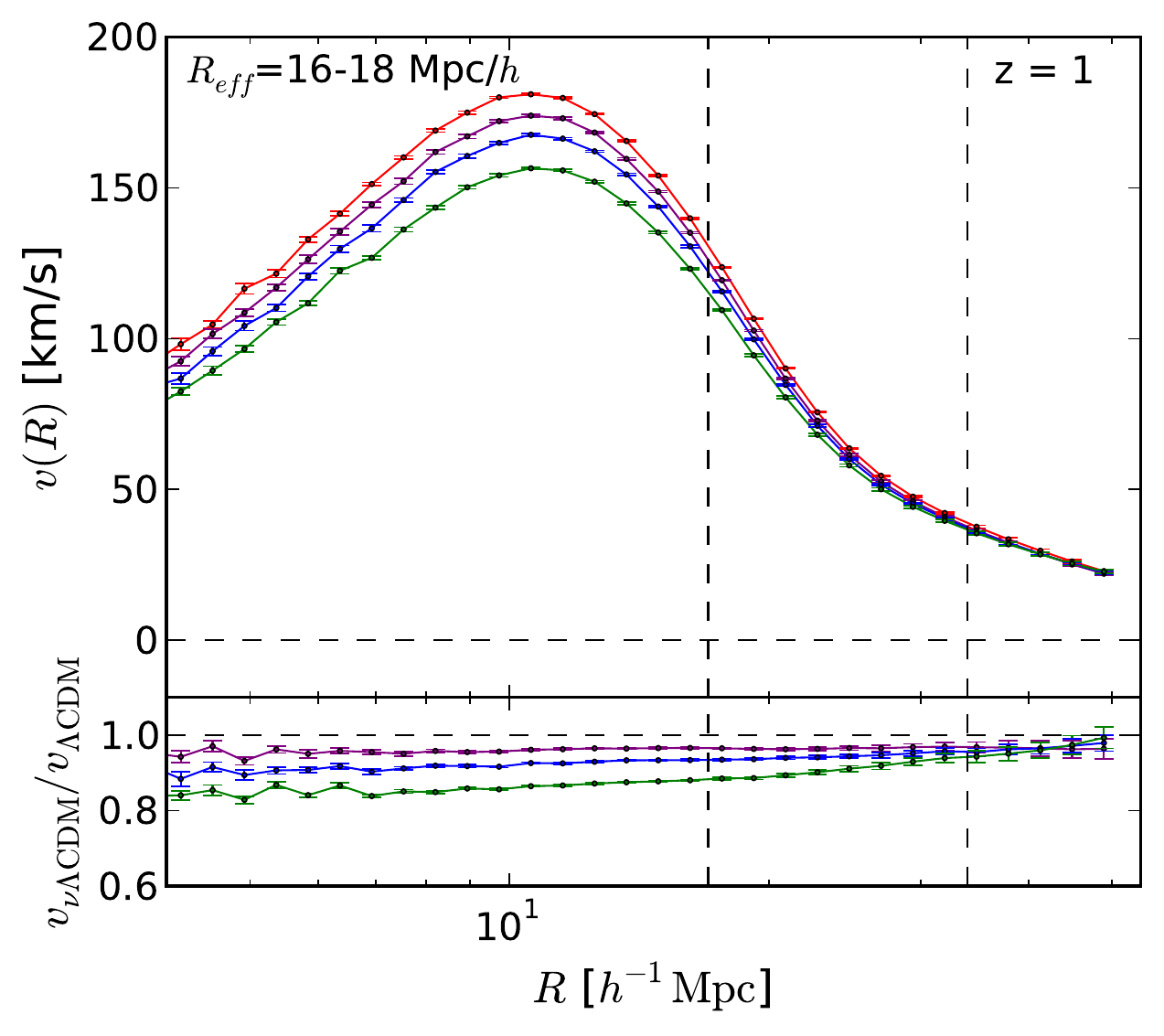}
\includegraphics[width=.45\textwidth,clip]{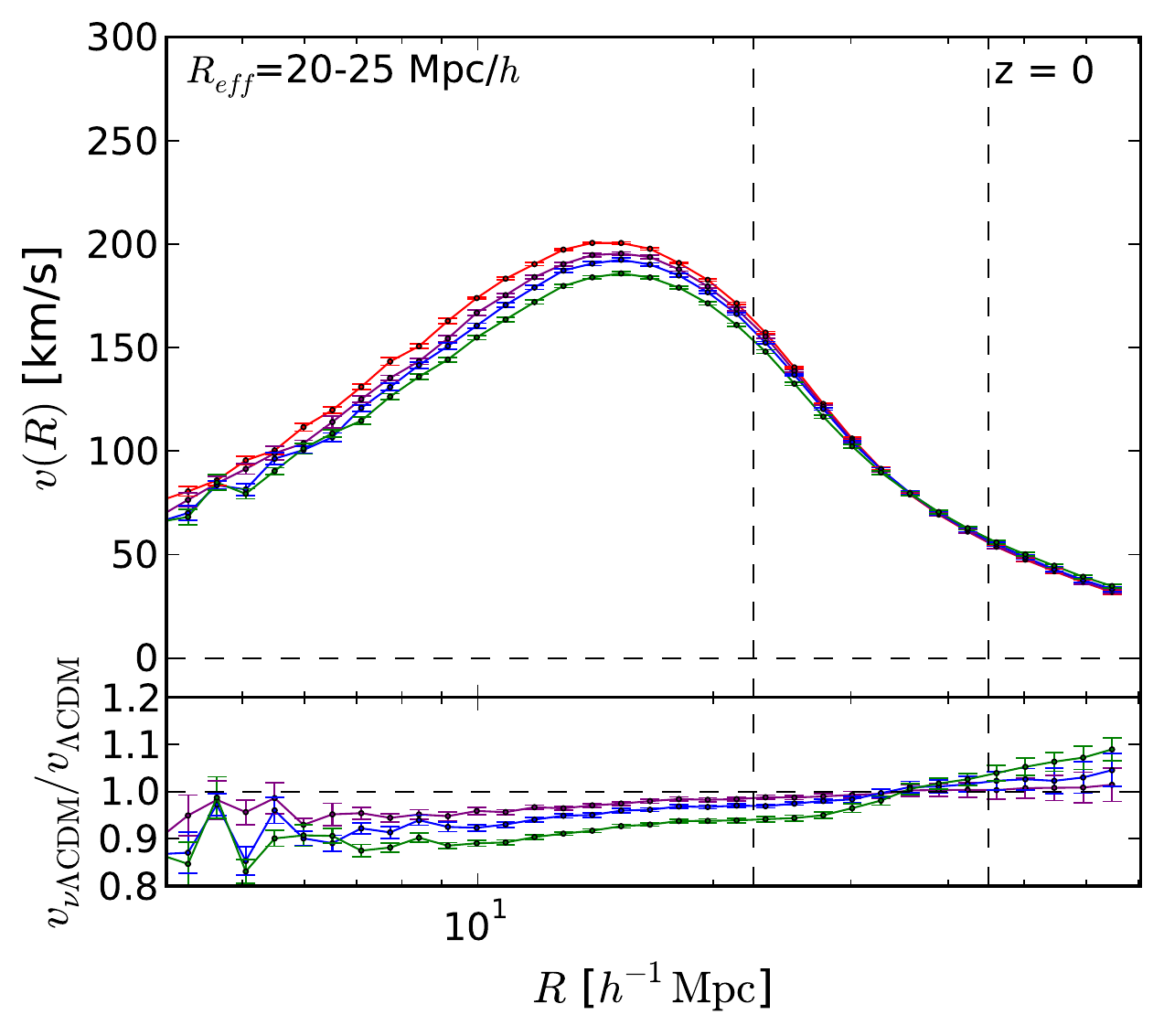}
\includegraphics[width=.45\textwidth,origin=c]{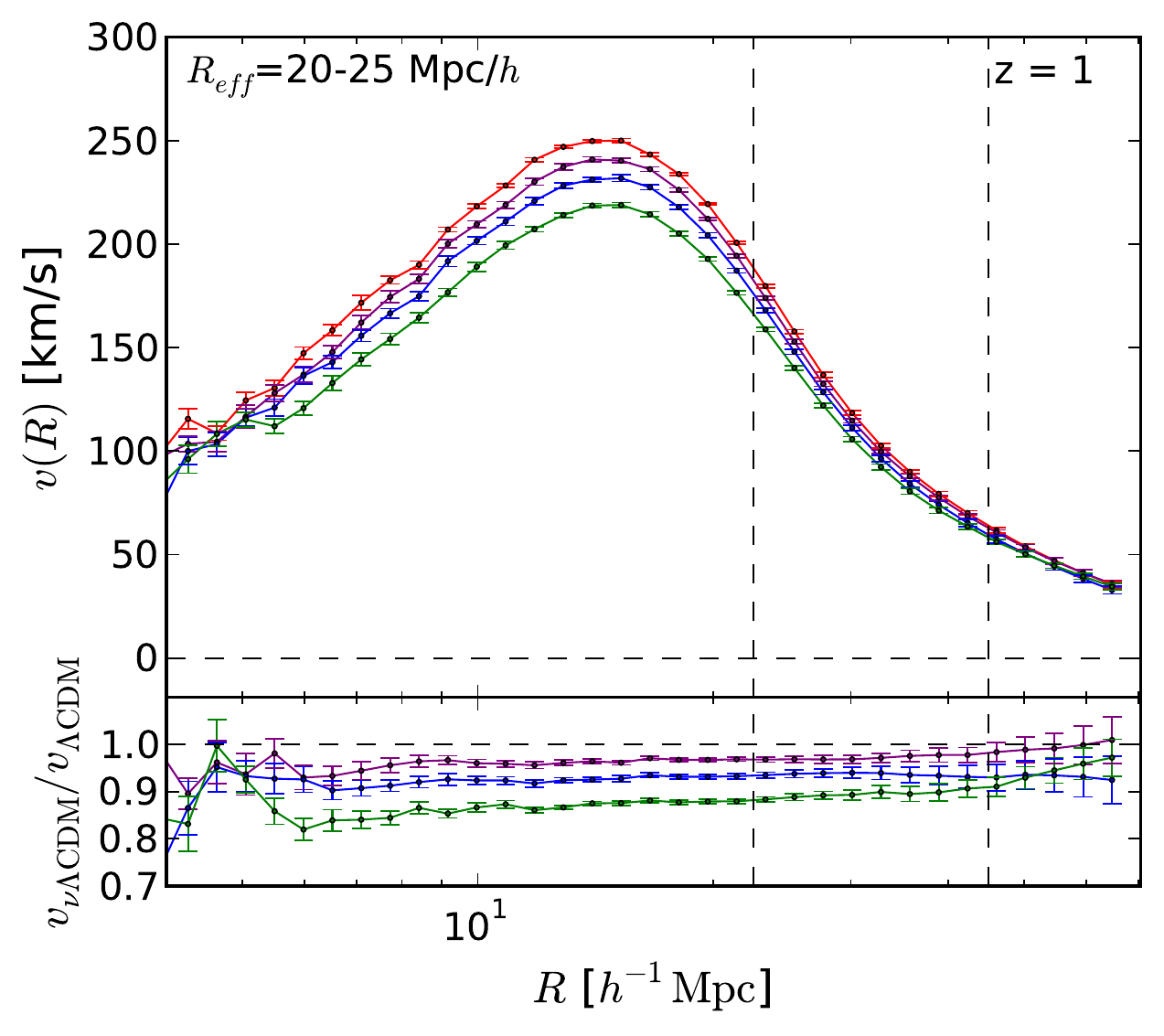}
\caption{\label{fig:velocity_matter} Total matter velocity profile measured from simulations around voids with different sizes: $R_{\rm eff}$=10-11 Mpc/$h$ (top), 16-18 Mpc/$h$ (center) and 20-25 Mpc/$h$ (bottom). Left and right panels display results at redshift z=0 and z=1, respectively. 
Red lines indicate the $\Lambda$CDM cosmology, purple, blue and green lines show the 0.15, 0.3 and 0.6 eV cosmologies. At the bottom of each panel there are the residuals between the massive neutrino cosmologies and the $\Lambda$CDM one. The dashed black lines indicate the mean value of the void radii in the selected range and two times the same quantity.}
\end{figure}

\begin{figure}[!tbp]
\centering %\begin{center}/\end{center} takes some additional vertical space
\includegraphics[width=.49\textwidth,clip]{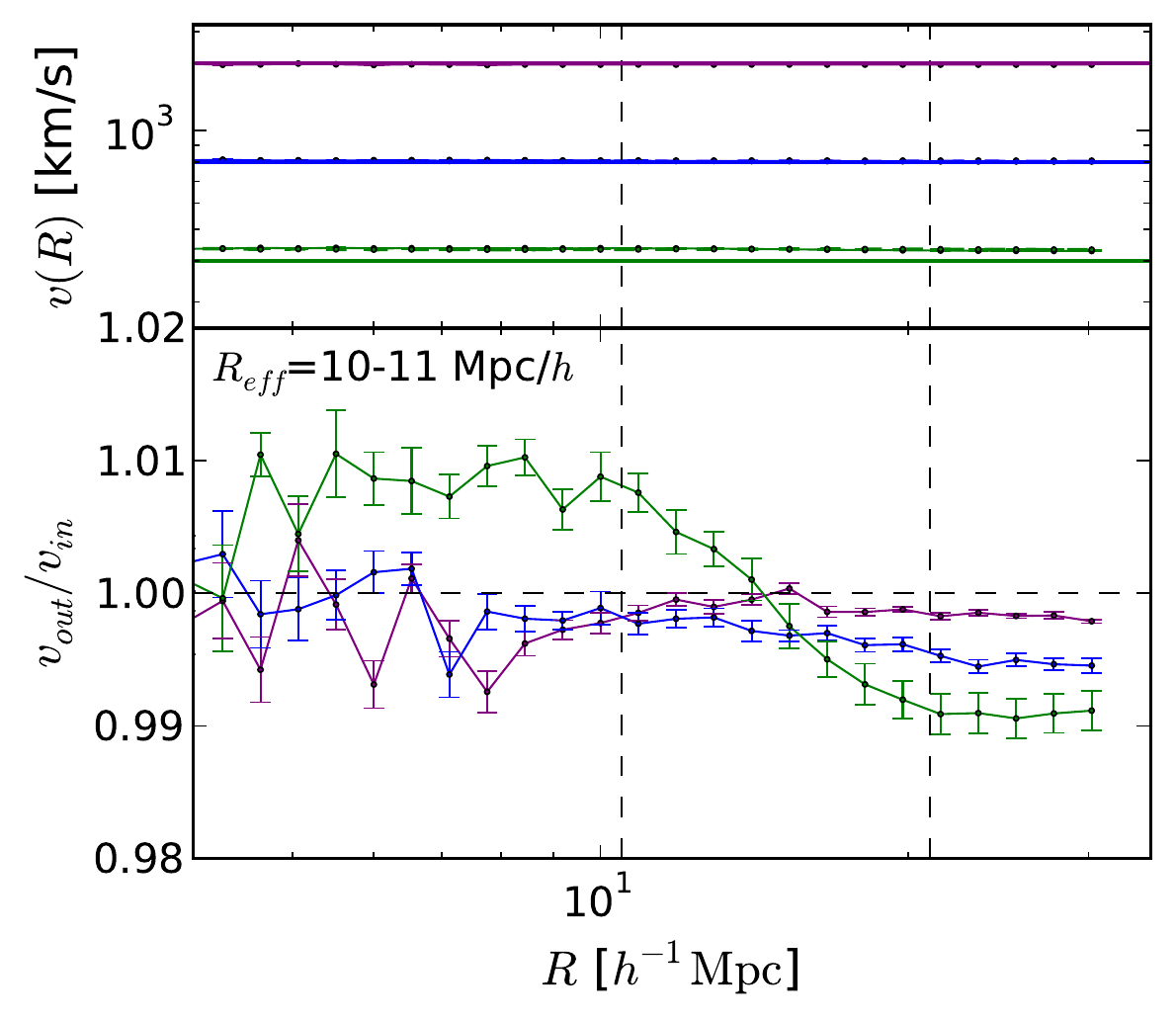}
%\hfill
\includegraphics[width=.49\textwidth,clip]{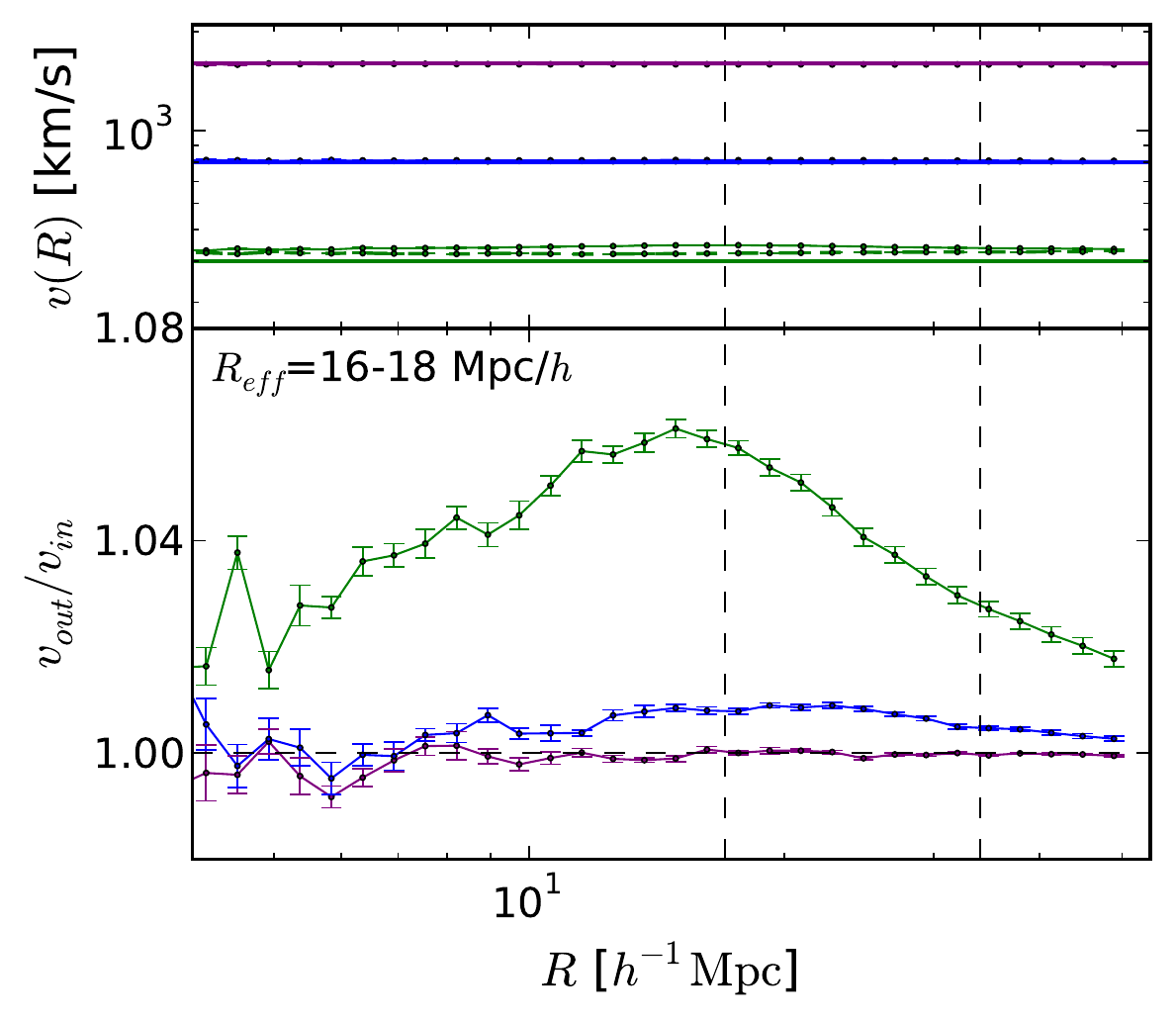}
%\hfill
\includegraphics[width=.49\textwidth,origin=c]{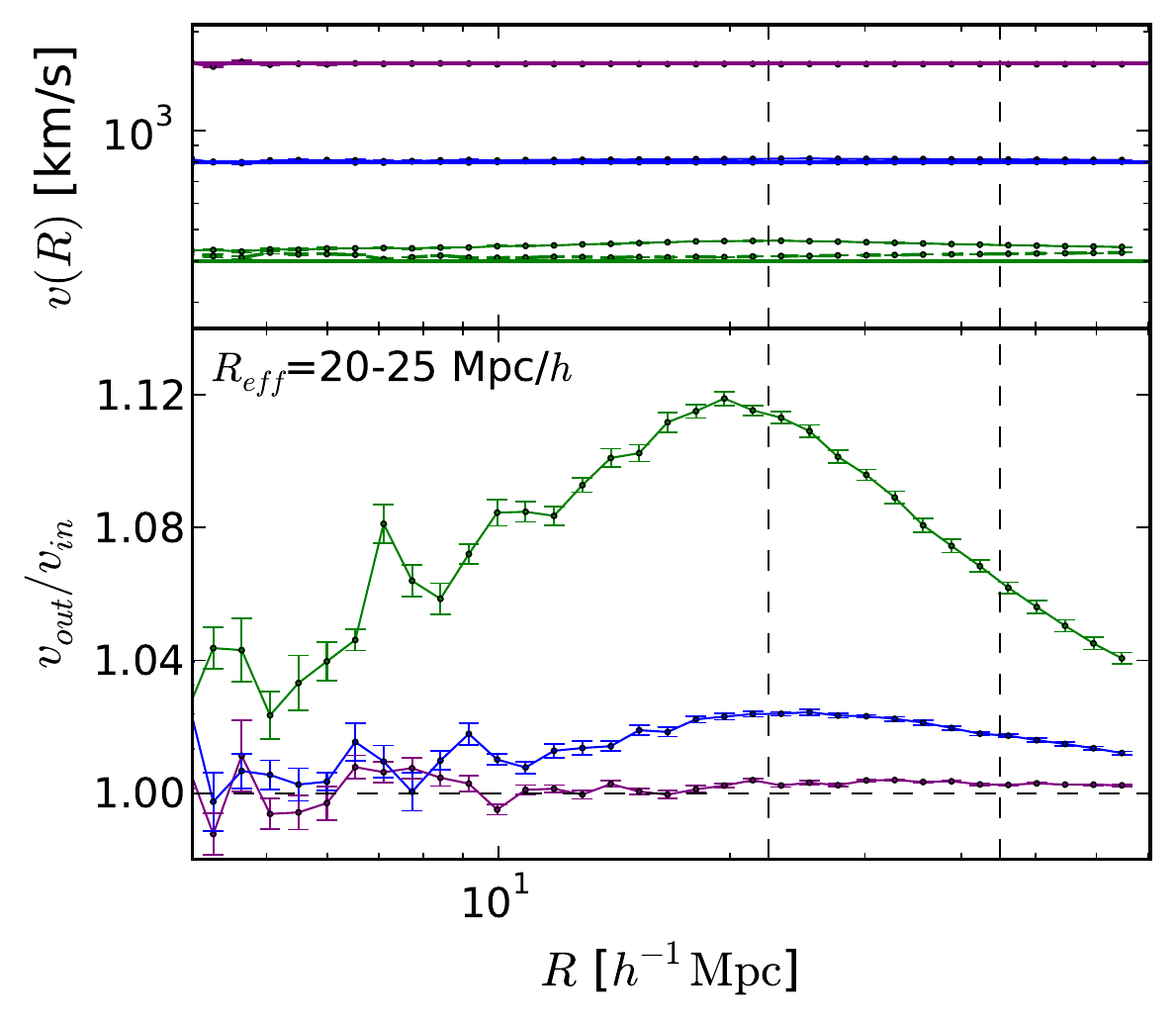}
\caption{\label{fig:velocity_neutrinos} Radial velocity profile of outgoing (solid lines) and incoming (dashed lines) neutrinos around voids with different sizes: $R_{\rm eff}$=10-11 Mpc/$h$ (top-left), 
16-18 Mpc/$h$ (top-left), 
and 20-25 Mpc/$h$ (bottom). The solid horizontal lines show the theoretical mean radial velocity; 
purple, blue and green indicate the 0.15, 0.3 and 0.6 eV cosmologies. At the bottom of each panel there are the ratios between the velocity of outgoing and the incoming neutrino for each cosmology. The dashed black lines indicate the mean value of the void radii in the selected range and two times the same quantity.}
\end{figure}

The results for the cold dark matter (solid lines) and the neutrino (dotted lines) fields are shown in figure~\ref{fig:velocity_cdm-nu}. As for the density profile, we consider three ranges for the void radius and we plot on the left/right the results at redshifts $z=0/z=1$. 

The cold dark matter particles have a positive radial velocity inside the effective radius (first vertical dashed lines from the left), meaning that the inner part of the voids is expanding and becoming more and more empty. The behavior outside the effective radius changes depending on the void size. Around very small voids (top panel), the cold dark matter velocity becomes negative, meaning that the particles are moving towards the void. Therefore, there is an ongoing construction of the wall around the edge of the void. The void will eventually shrink and later collapse because it is surrounded by an overdense region (this is the so-called void-in-cloud effect). Around bigger voids, the radial velocity is always positive; it presents a peak near half of the void radius and it decreases on large distances from the center. Therefore, the outer region of the void expands slower than the inner part, producing nevertheless a concentration of mass around the edge of the void. Looking at different cosmologies, the average radial velocity is higher in $\Lambda$CDM and it decreases as the sum of the neutrino masses increases. Going from redshift $z=0$ to $z=1$, the velocities increase for all void radius and cosmologies and differences between models become more pronounced.

As expected, we find that the neutrino radial velocity field follows the cold dark matter one. The velocity is positive inside the void radius and remains positive also outside, apart from the case of very small voids, where it becomes negative. However, the neutrino velocity appears to be smaller than cold dark matter one, which is only due to cancellation effects. Neutrinos have large thermal velocities that make them free-stream in every direction, therefore the average velocity is expected to be close to zero. To better understand this effect, Fig.~\ref{fig:velocity_neutrinos} shows the average measured positive and negative neutrino radial velocity profile, which correspond to the profiles for the outgoing and incoming neutrinos, and their theoretical prediction. The predicted mean thermal velocities of incoming and outgoing neutrinos are equal due to spherical symmetry and it is computed as $v_{in}=v_{out}\simeq 160/2 \,(3 {\rm eV}/\Sigma m_{\nu})$ km/s \cite{LesgourguesPastor}. The two measured velocities are similar for small voids and small neutrino masses, but they differ otherwise.
It is the gravitational interaction with the void and the surrounding matter which gives rise to this and to a positive (or negative) average neutrino velocity. At higher redshift ($z=1$), the neutrinos are faster and they feel less the dynamics of the underlying cold dark matter structure. Indeed, the average neutrino radial velocity profile is lower at $z=1$ than at $z=0$.

The results for the total matter radial velocity field are presented in Fig.~\ref{fig:velocity_matter}. The main features described for the cold dark matter  radial velocity also apply here; however, the differences between different cosmologies are more pronounced. In the bottom panel of each plot we show the ratio between  the results for the massive neutrino cosmologies to the $\Lambda$CDM case. At redshift $z=0$ and in the inner part of the voids there are differences at the order of 5-10-15\% for 0.15-0.3-0.6 eV cosmologies, respectively. Larger differences can be seen in the outer parts of the voids, in the ranges $R_{\rm eff}$=10-11 Mpc/$h$ (top) and $R_{\rm eff}$=16-18 Mpc/$h$ (center). At redshift $z=1$ the differences are slightly more pronounced.

\subsection{$\Omega_{\nu}$-$\sigma_8$ degeneracy}

\begin{figure}[tbp]
\centering %\begin{center}/\end{center} takes some additional vertical space
\includegraphics[width=.55\textwidth,clip]{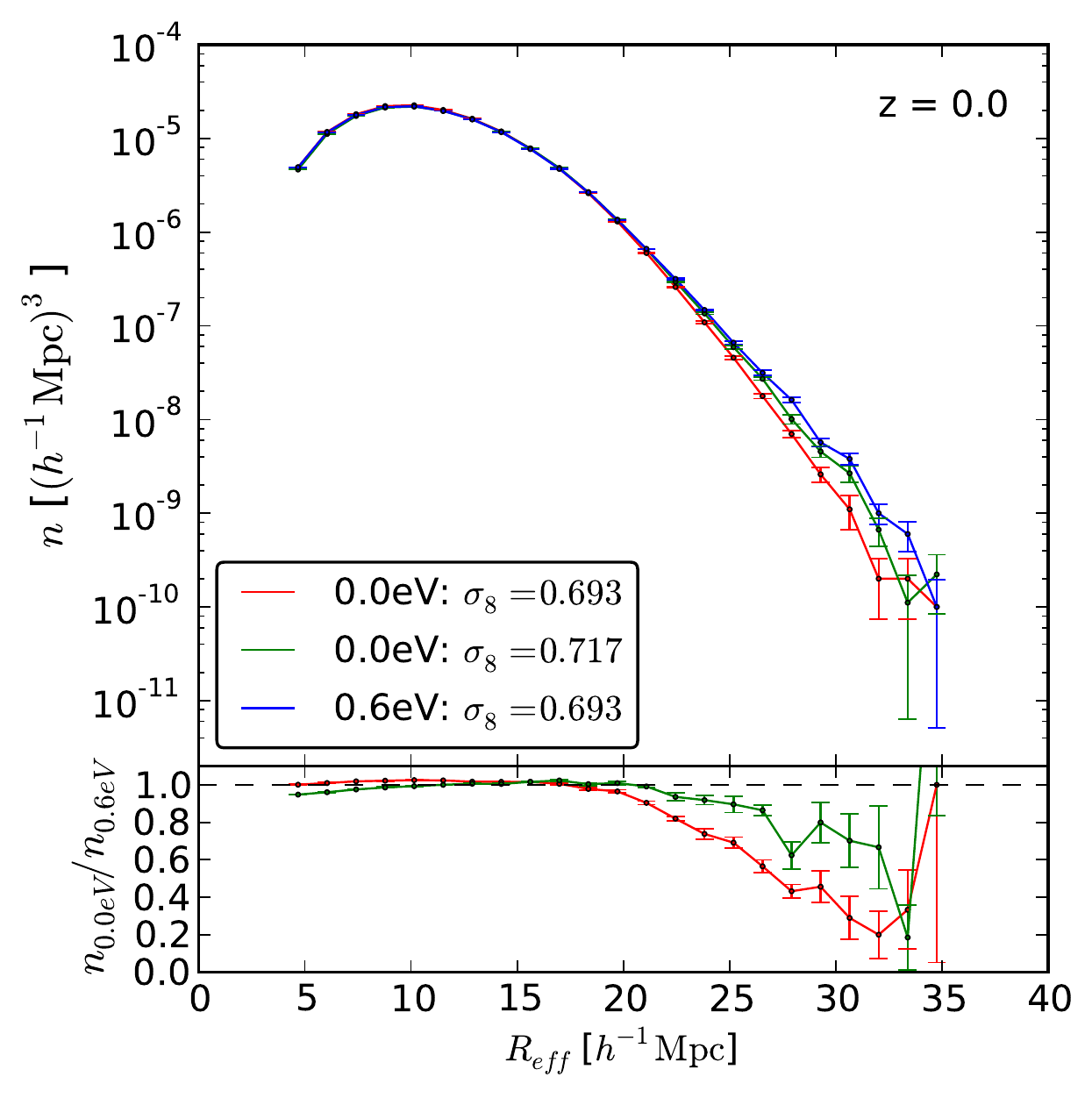}
%\hfill
% "\includegraphics" is very powerful; the graphicx package is already loaded
\caption{\label{fig:num_den_s8} Number density of voids as function of their effective radius. The top panel shows the results for different cosmologies: red and green lines are the $\Lambda$CDM cosmology with $\sigma_8$ equal to the one in the $\sum m_\nu=0.6$ eV case computed from the total matter and the CDM power spectra, respectively. The blue line correspond to the $\sum m_\nu=0.6$ eV cosmology. The bottom panels show the ratio between the void number density in massless and massive neutrino cases considered. }
\end{figure}
We also investigate the well known degeneracy between $\Omega_{\nu}$ and $\sigma_8$~\cite{Marulli_2011,Castorina_2013,Fontanot_2014}. For instance, the effects of massive neutrinos in many observables such as cluster number counts, can be mimicked by a cosmology with massless neutrinos but with a lower value of $\sigma_8$, which  is the r.m.s. value of the linear fluctuation in the mass distributions at $8$Mpc$/h$. We want to understand if this is the case for voids too. 

We use the simulations in $\Lambda$CDM with $\sigma_8$ taken from the $\sum m_\nu=0.6$ eV cosmology (L0s8 and L0s8CDM) to identify voids in the matter density field. If the degeneracy is present also in voids, these simulations should give catalogs very similar to the one of the 0.6 eV case. Moreover, we consider the two different $\sigma_8$ values because it has been shown that for halos the degeneracy arises when using the CDM field, since it is the fundamental one in driving the evolution of overdense structures~\cite{Castorina_2013}. We must check if this is the case for underdense structures as well.

In order to test all of this, we first compare the number density of voids and we plot the results in figure~\ref{fig:num_den_s8}. The $\Lambda$CDM cosmology with $\sigma_8$ computed from the total matter field (red line) matches quite well the predictions from the 0.6 eV case up to void radii equal to $\sim$17 Mpc/$h$. Above that, the departure is significant, giving a lack of big voids, and it reaches a disagreement at the level of 70-80\% for voids with size $\sim$30 Mpc/$h$. Things are different for the $\Lambda$CDM cosmology with $\sigma_8$ computed from the CDM field (green line). In this case there are fewer small voids and this translates into a worse match with the massive neutrino cosmology, which reach the 5\% for very small voids. On the other hand, the number density of big voids increases, mitigating the discrepancies at large $R_{\rm eff}$, which is here at the level of  $\sim$30\%.

In both cases, the results for the $\sum m_\nu=0.6$ eV cosmology cannot be reproduced, which means that the $\Omega_{\nu}$-$\sigma_8$ degeneracy is not present in the number density of voids. Voids, unlike halos, do not present this degeneracy because they are sensitive to different regions of the linear power spectrum. To be more precise, big voids should be influenced by scales in the $P(k)$ much larger than the ones important for the halos. On these large scales, the amplitude in the $\sum m_\nu=0.6$ eV case is higher than in the two considered massless neutrino cosmologies, and more power corresponds to a higher number density of voids.

We also perform the same comparison using the density and velocity profiles. The results are plotted in figure~\ref{fig:profiles_s8}, which we placed in Appendix~\ref{sec:degeneracy_profiles}.

%%%%%%%%%%%%%%%%%%%%%%%%%%%%%%%%%%%%%%%%%%%%%%
\section{Voids in the galaxy distribution}
\label{sec:voids_galaxies}
It has to be noticed that matter is not (unfortunately) directly observable and usually we rely on large scale structure tracers like galaxies. 
The distortion in the shape of galaxies caused by the weak gravitational effect can be used to constrain cosmology. The weak lensing signal will depend on the total matter density distribution and it would be interesting to answer the following two questions: 1) what is the matter density profile of the voids identified in the galaxy field? 2) what is the extent to which massive neutrinos impact the reconstructed profile? Recently some work have been done to address the first questions (see for example \cite{Krause_2012,Melchior_2013,Clampitt_2014}); here we focus on the second one. Therefore, in this section we present the analysis for voids identified in the galaxy distribution, which is obtained by post-processing $N$-body simulations. First, we explain how we populate halos with galaxies in order to generate a mock galaxy catalogue. Then we show the main results regarding the study of galaxy voids. Since the volume of our high-resolution simulations is relatively small and the number density of galaxy-voids is low, we cannot perform the analysis as for the voids in the cold dark matter field, studying all the void properties in different cosmologies. Therefore, we present only the number density of galaxy voids and the study of matter inside the identified galaxy voids. In particular, we show the matter density and velocity profiles around these objects.

\subsection{Halo Occupation Distribution}
In order to construct a mock galaxy catalog, we need to populate with galaxies the dark matter halos in $N$-body simulations. A commonly used tool is the so-called Halo Occupation Distribution (HOD) model. In this framework, the distribution of galaxies with respect to dark matter halos is described by a few parameters that can be calibrated in order to reproduce some particular features of the observed galaxy population. Here we calibrate the HOD parameters to reproduce the number density and the two-point clustering statistics of galaxies in the main sample of the Sloan Digital Sky Survey (SDSS) II Data Release 7~\cite{Zehavi:2010bh}.

The HOD model requires two ingredients: 1) the probability distribution $p(N|M)$ of having $N$ galaxies inside a halo of mass $M$ and 2) the way in which galaxies positions and velocities are related to the underlying matter particles. In our HOD model the first ingredient is described by three parameters: $M_{\rm min}$, $\alpha$ and $M_1$. We assume that halos with masses below $M_{\rm min}$ do not host any galaxies, whereas halos with masses above $M_{\rm min}$ host one central galaxy and a number of satellite galaxies following a Poissonian distribution with a mean equal to $(M/M_1)^\alpha$. We use the following equations:
\begin{equation}
\label{eq:54}
\langle N_c|M \rangle =
\begin{cases} 1 & \mbox{if } M\ge M_{\rm min} \\ 
0 & \mbox{if }M<M_{\rm min}\, 
\end{cases}
\; \; \; \; \; \; \; \; \; \; \; \; \;
\langle N_s|M \rangle =
\begin{cases} (M/M_1)^{\alpha} & \mbox{if } M\ge M_{\rm min} \\ 
\qquad 0 & \mbox{if }M<M_{\rm min}\, .
\end{cases}
\end{equation}
Our second HOD ingredient states that the central galaxy sits in the center of the corresponding halo and that the distribution and velocity of the satellites follow exactly the ones of the underlying cold dark matter particles inside the halo. This means that the galaxy bias and the velocity bias with respect to the cold dark matter field are both equal to 1. The values of the HOD parameters obtained for galaxies with magnitudes $M_r - 5 \log_{10} h = -21.0$ in the four different cosmologies are shown in table~\ref{tab:ii}.
\begin{table}
\begin{center}
\begin{tabular}{| c | c | c | c |}
\hline
$\sum m_{\nu}$ & $M_1$ & $\alpha$ & $M_{\rm min}$\\
(eV) &($M_{\odot}/h$) & &($M_{\odot}/h$)\\
\hline
$0.0$ & $1.22\times 10^{14}$ & $1.38$ & $4.92 \times 10^{12}$ \\
$0.15$ & $1.17 \times 10^{14}$ & $1.38$ & $4.79\times 10^{12}$\\
$0.3$ & $1.18 \times 10^{14}$ & $1.48$ & $4.50\times 10^{12}$\\
$0.6$ & $9.86 \times 10^{13}$ & $1.46$ & $4.16\times 10^{12}$\\
\hline
\end{tabular}
\caption{\label{tab:ii}Values of the HOD parameters, for four different cosmologies and for galaxies with magnitudes $M_r - 5 \log_{10} h = -21.0$.}
\end{center}
\end{table}

We focus our study at redshift $z=0$, where there are good measurements of the clustering of different galaxy populations. Now that we have created a mock galaxy catalog for each cosmology, we can run the void-finder VIDE on top of the galaxy field and identify the galaxy voids.

\subsection{Number density}
\begin{figure}[tbp]
\centering %\begin{center}/\end{center} takes some additional vertical space
\includegraphics[width=.55\textwidth,clip]{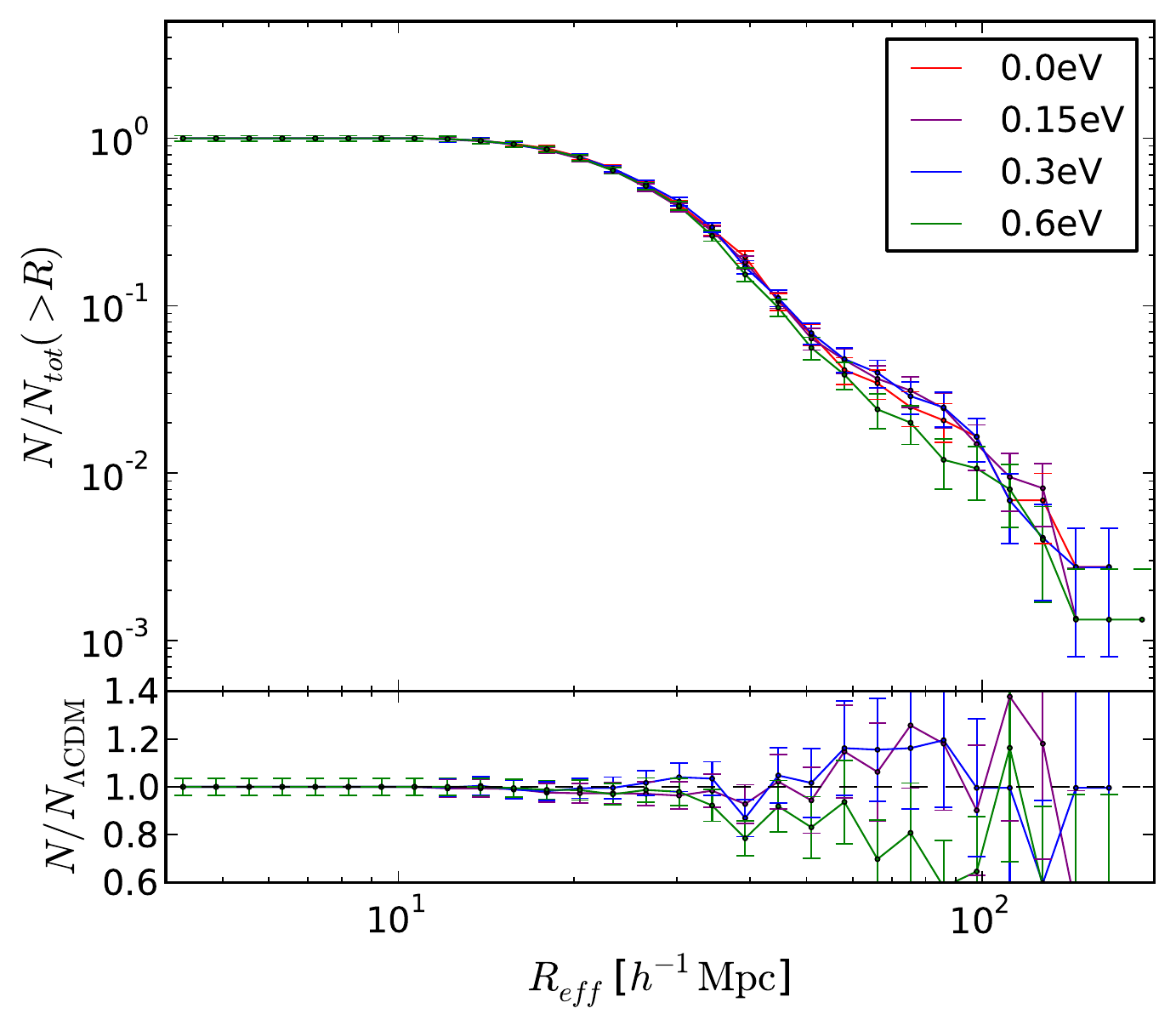}
%\hfill
% "\includegraphics" is very powerful; the graphicx package is already loaded
\caption{\label{fig:number_gvoid} Cumulative fraction of galaxy voids at $z=0$. The top panel shows results from different cosmologies: red line is the $\Lambda$CDM cosmology, and purple, blue and green lines correspond to 0.15, 0.3 and 0.6 eV cosmologies. The bottom panel shows the ratio between the same quantity in a massive and a massless neutrino cosmologies.}
\end{figure}

For each cosmology we compute the number of galaxy voids as a function of their size $R_{\rm eff}$
and we show the results in Fig.~\ref{fig:number_gvoid}. On the upper panel we plot the cumulative number of galaxy voids normalized by the total number and the associated errors, which are computed using error propagation and assuming a Poisson distribution in $R_{\rm eff}$ for the galaxy voids. Different colors indicate different cosmologies. On the bottom panel we display the ratio between the results for the massive neutrino cosmologies and the ones for the $\Lambda$CDM model. We can notice some differences among the cosmologies only for voids with $R_{\rm eff}>30$ Mpc$/h$, where anyway the errors are very large. Moreover, it is interesting to notice that the size of the voids identified in the galaxy field are larger than the ones found in the cold dark matter field. This is just a consequence of the number density of the tracers used to identify voids: a smaller value for the number density of tracers brings to larger voids.

\subsection{Density profile}

\begin{figure}[!tbp]
\centering %\begin{center}/\end{center} takes some additional vertical space
\includegraphics[width=.49\textwidth,clip]{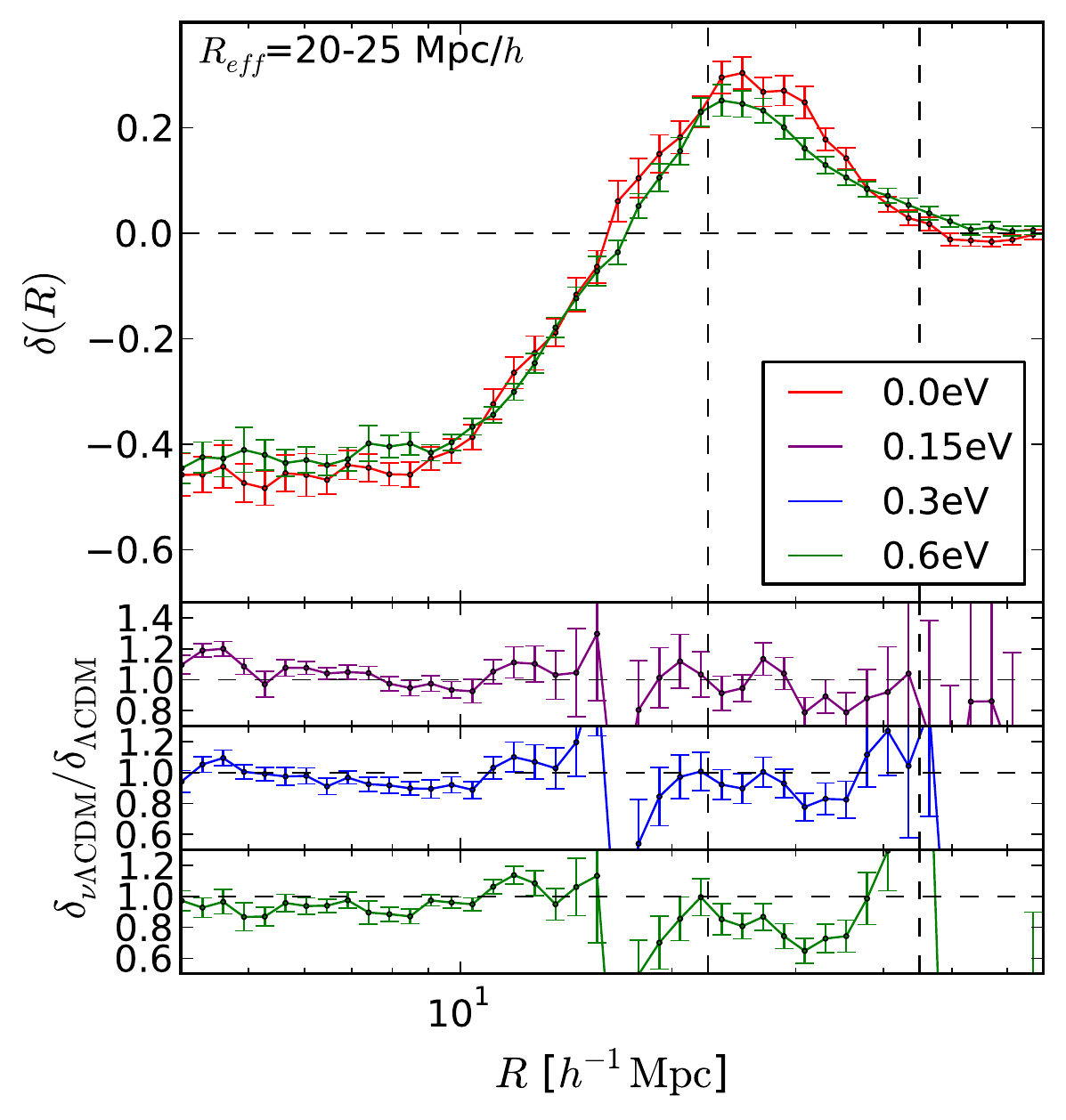}
%\hfill
\includegraphics[width=.49\textwidth,origin=c]{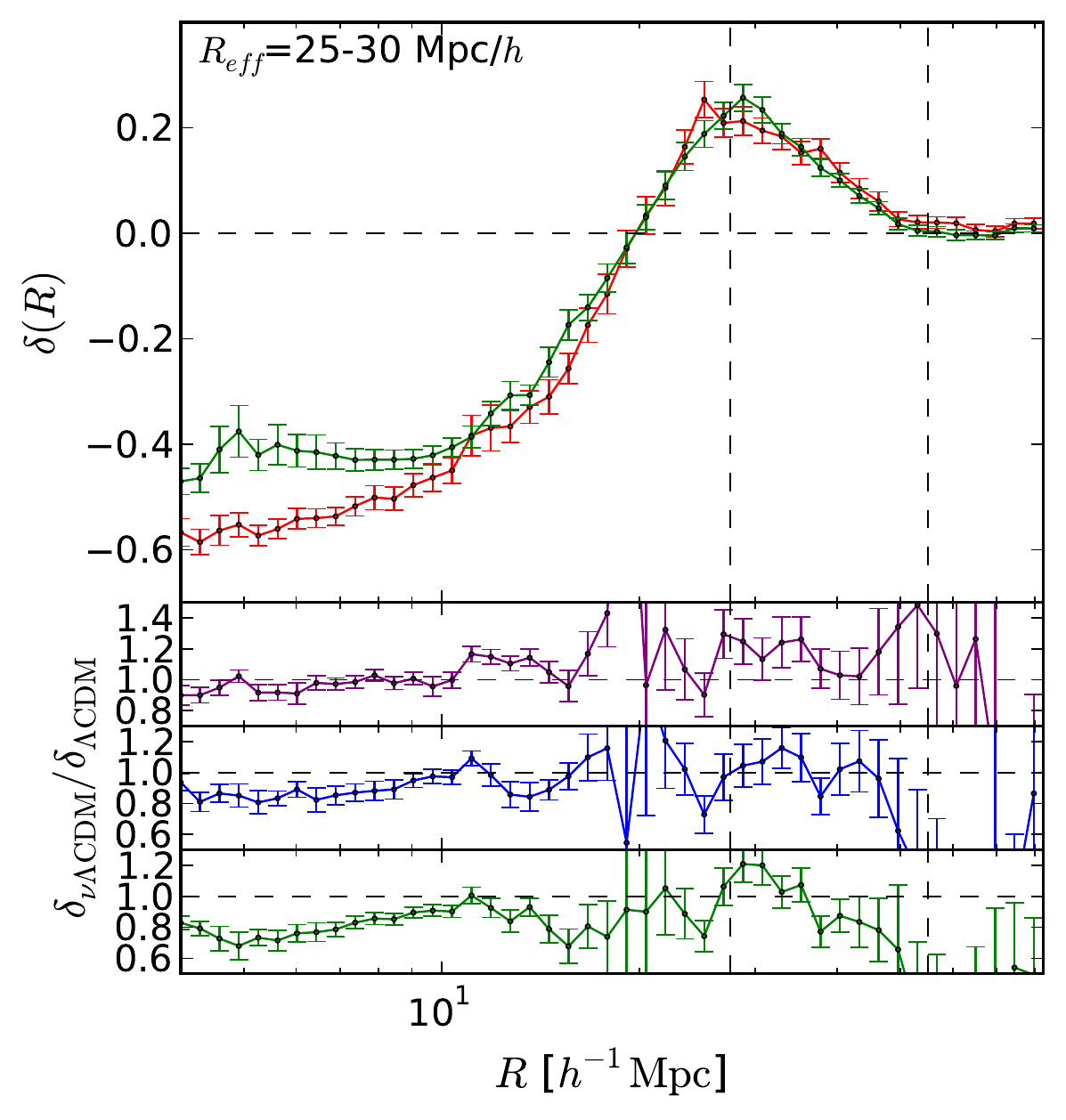}
\includegraphics[width=.49\textwidth,clip]{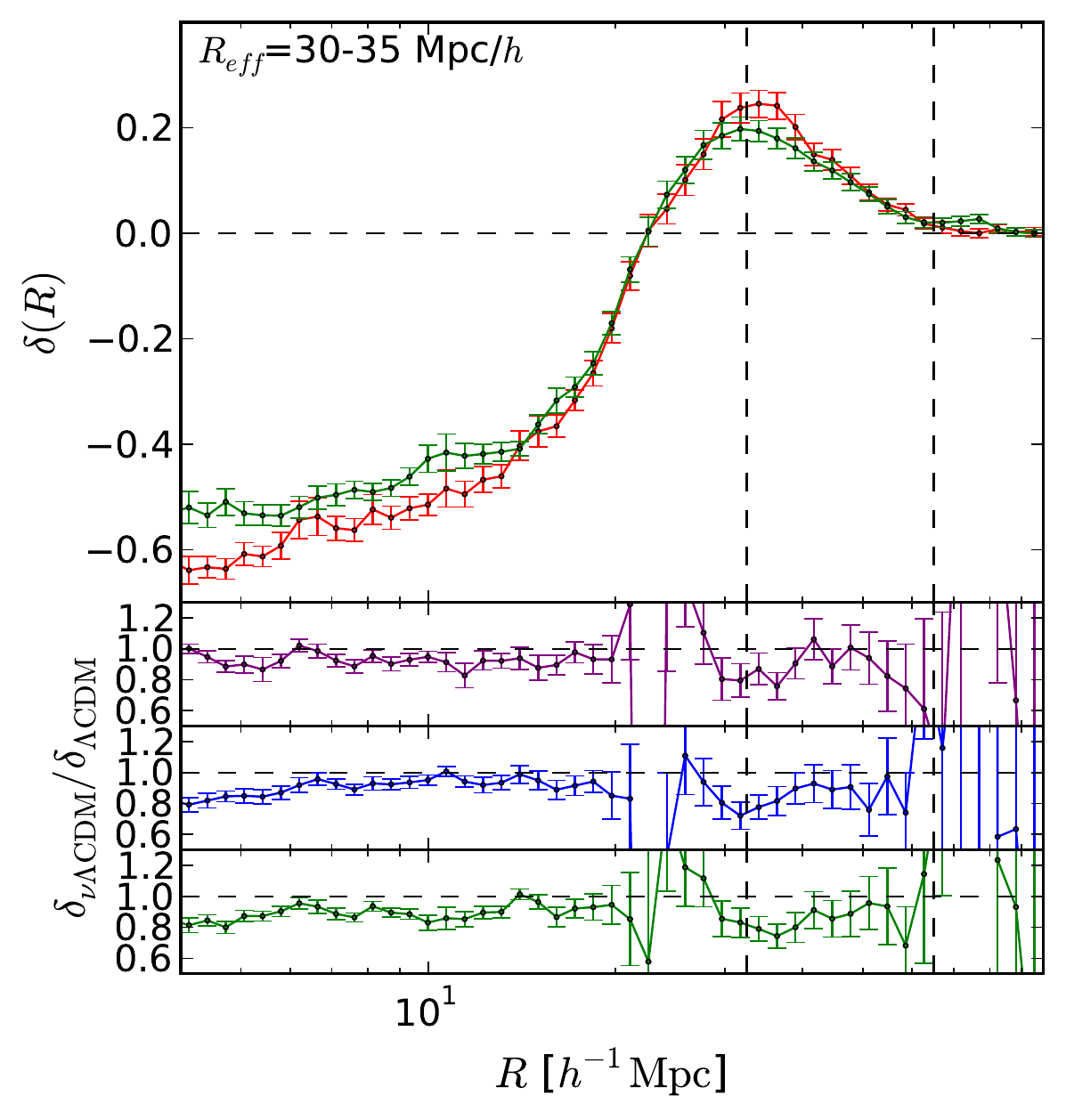}
%\hfill
\includegraphics[width=.49\textwidth,origin=c]{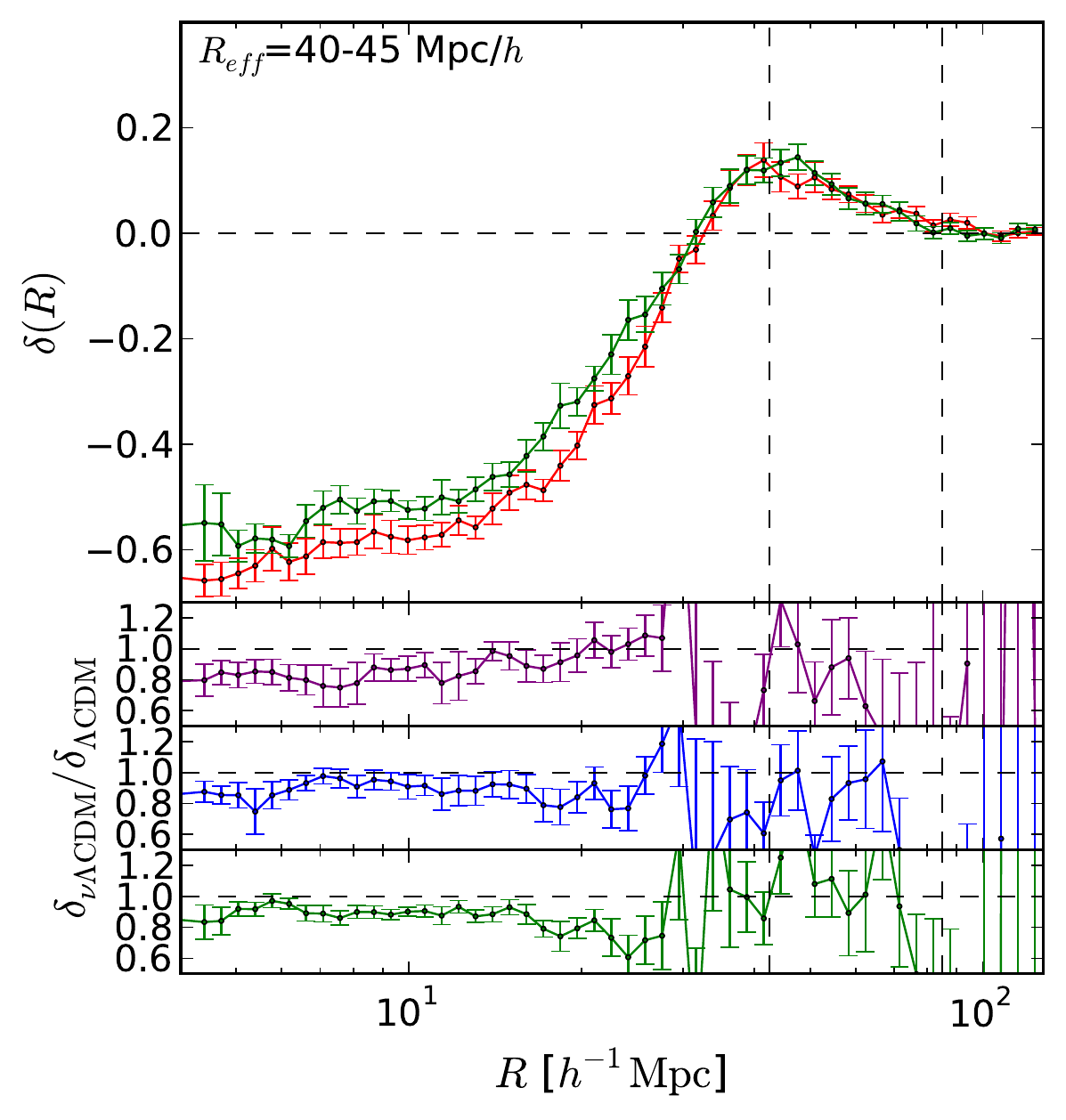}
\caption{\label{fig:density_matter_InGVoid} Density profile of the total matter field (cdm + neutrinos)  around galaxy voids with different sizes: $R_{\rm eff}=20-25$ Mpc/$h$ (top-left), $R_{\rm eff}=25-30$ Mpc/$h$ (top-right), $R_{\rm eff}=30-35$ Mpc/$h$ (bottom-left), and $R_{\rm eff}=40-45$ Mpc/$h$ (bottom-right). In each plot, the main panel shows the result for the 0.0 eV (red) and the 0.6 eV (green) cosmologies. The bottom panels display the ratios between the results from the massive neutrino cosmologies (0.6 eV but also 0.15 eV in purple and 0.3 eV in blue) to the $\Lambda$CDM one. The vertical dashed black lines indicate the mean value of the void radii in the selected range and two times the same quantity.}
\end{figure}

Here we compute the total (cold dark matter + neutrinos) matter density profile in the identified galaxy voids, once they are divided into four groups, depending on their effective radius: $R_{\rm eff}$=20-25, 25-30, 30-35, 40-45 Mpc/$h$. Figure~\ref{fig:density_matter_InGVoid} shows the profiles (top panel of each plot) only for 0.0 eV and 0.6 eV cosmologies, while the residuals between massive and massless cases (bottom panel of each plot) for all the massive neutrino cosmologies are displayed. As in the previous section, the vertical dashed lines indicate the effective radius and two times the same quantity. 

We  immediately notice that the matter profile here, in galaxy voids, is very different from the one around voids identified in the cold dark matter field (see figure~\ref{fig:matter_dens_profile}). First of all, galaxy voids are less empty. Secondly, there is a compensated wall even around very large voids. These differences are due to the fact that galaxy voids incorporate within them many small cold dark matter voids, together with small halos and filaments. Their edge presents an overdensity in the galaxy field, which can sit only in very big overdensities in the cold dark matter field. In other words, this should be due to the galaxy bias $b$~\cite{Elena_void_2015}. If we assume that voids are patches where the enclosed number density contrast of the tracers is equal to a certain threshold $\delta_t = -0.8$, and that the tracers are galaxies with a density field $\delta_g = b\, \delta_m$ biased with respect to the matter field, then the corresponding enclosed matter density will be $\delta_m = \delta_t/b$. If the galaxy bias is $b\sim 2$, the enclosed matter density will be 2 times denser then the galaxy one. 

Focusing on the comparison among cosmologies, we can notice the same trend as in the cold dark matter voids: the matter profiles of $\Lambda$CDM are emptier and they present a higher wall. However, here the small galaxy void statistics makes this trend less clear. Moreover, the differences among $\Lambda$CDM and massive neutrinos cosmologies are more noisy, but not less pronounced. In small voids there are smaller differences, whereas in large voids the differences are of the order of 10-20\% in the inner part of voids and for all the massive neutrino cosmologies; the discrepancies are less pronounced near the compensated wall.

\subsection{Velocity profile}
\begin{figure}[!tbp]
\centering %\begin{center}/\end{center} takes some additional vertical space
\includegraphics[width=.49\textwidth,clip]{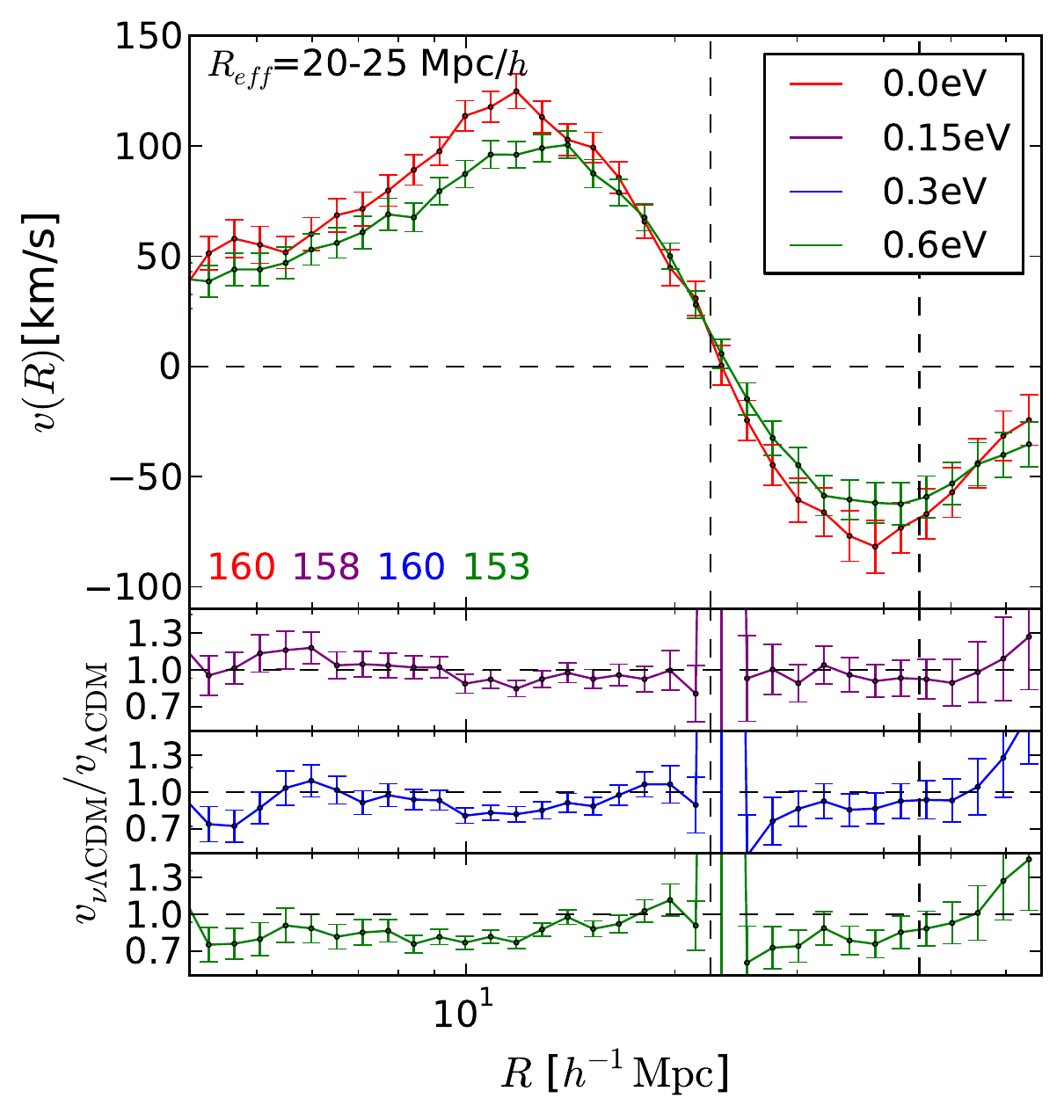}
%\hfill
\includegraphics[width=.49\textwidth,origin=c]{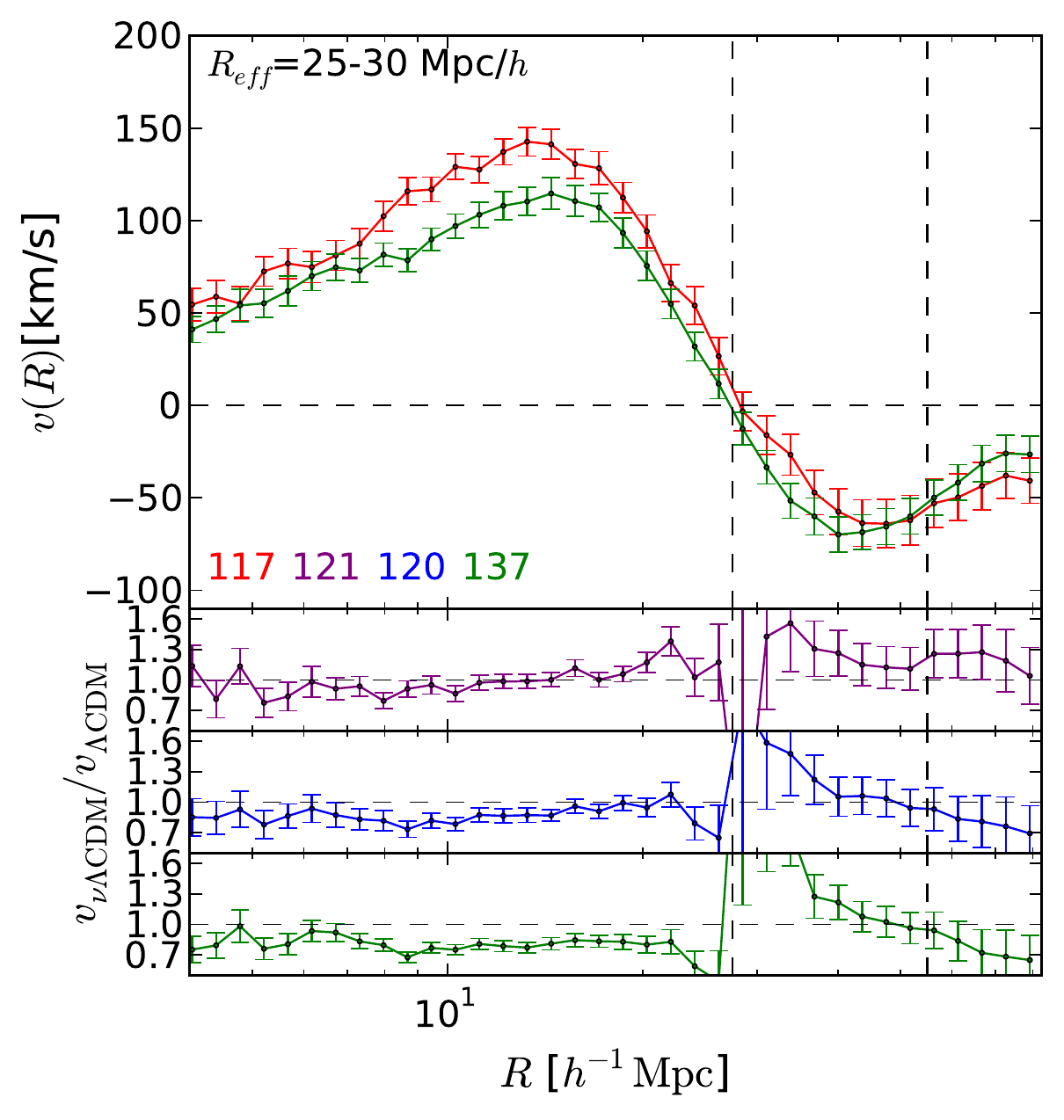}
\includegraphics[width=.49\textwidth,clip]{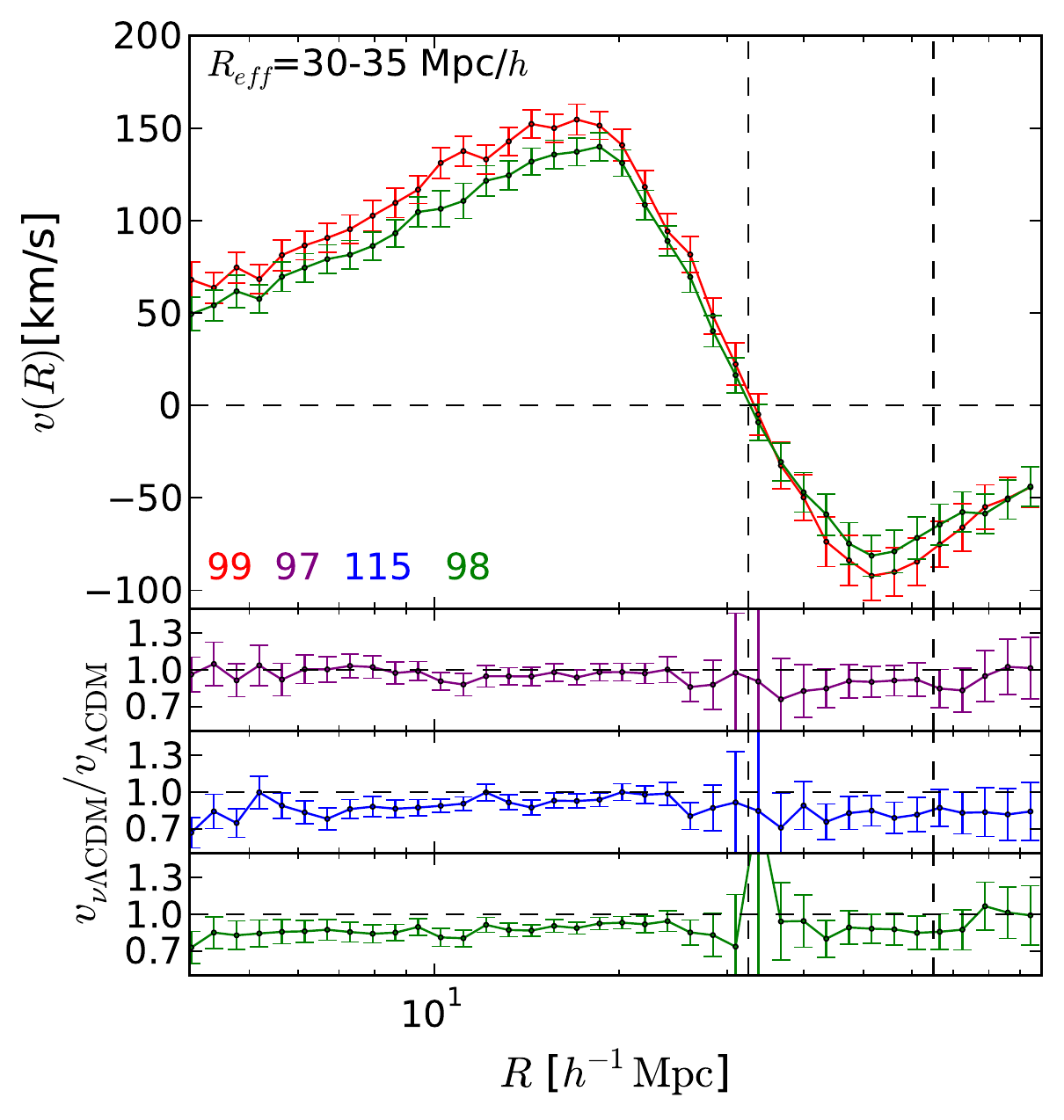}
%\hfill
\includegraphics[width=.49\textwidth,origin=c]{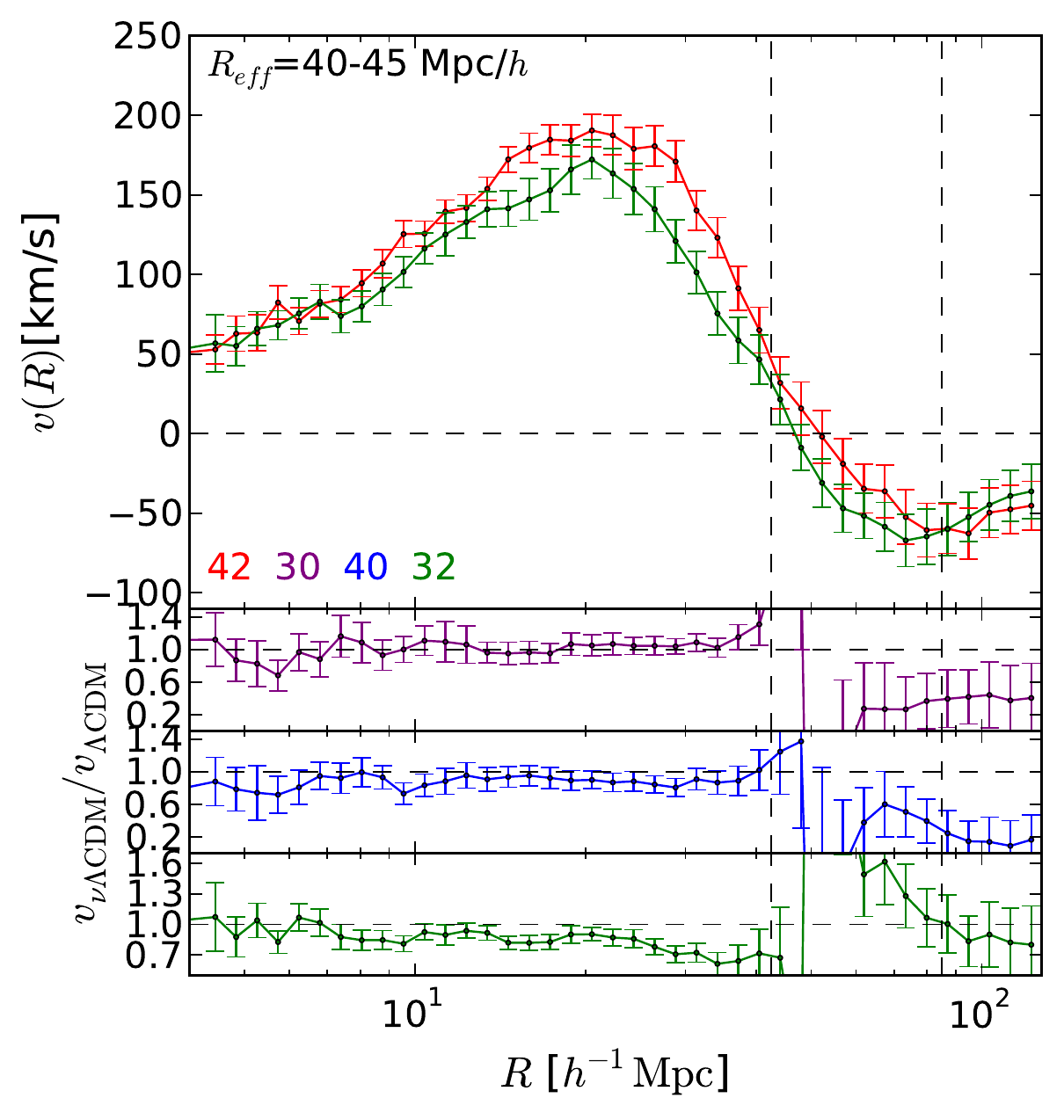}
\caption{\label{fig:velocity_matter_InGVoid} Radial velocity profile of the total matter field (cdm + neutrinos)  around galaxy voids with different sizes: $R_{\rm eff}=20-25$ Mpc/$h$ (top-left), $R_{\rm eff}=25-30$ Mpc/$h$ (top-right), $R_{\rm eff}=30-35$ Mpc/$h$ (bottom-left), and $R_{\rm eff}=40-45$ Mpc/$h$ (bottom-right). In each plot, the main panel shows the results for the 0.0 eV (red) and the 0.6 eV (green) cosmologies. The bottom panels display the ratios between the results from the massive neutrino cosmologies (0.6 eV but also 0.15 eV in purple and 0.3 eV in blue) to the $\Lambda$CDM one. The vertical dashed black lines indicate the mean value of the void radii in the selected range and two times the same quantity, whereas the colored numbers indicate the number of stack voids in each simulation.}
\end{figure}
We present here the analysis for the velocity profile of matter around galaxy voids. We select the galaxy voids in the same size ranges as for the density profile. We compute the radial velocity of matter as a density weighted average between the velocity profiles of cold dark matter and neutrino fields, as shown in equation~\eqref{eq:velocity}. The associated errors are computed via error propagation, using the dispersion around the mean values of  density and velocity profiles of cold dark matter and neutrinos obtained from the stack of the selected voids in the simulation. 

The results are shown in Fig.~\ref{fig:velocity_matter_InGVoid}. Again we plot the profiles for $\Lambda$CDM and 0.6 eV cosmologies only, but we present the residuals with respect to $\Lambda$CDM for all the massive neutrino cosmologies considered in this paper. The main feature to notice is the particular shape of the profiles. They are positive in the inner part of the galaxy voids, they reach zero around the effective radius and they are negative outside. Even the profiles around very big voids share the same shape, contrary to what happens in cold dark matter voids (see figure~\ref{fig:velocity_matter}). It seems that in these regions the cold dark matter is building a wall near the effective radius, with the additional effect of shrinking the underdensity. 

If we compare the cosmologies we notice that inside galaxy voids the velocity is higher in $\Lambda$CDM, whereas in outer regions the trend is not unique for all the void sizes. However, for most of the cases, the massive neutrino cosmologies have slower motion towards the center of the galaxy voids.

%%%%%%%%%%%%%%%%%%%%%%%%%%%%%%%%%%%%%%%%%%%%%%
\section{Conclusions}
\label{sec:conclusions}

The aim of this work is to analyze the relative differences in the void properties due to the presence of massive neutrinos.
We have run two sets of $N$-body simulations, one with low resolution and another with higher resolution. Each set of simulations have been computed in four different cosmologies, i.e. with neutrino masses equal to 0.0 eV, 0.15 eV, 0.3 eV and 0.6 eV. We have used the low resolution simulations to identify voids in the CDM particle distribution and the high resolution ones for studying voids in the galaxy field. 

For the voids in the CDM field, we have initially studied the number density of voids in the four different cosmologies at redshift $z = 0$ and $z = 1$. In both cases, the number density of small voids is higher in massive neutrino cosmologies, whereas the number density of big voids is lower. The difference between $\Lambda$CDM and cosmologies with massive neutrinos increases with the neutrino mass, but it decreases with the redshift. This can be understood in terms of the total mass contained in the void. Indeed, neutrinos have high thermal velocity which prevents them to feel the void dynamics and to create a deep underdensity around the void center. This translates into an extra mass within the void that will evolve slowly. Therefore voids in the massive neutrino universe are smaller. However, the additional mass has a smaller effect at high redshift, where voids are denser, thus resulting in smaller differences at $z=1$.

Secondly we studied the distribution of void ellipticities. In $\Lambda$CDM we found that the fraction of voids with low ellipticity is smaller than in the massive neutrino cases. On the other hand, the fraction with high ellipticity is larger. This can be interpreted in term of evolutionary stages of voids in the different cosmologies: voids in massive neutrinos universes are younger than in their massless neutrino cosmology corresponding model.

We computed the correlation function of voids having radius within specific ranges. The results showed the exclusion effect on very small scales, which is due to the fact that voids are extended objects that cannot overlap. This effect, in addition to the presence of void clustering on small-intermediate scales, creates a positive peak, which is placed at different distances depending on the void size considered. The height of the peak changes with the cosmology: it increases with lower neutrino masses. Looking at  the differences between redshift $z = 0$ and $z = 1$, we can understand that this is due to the fact that voids in massive neutrino cosmologies are effectively younger. Moreover, differences between cosmologies become less important with redshift.

Another interesting property of voids is their density profile. We computed the CDM, the neutrino, and the total matter density profiles for different void sizes and in the four different cosmologies. We found that small voids present a compensated wall at the edge, whereas large voids have a non-positive profile. Again, we can understand the differences between massive and massless cosmologies once we consider that voids are effectively less evolved in the presence of massive neutrinos. Indeed, voids evolve by evacuating particles, becoming progressively emptier and building the wall. For each void size in both CDM and matter density field, $\Lambda$CDM cosmology presents higher walls and emptier cores, in agreement with this explanation. Instead, the neutrinos profile is flatter, given the neutrinos' high thermal velocities. They also tend to follow the CDM one: around small voids there is a positive overdensity of neutrinos in correspondence with the wall, whereas the neutrino profile presents usually an underdensity in the core of big voids. All departures from the mean background get more pronounced as the neutrino mass increases. Focusing on the total matter profile, which is the most important one since most of the observables depend on the total matter distribution, we obtained the following results. In the void core, the differences between $\Lambda$CDM and the 0.15 eV cosmology are at the level of 1-3\%, depending on the void radius, and they reach the 10\% for the 0.6 eV case. Near the wall and for the very small voids, the difference is at the 5\% level for 0.15 eV and at the 20\% level for 0.6 eV cosmologies. The differences increase at redshift $z=1$.

The last investigated property of voids identified in the CDM field is the radial velocity profile of CDM, neutrino, and total matter around voids. The CDM and total matter profiles have the following behavior: the radial velocity is positive inside the void radius around which it reaches the maximum, then it decreases and it becomes negative for very small voids. It remains positive, but at a small value, for large ones. This means that particles inside voids move towards the edge, evacuating and expanding the voids, whereas the outer particles have different behaviors depending on the void size. In small voids they move towards the center participating in the building up of the wall and eventually making the void collapse (the void-in-cloud effect). Instead, in big voids they go far away, but at somewhat smaller velocity than the ones in the inner parts, giving rise to a concentration of mass around the edge of the void. The neutrino radial velocity follows the same behavior, but it has a smaller magnitude, and this is due only to a cancellation effect. Indeed, neutrinos have high thermal velocities that make them free-stream in every direction and the average velocity is close to zero.

We also studied the degeneracy between $\sigma_8$ and $\Omega_{\nu}$ running two set of 10 simulations in $\Lambda$CDM with $\sigma_8$ taken from the $\sum m_\nu=0.6$ eV cosmology. We saw that this degeneracy is not present when focusing on void properties since their number density, but also their density profile and velocity, is different in massless and massive cosmologies with same $\sigma_8$.

In the second part of this work we have analyzed the high resolution N-body simulations. We have populated these simulations with galaxies, following the HOD prescriptions, and we have identified voids in the galaxy distribution. We have shown the cumulative fraction of voids at redshift $z=0$. There are some differences among the cosmologies only for voids with $R_{\rm eff} > 30$ Mpc$/h$, where the errors are very large due to the small volume probed by our high resolution simulations. Moreover, the voids are bigger than the ones selected in the CDM field, since the number of tracers here is low given the small box-size. Current surveys can span a larger volume: for example the Sloan Digital Sky Survey\footnote{http://www.sdss.org} III (SDSS-III) with the Data Release 9 (DR9) CMASS sample can covers an effective volume of nearly $1.5$ (Gpc$/h$)$^3$, which is about $20$ times bigger than our box-size.

Finally we studied the total matter density and velocity profiles around the galaxy voids, since future surveys like Euclid\footnote{www.euclid-ec.org} and DESI\footnote{desi.lbl.gov} are expected to significantly increase the number of observed galaxies. Furthermore, the weak lensing signal will allow to measure the total matter density field. The matter density profiles present some differences with respect to the ones in the CDM voids: they are less underdense and they present the wall even around very big voids. We argued that this is probably due to the galaxy bias with respect to the matter density field. Instead, for what regards comparison among cosmologies, we found the same behavior as in the CDM voids. The velocity profiles show a peculiar shape shared by all the void sizes analyzed, despite of what happens for the CDM voids. The profiles are positive in the inner part of the galaxy voids, they reach zero around the effective radius and they are negative outside. In the inner regions the velocity is higher in $\Lambda$CDM, whereas in the outer part of the voids the trend is not unique.

\acknowledgments We thank Ravi Sheth, Emanuele Castorina, Marcello Musso, Emiliano Sefusatti and Susana Planelles for useful conversations. N-body simulations have been run in the Zefiro cluster (Pisa, Italy).  FVN and MV are supported by the ERC
Starting Grant ``cosmoIGM'' and partially supported by INFN IS PD51
"INDARK".  We acknowledge partial support from "Consorzio per la
Fisica - Trieste".

\appendix
\section{$\Omega_{\nu}$-$\sigma_8$ degeneracy: profiles}
\label{sec:degeneracy_profiles}

\begin{figure}[!tbp]
\centering %\begin{center}/\end{center} takes some additional vertical space
\includegraphics[width=.45\textwidth,clip]{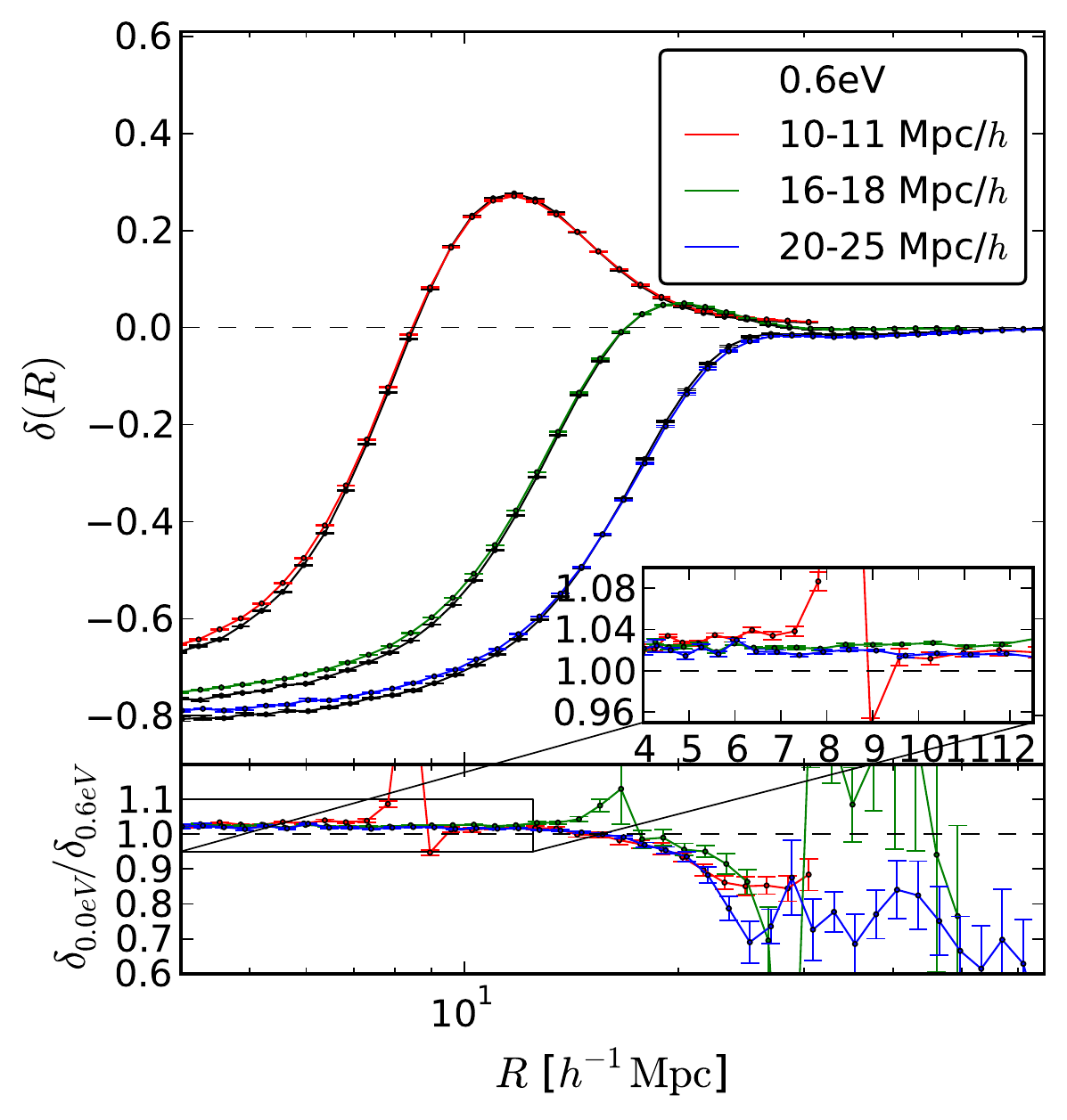}
%\hfill
\includegraphics[width=.45\textwidth,origin=c]{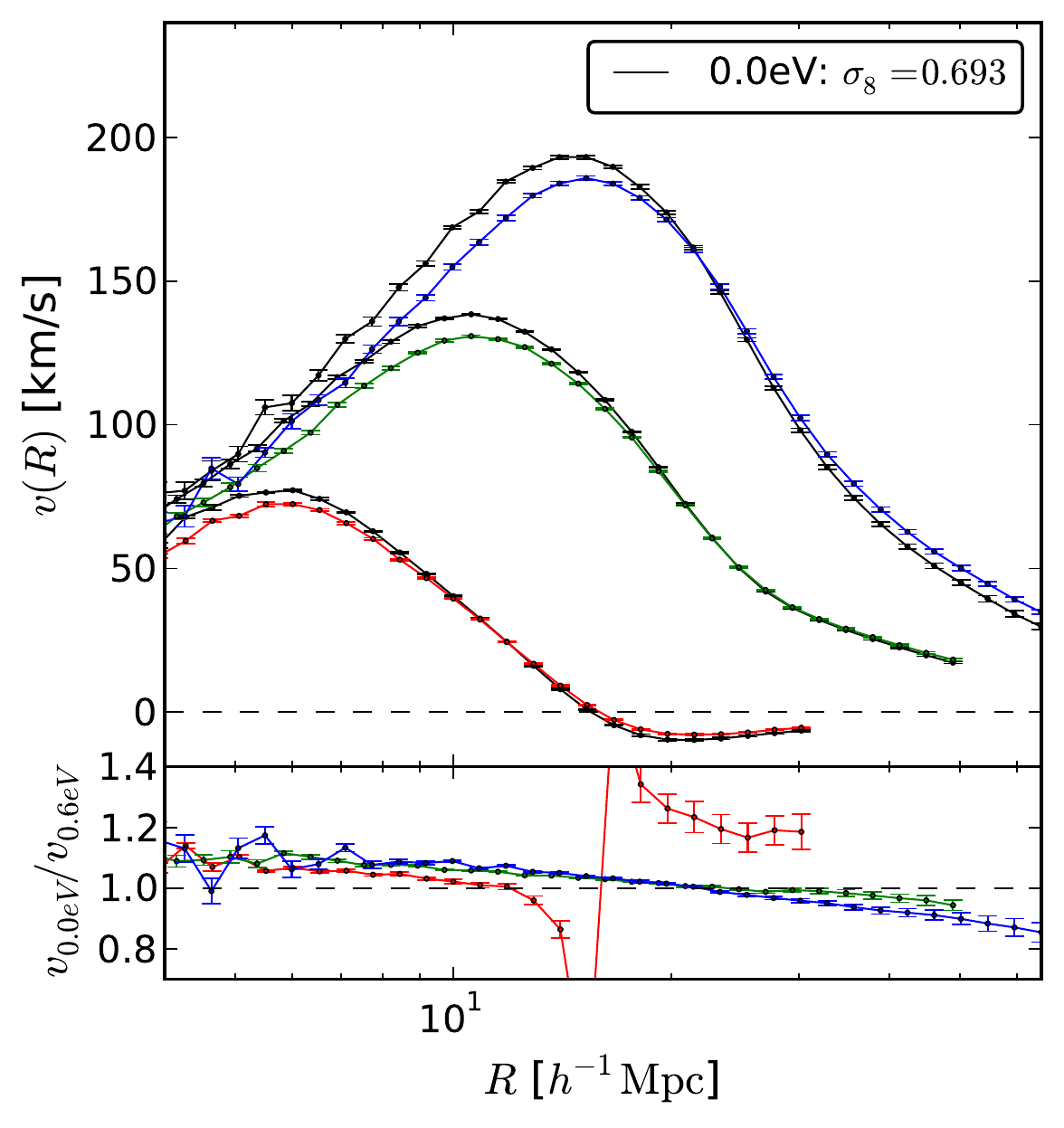}
\caption{\label{fig:profiles_s8} Left panel: matter density profile. Right panel: radial velocity profile. In both cases colored lines show the profiles for the $\sum m_\nu=0.6$ eV cosmology, whereas black lines indicate $\Lambda$CDM cosmology with the same $\sigma_8$ of the massive neutrino case, when it is computed using the total matter density field. The bottom panels show the ratios between the results from the two cosmologies. }
\end{figure}

We present the second part of the analysis for investigating the $\Omega_{\nu}$-$\sigma_8$ degeneracy using voids. Here, we focus on the comparison between the $\sum m_\nu=0.6$ eV cosmology and the massless case with same $\sigma_8$ computed from the total matter linear power spectrum. Figure~\ref{fig:profiles_s8} shows the density and velocity profiles of the matter field around voids selected in ranges of $R_{\rm eff}$. The colored lines show the results for the massive neutrino case and black lines indicate the massless one. The mismatch in the density profile is around 2-4\% in the inner part of voids and it increases in the outer part, whereas for the velocity profiles it reaches the 10\% in the void core.

\section{Void-by-void comparison among different cosmologies}
\label{sec:profiles_sp}
\begin{figure}%[!tbp]
\centering %\begin{center}/\end{center} takes some additional vertical space
\includegraphics[width=.41\textwidth,clip]{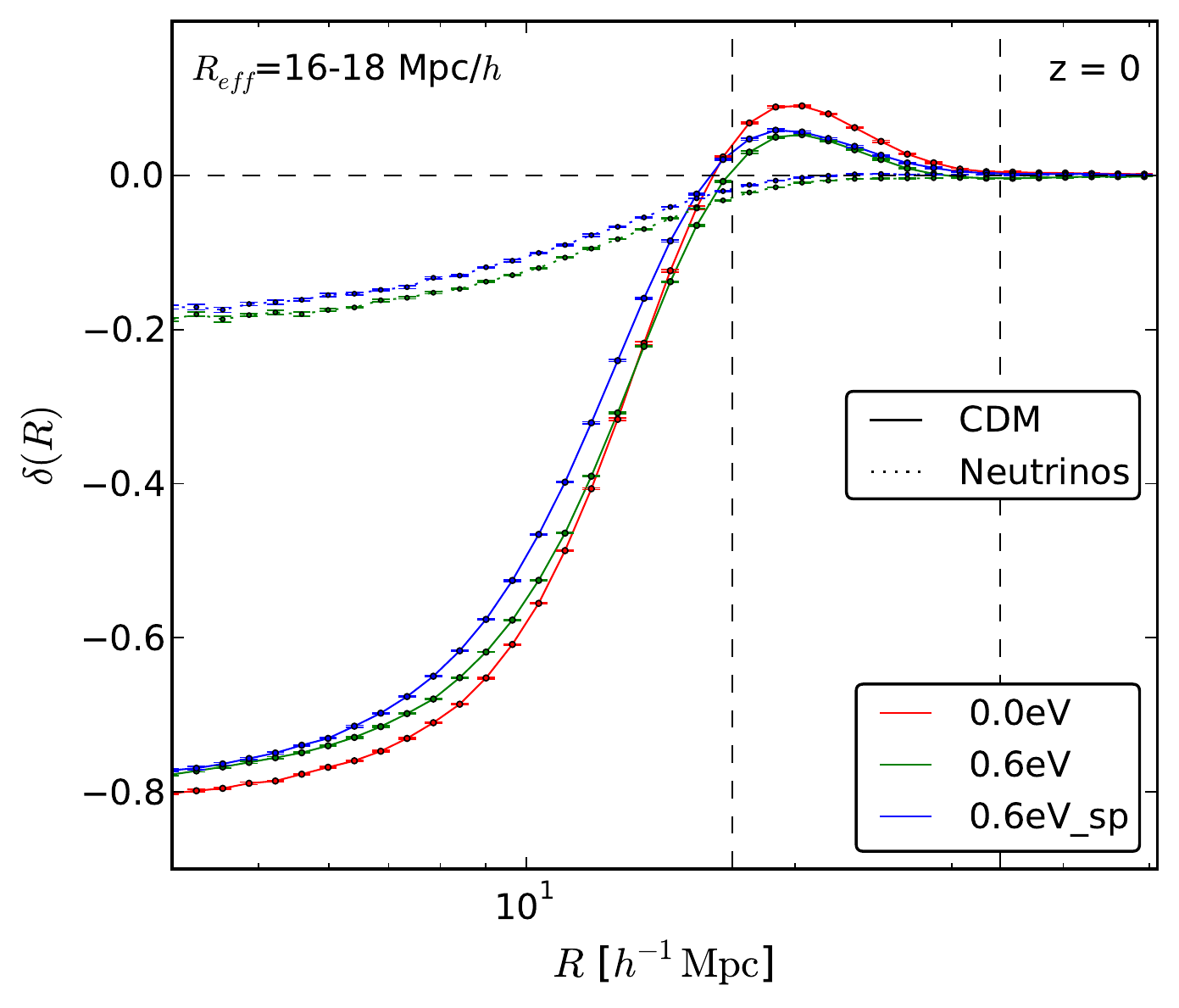}
%\hfill
\includegraphics[width=.41\textwidth,origin=c]{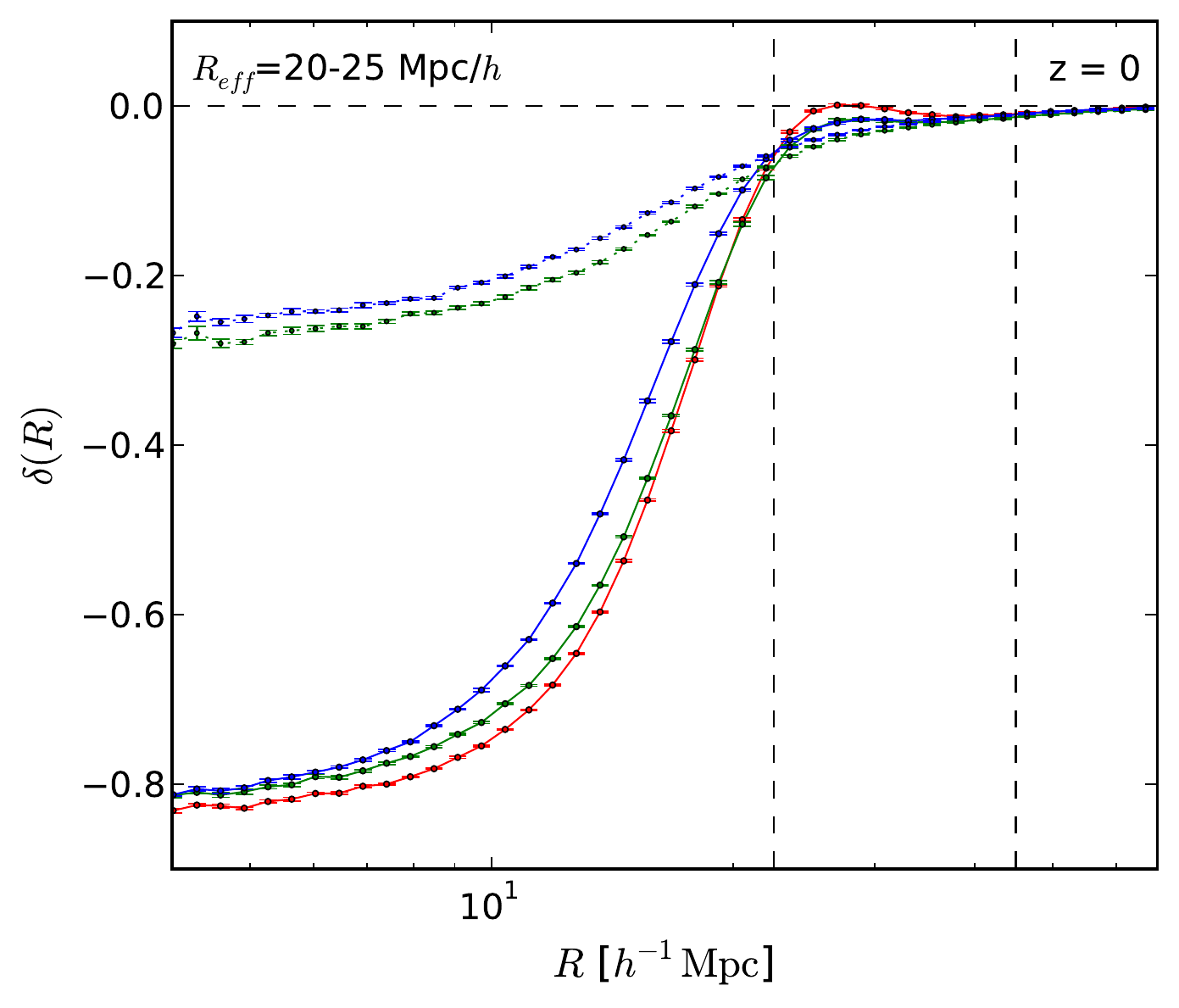}
\includegraphics[width=.41\textwidth,clip]{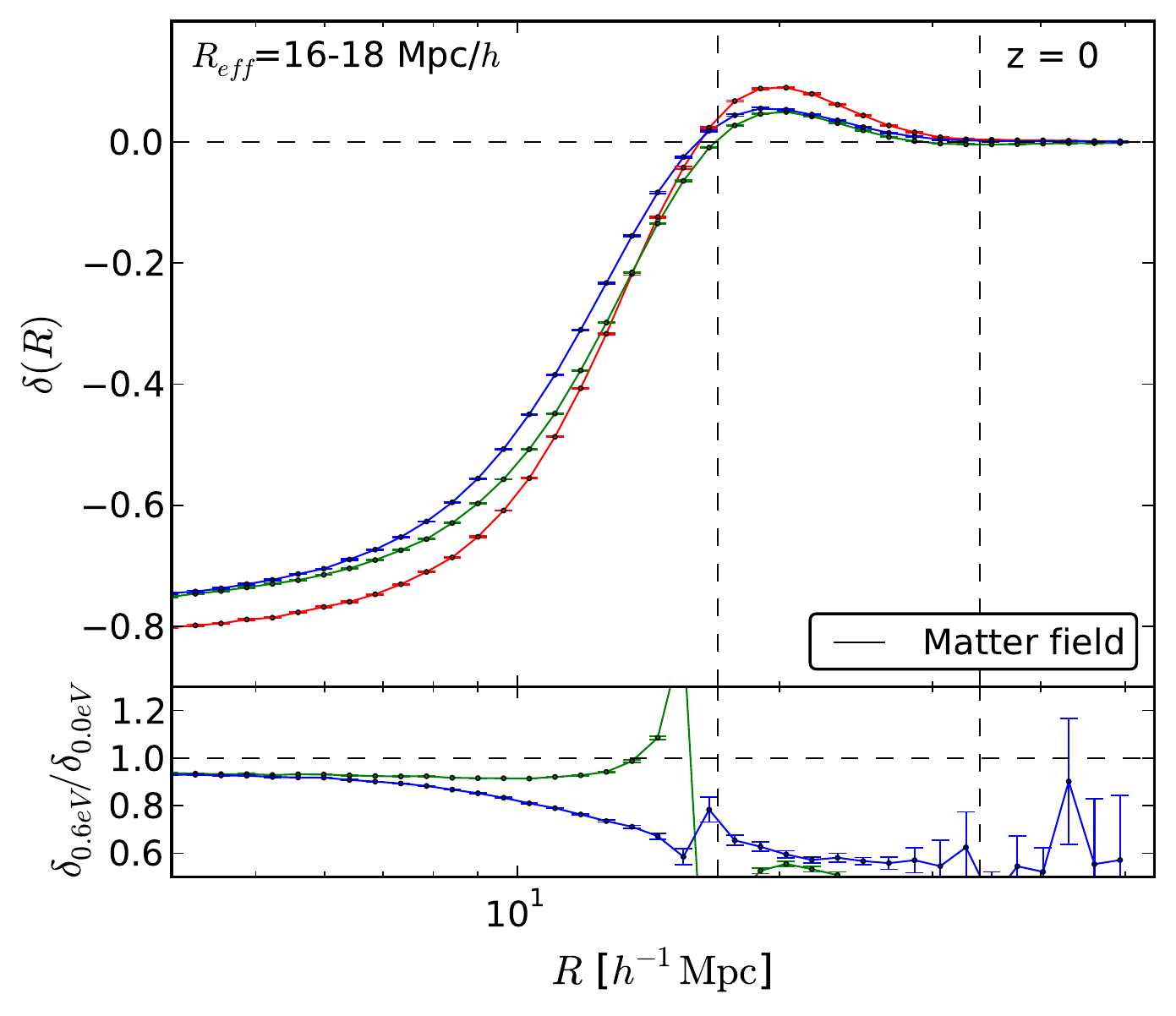}
%\hfill
\includegraphics[width=.41\textwidth,origin=c]{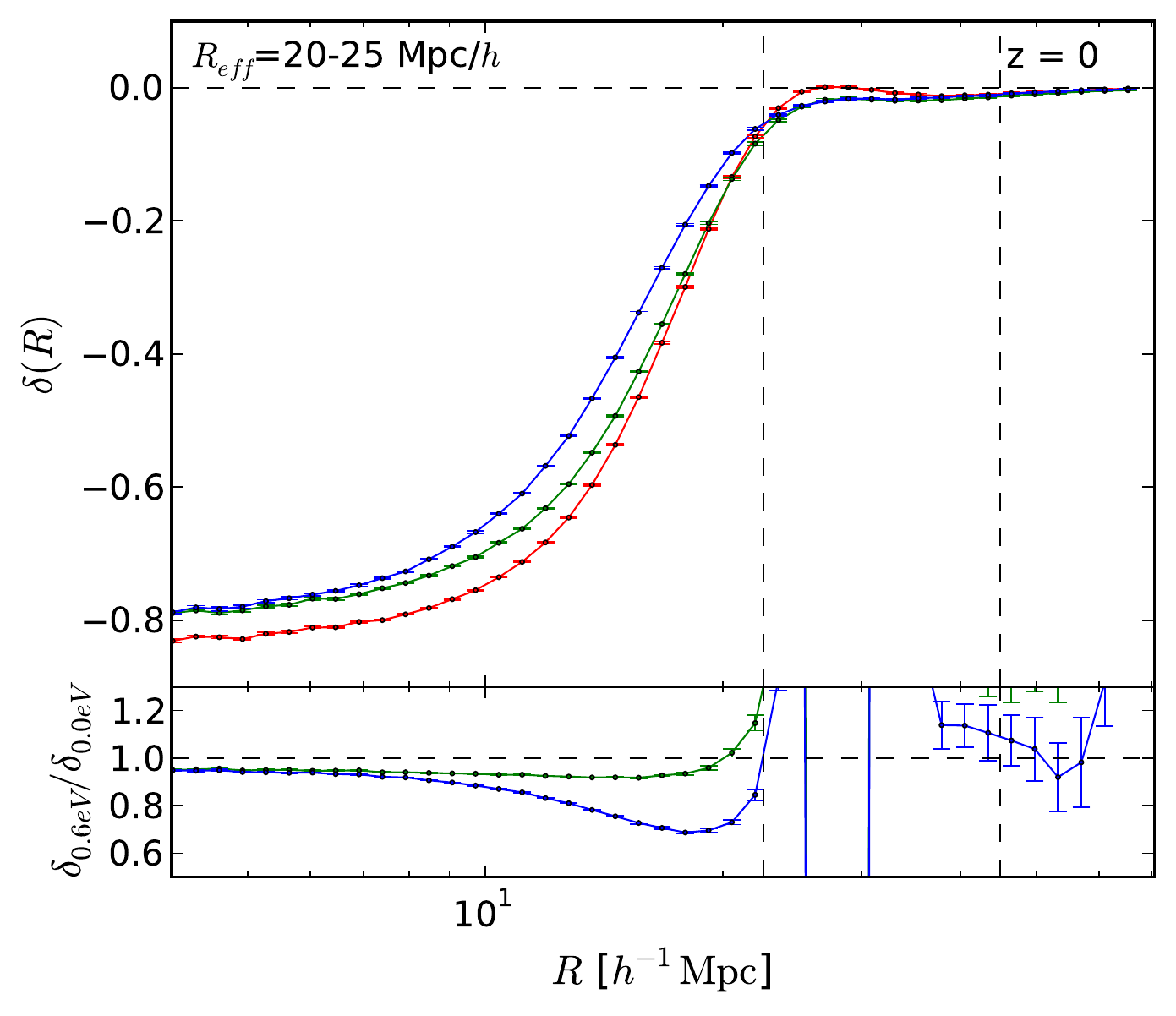}
\includegraphics[width=.41\textwidth,clip]{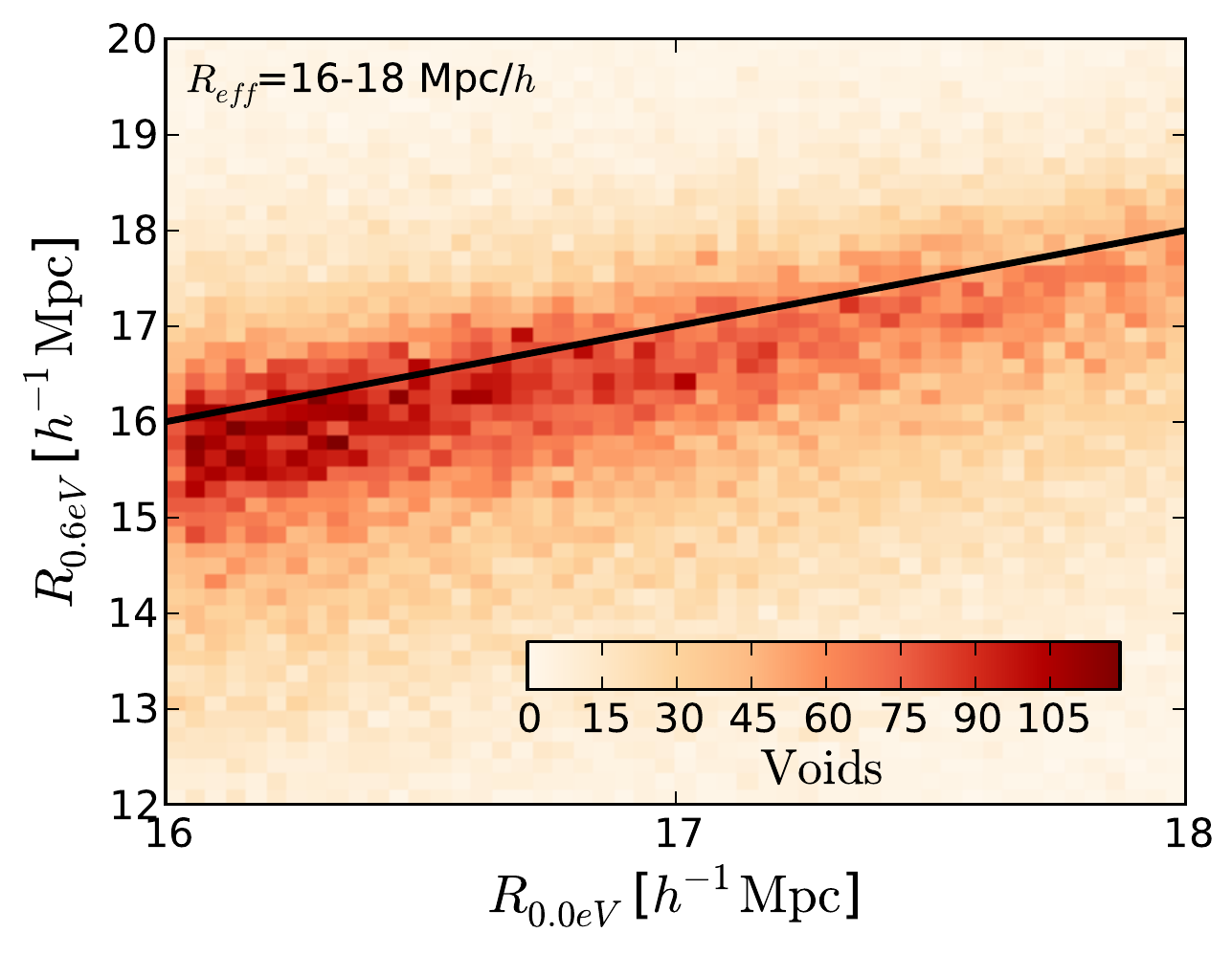}
\includegraphics[width=.41\textwidth,origin=c]{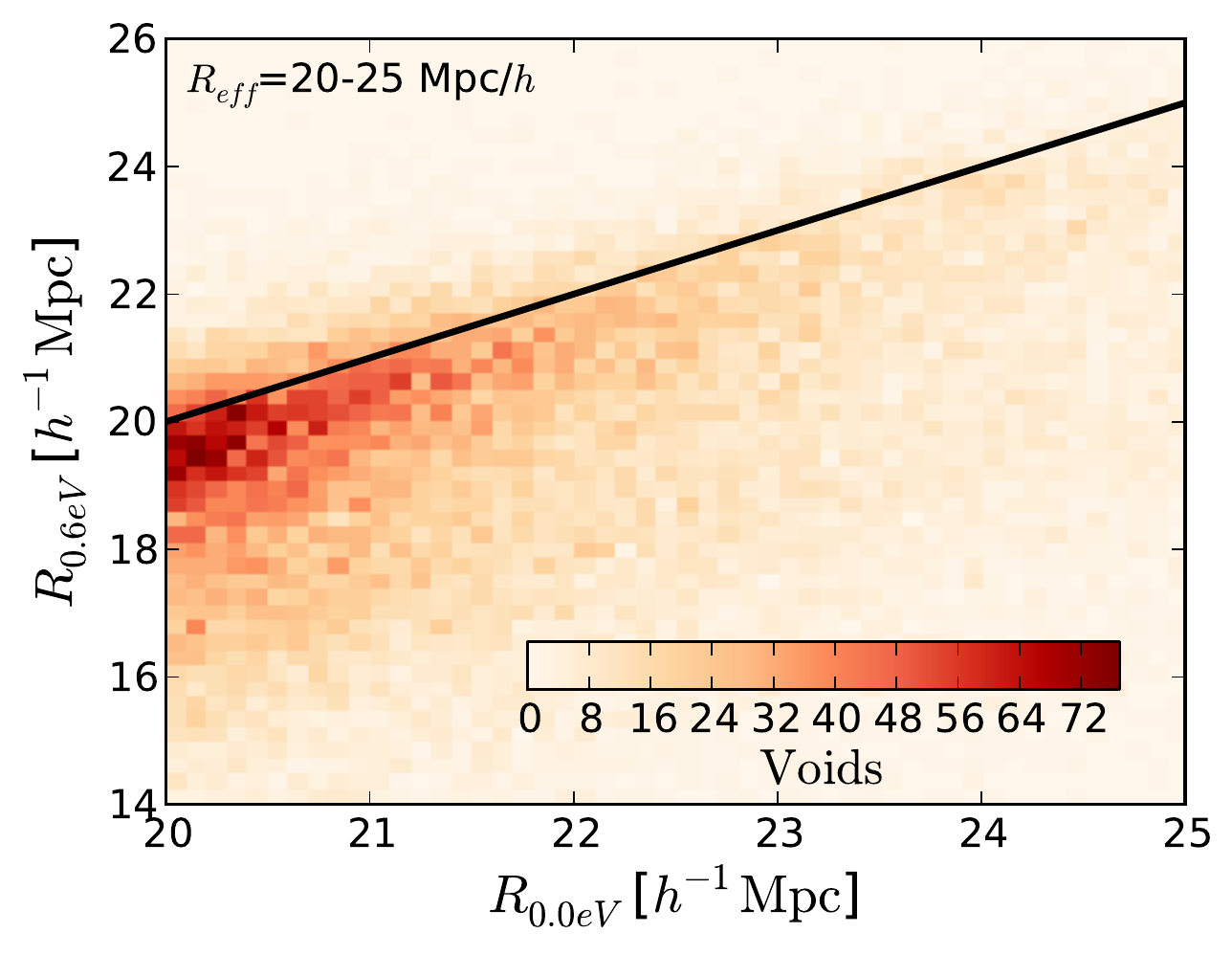}
\caption{\label{fig:samePos_dens_profile} Comparison void-by-void between the $\Lambda$CDM and the $\sum m_\nu=0.6~{\rm eV}$ cosmologies. We have selected all voids in the $\Lambda$CDM cosmology with radii $R_{\rm eff}$=16-18 Mpc/$h$ (left column) and $R_{\rm eff}$=20-25 Mpc/$h$ (right column) and we plot their density profiles in red. The upper panels show the CDM and neutrino density profiles whereas middle panels display the results for the total matter density profiles. For the 0.6 eV cosmology we have chosen the voids using two different procedures: we select all voids in above radii range (green lines) and we choose the voids by matching their positions to these of the $\Lambda$CDM cosmology (blue lines). The vertical dashed black lines indicate the mean value of the void radii in the selected range and two times the same quantity. The lower panels show the void-by-void comparison between the radii of the $\Lambda$CDM and the 0.6 eV voids. Each pixel in those panels displays the number of void pairs as a function of the void radius $R_{\rm 0.0eV}$ in the $\Lambda$CDM (x-axis) and the radius of the corresponding void $R_{\rm 0.6eV}$ in the 0.6 eV cosmology (y-axis). The solid black lines represent the curve $R_{\rm 0.6eV}=R_{\rm 0.0eV}$.}
\end{figure}

In section~\ref{sec:density_profile} we have compared the density profiles of voids with the same fixed size. It is also interesting, mainly for model building purposes, to study the differences among different cosmologies by performing a void-by-void comparison. The procedure we use to perform a void-by-void comparison is as follows. We first select all voids of a given size in the $\Lambda$CDM cosmology. Then, for each of the selected void, we find the corresponding one in a massive neutrino cosmology by searching the void whose center lie closest to the center of the $\Lambda$CDM void. Finally, we use these voids to compute the mean density profile in the massive neutrino cosmology. We have explicitly checked that this simple procedure works well when dealing with large voids, whereas for small voids the comparison is not always satisfactory. Therefore, we limit our analysis to big voids. In figure \ref{fig:samePos_dens_profile} we compare the density profiles for CDM, neutrinos and total matter for voids in $\Lambda$CDM and $\sum m_\nu=0.6~{\rm eV}$ cosmologies. For the latter cosmology we use both procedures, i.e. voids with radii within a given range (green lines) and voids selected by matching them with those of the $\Lambda$CDM cosmology (blue line).

The comparison between the density profiles of voids in the $\Lambda$CDM cosmology and the corresponding voids in the massive neutrino cosmology confirms that the presence of massive neutrinos makes the voids matter evacuation slower and therefore they are less empty in the center and they present a lower wall at the edge. The comparison between void profiles in the same cosmology ($\sum m_\nu=0.6~{\rm eV}$) but found using the two different ways (blue and green lines) is enlightening. Indeed, voids selected using the $\Lambda$CDM catalogue appear to be smaller than the one selected looking at the void size, since their slope is closer to the center. This is even more clear when plotting the void radii in the void-by-void selection for $\Lambda$CDM and $\sum m_\nu=0.6~{\rm eV}$ cosmology (lower row in figure~\ref{fig:samePos_dens_profile}). The black line indicates the equality between the radii in the two cosmologies and most of the points are below that line, which means that majority of voids in massive neutrino cosmology are smaller that the ones in the massless case. This is even stronger evidence of what we expected: the presence of massive neutrinos slow down the evolution (expansion) of voids.

\bibliographystyle{JHEP}
\bibliography{Bibliography}

\end{document}